\documentclass[prd,%
preprint,
tightenlines,superscriptaddress,showpacs,%
nofootinbib,eqsecnum,amsfonts,amsmath]{revtex4}
\usepackage{bm}

\allowdisplaybreaks

\begin{document}

\title{The third post-Newtonian gravitational wave polarisations and
associated spherical harmonic modes for
inspiralling \\ compact binaries in quasi-circular orbits}

\date{\today} \author{Luc Blanchet} \email
{blanchet@iap.fr}\affiliation{${\mathcal{G}}{\mathbb{R}}
\varepsilon{\mathbb{C}}{\mathcal{O}}$, Institut d'Astrophysique de Paris ---
C.N.R.S., 98$^{\text{bis}}$ boulevard Arago, 75014 Paris, France} \author
{Guillaume Faye}\email{faye@iap.fr}\affiliation{${\mathcal{G}}{\mathbb{R}}
\varepsilon{\mathbb{C}}{\mathcal{O}}$, Institut d'Astrophysique de Paris ---
C.N.R.S., 98$^{\text{bis}}$ boulevard Arago, 75014 Paris, France} \author{Bala
R. Iyer} \email{bri@rri.res.in} \affiliation{Raman Research Institute,
Bangalore 560 080, India} \author{Siddhartha Sinha}
\email{p_siddhartha@rri.res.in} \affiliation{Raman Research Institute,
Bangalore 560 080, India}\affiliation{Department of Physics, Indian Institute
of Science, Bangalore 560 012, India}

\begin{abstract}
The gravitational waveform (GWF) generated by inspiralling compact binaries
moving in quasi-circular orbits is computed at the third post-Newtonian (3PN)
approximation to general relativity. Our motivation is two-fold: (i) To
provide accurate templates for the data analysis of gravitational wave
inspiral signals in laser interferometric detectors; (ii) To provide the
associated spin-weighted spherical harmonic decomposition to facilitate
comparison and match of the high post-Newtonian prediction for the inspiral
waveform to the numerically-generated waveforms for the merger and ringdown.
This extension of the GWF by half a PN order (with respect to previous work at
2.5PN order) is based on the algorithm of the multipolar post-Minkowskian
formalism, and mandates the computation of the relations between the
radiative, canonical and source multipole moments for general sources at 3PN
order. We also obtain the 3PN extension of the source multipole moments in the
case of compact binaries, and compute the contributions of hereditary terms
(tails, tails-of-tails and memory integrals) up to 3PN order. The end results
are given for both the complete plus and cross polarizations and the separate
spin-weighted spherical harmonic modes.
\end{abstract}

 \preprint{} \maketitle

\widetext
\baselineskip22pt

\section{Introduction}\label{secI}

One of the most important sources of gravitational radiation for the laser
interferometric detectors LIGO, VIRGO \cite{ligo,virgo} and LISA \cite{lisa}
is the inspiralling and merging compact binary system. Until the late
inspiral, prior to merger, the gravitational waves are accurately described by
the post-Newtonian (PN) approximation to general relativity \cite{Bliving},
while the late inspiral and subsequent merger and ringdown phases are computed
by a full-fledged numerical integration of the Einstein field equations
\cite{Pretorius05,Baker,Campanelli,BCPZ}. A new field has emerged recently
consisting of high-accuracy comparisons between the PN predictions and the
numerically-generated waveforms. Such comparisons and matching to the PN
results have proved currently to be very successful
\cite{BCP07,Berti,Jena,Boyle}. They clearly show the need to include high PN
corrections not only for the evolution of the binary's orbital phase but also
for the modulation of the gravitational amplitude.

The aim of this paper is to compute the full gravitational waveform generated
by inspiralling compact binaries moving in quasi-circular orbits at the third
post-Newtonian (3PN) order\footnote{As usual, we refer to $n$PN as the order
equivalent to terms $\sim (v/c)^{2n}$ in the asymptotic waveform (beyond the
Einstein quadrupole formula), where $v$ denotes the binary's orbital velocity
and $c$ is the speed of light.}. By the full waveform (FWF) at a certain PN
order, we mean the waveform including all higher-order amplitude corrections
and hence all higher-order harmonics of the orbital frequency consistent with
that PN order. The FWF is to be contrasted with the so-called restricted
waveform (RWF) which retains only the leading-order harmonic at twice the
orbital frequency. In applications to data analysis both the FWF and RWF
should incorporate the orbital phase evolution up to the maximum available
post-Newtonian order which is currently 3.5PN
\cite{BIJ02,BFIJ02,BDEI04}. Previous investigations\cite{BIWW96,ABIQ04,KBI07}
have obtained the FWF up to 2.5PN order\footnote{The computation of the FWF is
more demanding than that of the phase because it not only requires multipole
moments with higher multipolarity but also higher PN accuracy in many of these
multipole moments. This is why the FWF is known to a lower PN order than the
phase.}. Recently Kidder \cite{K07} pointed out that there is already enough
information in the existing PN results \cite{ABIQ04} to control the
\textit{dominant} mode of the waveform, in a spin-weighted spherical harmonic
decomposition, at the 3PN order. This mode, having $(\ell,m)=(2,2)$, is the
one which is computed in most numerical simulations, and which is therefore
primarily needed for comparison with the PN waveforms. In the present paper we
shall extend the works \cite{BIWW96,ABIQ04,KBI07,K07} by computing all the
spin-weighted spherical harmonic modes $(\ell,m)$ consistent with the 3PN
gravitational polarisations.

The data analysis of ground-based and space-based detectors has traditionally
been based on the RWF approximation
\cite{3mn,CF94,SathyaFilter94,PW95,BalSatDhu96,DIS98,BCV03a}. However, the need to
consider the FWF as a more powerful template has been emphasized, not only for
performing a more accurate parameter estimation
\cite{HM03,MH02,SinVecc00a,SinVecc00b}, but also for improving the mass reach and the
detection rate \cite{Chris06,ChrisAnand06,AISS07}. Another motivation for considering
the FWF instead of the RWF is to perform cosmological measurements of the
Hubble parameter and dark energy using supermassive inspiralling black-hole
binaries which are known to constitute standard gravitational-wave candles (or
sirens) in cosmology \cite{HolzHugh05,DaHHJ06}. Indeed it has been shown that using the
FWF in the data analysis of LISA will yield substantial improvements (with
respect to the RWF) of the angular resolution and the estimation of the
luminosity distance of gravitational-wave sirens \cite{AISSV,Trias}. This means
that LISA may be able to uniquely identify the galaxy cluster in which the
supermassive black-hole coalescence took place, and thereby permit the
measurement of the red-shift of the source which is crucially needed for
investigating the equation of state of dark energy \cite{AISSV}.

It turns out that in order to control the FWF at the 3PN order we need to
further develop the multipolar post-Minkowskian (MPM) wave generation
formalism \cite{BD86,B87,BD92,B95,B96,B98mult}. The MPM formalism describes
the radiation field of any isolated post-Newtonian source and constitutes the
basis of current PN calculations\footnote{An alternative formalism called
DIRE has been developed by Will and collaborators \cite{WWi96,PW00,PW02}.}. In
this formalism, the radiation field is first of all parametrized by means of
two sets of radiative multipole moments \cite{Th80}. These moments are then
related (by means of an algorithm for solving the non-linearities of the field
equations) to the so-called canonical moments which constitute some useful
intermediaries for describing the external field of the source. Finally, the
canonical moments are expressed in terms of the operational source moments
which are given by explicit integrals extending over the matter source and
gravitational field. In previous studies \cite{BDI95,BIJ02,ABIQ04,BI04mult}
most of the required source moments in the case of compact binaries were
computed, or techniques were developed to compute them. The important step
which remains here is to refine, by applying the MPM framework, the
relationships between the radiative and canonical moments --- this means
taking into account more non-linear interactions between multipole moments ---
and between the canonical and source moments. The latter relationship involves
controlling the coordinate transformation between two MPM algorithms
respectively defined from the sets of canonical and source moments.

The plan of this paper is as follows. In Section \ref{secII} we recall the
basic formulas for defining the FWF in terms of radiative multipole
moments. Sections \ref{secIII} and \ref{secIV} apply the MPM formalism to
obtain general formulas for relating the radiative moments to the source
moments \textit{via} the canonical moments. Section \ref{secV} summarises the
results for all the relevant moments parametrizing the FWF at 3PN order. The
time derivatives of source moments are investigated in Section \ref{secVI} and
the various hereditary contributions are computed in Section \ref{secVII}.
The complete polarization waveforms at 3PN order are given in Section
\ref{secVIII} for data analysis applications. Finally, the spin-weighted
spherical harmonic modes of the 3PN waveform are provided in Section
\ref{secIX} for use in numerical relativity.

For the benefits of readers we provide in Appendix \ref{appA} a list of
symbols used in the paper together with their main meaning.

\section{The polarization waveforms}\label{secII}

The full waveform (FWF) propagating in the asymptotic regions of an isolated
source, $h_{ij}^\mathrm{TT}$, is the transverse-traceless (TT) projection of
the metric deviation at the leading-order $1/R$ in the distance
$R=\vert\mathbf{X}\vert$ to the source, in a radiative-type coordinate system
$X^\mu=(c\,T,\mathbf{X})$. The FWF can be uniquely decomposed \cite{Th80} into
radiative multipole components parametrized by symmetric-trace-free (STF)
mass-type moments $U_L$ and current-type ones $V_L$.\footnote{The notation
is: $L=i_1\cdots i_\ell$ for a multi-index composed of $\ell$ multipolar
spatial indices $i_1, \cdots, i_\ell$ (ranging from 1 to 3); similarly
$L-1=i_1\cdots i_{\ell-1}$ and $aL-2=a i_1\cdots i_{\ell-2}$; $N_L =
N_{i_1}\cdots N_{i_\ell}$ is the product of $\ell$ spatial vectors $N_i$
(similarly for $x_L = x_{i_1}\cdots x_{i_\ell}$); $\partial_L =
\partial_{i_1}\cdots \partial_{i_\ell}$ and say $\partial_{aL-2} =
\partial_{a}\partial_{i_1}\cdots \partial_{i_{\ell-2}}$ denote the product of
partial derivatives $\partial_i=\partial/\partial x^i$; in the case of
summed-up (dummy) multi-indices $L$, we do not write the $\ell$ summations
from 1 to 3 over their indices; the STF projection is indicated using
brackets, $T_{\langle L\rangle}=\mathrm{STF}[T_L]$; thus $U_L=U_{\langle
L\rangle}$ and $V_L=V_{\langle L\rangle}$ for STF moments; for instance we
write $x_{\langle i}v_{j\rangle}=\frac{1}{2}(x_iv_j+x_jv_i)
-\frac{1}{3}\delta_{ij}\mathbf{x}\cdot\mathbf{v}$; $\varepsilon_{abc}$ is the
Levi-Civita antisymmetric symbol such that $\varepsilon_{123}=1$; time
derivatives are denoted with a superscript $(n)$.}  The radiative moments are
functions of the retarded time $T_R=T-R/c$ in radiative coordinates. By
definition we have, up to any multipolar order $\ell$,
\begin{align}\label{hijTT}
h^\mathrm{TT}_{ij} &= \frac{4G}{c^2R} \,\mathcal{P}^\mathrm{TT}_{ijkl}
(\mathbf{N}) \sum^{+\infty}_{\ell=2}\frac{1}{c^\ell \ell !} \biggl\{ N_{L-2}
\,U_{klL-2}(T_R) - \frac{2\ell}{c(\ell+1)} \,N_{aL-2} \,\varepsilon_{ab(k}
\,V_{l)bL-2}(T_R)\biggr\} \nonumber\\&+
\mathcal{O}\left(\frac{1}{R^2}\right)\,.
\end{align}
Here $\mathbf{N} = \mathbf{X}/R = (N_i)$ is the unit vector pointing from the
source to the far away detector. The TT projection operator in \eqref{hijTT}
reads $\mathcal{P}^\mathrm{TT}_{ijkl} =
\mathcal{P}_{ik}\mathcal{P}_{jl}-\frac{1}{2}\mathcal{P}_{ij}\mathcal{P}_{kl}$
where $\mathcal{P}_{ij}=\delta_{ij}-N_iN_j$ is the projector orthogonal to the
unit direction $\mathbf{N}$. We introduce two unit polarisation vectors
$\mathbf{P}$ and $\mathbf{Q}$, orthogonal and transverse to the direction of
propagation $\mathbf{N}$ (hence $\mathcal{P}_{ij}=P_iP_j+Q_iQ_j$). Our
convention for the choice of $\mathbf{P}$ and $\mathbf{Q}$ will be clarified
in Section \ref{secVIII}. Then the two ``plus'' and ``cross'' polarisation
states of the FWF are defined by
\begin{align}\label{hpc}
\left(\begin{array}{l}h_+\\[0.5cm]h_\times
\end{array}\right) &= \frac{4G}{c^2R}
\left(\begin{array}{l}\frac{P_iP_j-Q_iQ_j}{2}\\[0.5cm]\frac{P_iQ_j+P_jQ_i}{2}
\end{array}\right)\sum^{+\infty}_{\ell=2}\frac{1}{c^\ell \ell !} \biggl\{ N_{L-2}
U_{ijL-2}(T_R) - \frac{2\ell}{c(\ell+1)} N_{aL-2} \varepsilon_{ab(i}
V_{j)bL-2}(T_R)\biggr\}\nonumber\\&+ \mathcal{O}\left(\frac{1}{R^2}\right)\,.
\end{align}

Although the multipole decompositions \eqref{hijTT} and \eqref{hpc} are all
what we need for our purpose, it will also be important, having in view the
ongoing comparisons between the PN and numerical results
\cite{BCP07,Berti,Jena,Boyle}, to consider separately the various modes
$(\ell,m)$ of the FWF as defined with respect to a basis of spin-weighted
spherical harmonics. To this end we decompose $h_+$ and $h_\times$ in the
standard way as (see \textit{e.g.} \cite{BCP07,K07})
\begin{equation}\label{spinw}
h_+ - i h_\times = \sum^{+\infty}_{\ell=2}\sum^{\ell}_{m=-\ell} h^{\ell
m}\,Y^{\ell m}_{-2}(\Theta,\Phi)\,,
\end{equation}
where the spin-weighted spherical harmonics of weight $-2$ is function of the
spherical angles $(\Theta,\Phi)$ defining the direction of propagation
$\mathbf{N}$,\footnote{For the data analysis of compact binaries in Section
\ref{secVIII} the direction of propagation will be defined by the angles
$(\Theta,\Phi)=(i,\frac{\pi}{2})$ where $i$ is the inclination angle of the
orbit over the plane of the sky.} and is given by
\begin{subequations}\label{harm}\begin{align}
Y^{\ell m}_{-2} &= \sqrt{\frac{2\ell+1}{4\pi}}\,d^{\,\ell
m}_{\,2}(\Theta)\,e^{i \,m \,\Phi}\,,\\d^{\,\ell m}_{\,2} &=
\sum_{k=k_1}^{k_2}\frac{(-)^k}{k!}
\frac{\sqrt{(\ell+m)!(\ell-m)!(\ell+2)!(\ell-2)!}}
{(k-m+2)!(\ell+m-k)!(\ell-k-2)!}\left(\cos\frac{\Theta}{2}\right)^{2\ell+m-2k-2}
\!\!\!\left(\sin\frac{\Theta}{2}\right)^{2k-m+2}\,.
\end{align}\end{subequations}
Here $k_1=\mathrm{max}(0,m-2)$ and $k_2=\mathrm{min}(\ell+m,\ell-2)$. Using
the orthonormality properties of these harmonics we obtain the separate modes
$h^{\ell m}$ from the surface integral
\begin{equation}\label{decomp}
h^{\ell m} = \int d\Omega \,\Bigl[h_+ - i h_\times\Bigr] \,\overline{Y}^{\,\ell
m}_{-2} (\Theta,\Phi)\,,
\end{equation}
where the bar or overline denotes the complex conjugate. On the other hand, we
can also, following \cite{K07}, relate $h^{\ell m}$ directly to the multipole
moments $U_L$ and $V_L$. The result is\footnote{We have an overall sign
difference with \cite{K07} due to a different choice for the polarization
triad $(\mathbf{N},\mathbf{P},\mathbf{Q}$).}
\begin{equation}\label{inv}
h^{\ell m} = -\frac{G}{\sqrt{2}\,R\,c^{\ell+2}}\left[U^{\ell
m}-\frac{i}{c}V^{\ell m}\right]\,,
\end{equation}
where $U^{\ell m}$ and $V^{\ell m}$ are the radiative mass and current moments
in standard (non-STF) guise \cite{K07}.  These are related to the STF moments
by
\begin{subequations}\label{UV}\begin{align}
U^{\ell m} &= \frac{4}{\ell!}\,\sqrt{\frac{(\ell+1)(\ell+2)}{2\ell(\ell-1)}}
\,\alpha_L^{\ell m}\,U_L\,,\\ V^{\ell m} &=
-\frac{8}{\ell!}\,\sqrt{\frac{\ell(\ell+2)}{2(\ell+1)(\ell-1)}}
\,\alpha_L^{\ell m}\,V_L\,.
\end{align}\end{subequations}
Here $\alpha_L^{\ell m}$ denotes the STF tensor connecting together the usual
basis of spherical harmonics $Y^{\ell m}$ to the set of STF tensors
$N_{\langle L\rangle}=N_{\langle i_1}\cdots N_{i_\ell\rangle}$ (where the
brackets indicate the STF projection). Indeed both $Y^{\ell m}$ and
$N_{\langle L\rangle}$ are basis of an irreducible representation of weight
$\ell$ of the rotation group. They are related by
\begin{subequations}\label{NY}\begin{align}
N_{\langle L\rangle}(\Theta,\Phi) &= \sum_{m=-\ell}^{\ell}
\alpha_L^{\ell m}\,Y^{\ell m}(\Theta,\Phi)\,,\\Y^{\ell m}(\Theta,\Phi) &=
\frac{(2\ell+1)!!}{4\pi \ell!}\,\overline{\alpha}_L^{\ell m}\,N_{\langle
L\rangle}(\Theta,\Phi)\,,
\end{align}\end{subequations}
with the STF tensorial coefficient being\footnote{The notation used in
\cite{Th80,K07} is related to ours by $\mathcal{Y}_L^{\ell
m}=\frac{(2\ell+1)!!}{4\pi \ell!}\,\overline{\alpha}_L^{\ell m}$.}
\begin{align}\label{alpha}
\alpha_L^{\ell m} &= \int d\Omega\,N_{\langle L\rangle}\,\overline{Y}^{\,\ell
m}\,.
\end{align}
As observed in \cite{K07} this is especially useful if some of the radiative
moments are known to higher PN order than others. In this case the comparison
with the numerical calculation for these individual modes can be made at
higher PN accuracy.

\section{Relation between the radiative and canonical 
moments} \label{secIII}

The basis of our computation of the radiative moments is the
multipolar-post-Minkowskian (MPM) formalism
\cite{BD86,B87,BD92,B95,B96,B98mult} which iterates the general solution of
the Einstein field equations outside an isolated matter system in the form of
a post-Minkowskian or non-linearity expansion. The formalism is then
supplemented by a matching to the PN gravitational field valid in the near
zone of the source. In this Section and the next one we sketch the main
features of the MPM iteration of the exterior field while limiting ourselves
to quadratic non-linear order because this is what we need for the new terms
required in the FWF at 3PN order\footnote{Cubic non-linearities do
contribute at the 3PN order in the form of ``tail-of-tails'' but those have
already been computed \cite{B98tail}.}. We shall work with harmonic
coordinates $x^\mu=(c\,t,\mathbf{x})$, which means that
\begin{equation}\label{dh}
\partial_\mu h^{\alpha\mu}=0
\,,\end{equation}
where the ``gothic'' metric deviation reads
$h^{\alpha\beta}=\sqrt{-g}\,g^{\alpha\beta}-\eta^{\alpha\beta}$, with $g$ the
determinant and $g^{\alpha\beta}$ the inverse of the usual covariant metric,
and with $\eta^{\alpha\beta}=\mathrm{diag}(-1,1,1,1)$ being an auxiliary
Minkowskian metric\footnote{Beware of the fact that the TT waveform defined
by \eqref{hijTT} differs by a sign from the spatial components of the gothic
metric deviation,
$h^\mathrm{TT}_{ij}=-\mathcal{P}^\mathrm{TT}_{ijkl}h_{kl}+\mathcal{O}(h^2)$.}.
Up to quadratic non-linear order the vacuum Einstein field equations take the
form
\begin{equation}\label{EE}
\Box h^{\alpha\beta} = N_2^{\alpha\beta}(h) + \mathcal{O}(h^3)
\,,\end{equation}
where $\Box=\eta^{\mu\nu}\partial_\mu\partial_\nu$ is the flat space-time
d'Alembertian operator, and where $N_2^{\alpha\beta}$ denotes the quadratic
part of the gravitational source term in harmonic coordinates --- a quadratic
functional of $h$ and its first and second space-time derivatives given
explicitly by
\begin{align}\label{N2}
N_2^{\alpha\beta}(h) =& - h^{\mu\nu} \partial_\mu \partial_\nu h^{\alpha\beta}
+ {1\over 2} \partial^\alpha h_{\mu\nu} \partial^\beta h^{\mu\nu} - {1\over 4}
\partial^\alpha h \partial^\beta h \nonumber\\ & -2 \partial^{(\alpha}
h_{\mu\nu} \partial^\mu h^{\beta)\nu} +\partial_\nu h^{\alpha\mu}
(\partial^\nu h^\beta_\mu + \partial_\mu h^{\beta\nu}) \nonumber \\ & +
\eta^{\alpha\beta} \biggl[ -{1\over 4}\partial_\rho h_{\mu\nu} \partial^\rho
h^{\mu\nu} +{1\over 8}\partial_\mu h \partial^\mu h +{1\over 2}\partial_\mu
h_{\nu\rho} \partial^\nu h^{\mu\rho}\biggr] \,,
\end{align} 
with $h=\eta^{\mu\nu}h_{\mu\nu}$. The four-divergence of this source term
reads
\begin{equation}\label{dN2}
\partial_\mu N_2^{\alpha\mu} = - {1\over 4} \partial^\alpha h \,\Box h + {1\over
2}\Bigl[\partial^\alpha h_{\mu\nu} - 2 \partial_\mu h^{\alpha}_\nu \Bigr]
\Box h^{\mu\nu}
\,.\end{equation}

In this paper we shall consider two explicit constructions of the
quadratic-order external metric following the MPM formalism. The first
construction, dealt with in this Section, will be parametrized by two (and
only two) sets of moments, mass moments $M_L$ and current moments $S_L$, which
are referred to as the \textit{canonical multipole moments}. The canonical
moments are crucially distinct from the radiative moments $U_L$ and $V_L$, and
the MPM construction will provide the relations linking them to $U_L$,
$V_L$. The second construction (in Section \ref{secIV}) will deal with the
link between $M_L$, $S_L$ and \textit{six} sets of moments $I_L$, $J_L$,
$W_L$, $X_L$, $Y_L$ and $Z_L$ collectively named the \textit{source multipole
moments}. Among these, the moments $I_L$ (mass-type) and $J_L$ (current-type)
play the most important role, while for reasons explained below the other
moments $W_L$, $X_L$, $Y_L$ and $Z_L$ are called the \textit{gauge multipole
moments} and will appear to be subdominant.

Armed with such definitions, the computation of the radiative field
\eqref{hijTT}--\eqref{hpc} proceeds in a modular way (see Section 6 of
\cite{B98mult} for further discussion). We start with relating the radiative
moments $\{U_L, V_L\}$ to the canonical moments $\{M_L, S_L\}$ which are to be
viewed as convenient intermediate constructs relating the radiation field and
the matter source. The canonical moments are then in turn connected to the
actual multipole moments of the source $\{I_L, J_L, W_L, X_L, Y_L, Z_L\}$. The
point of the above strategy is that the source moments admit closed-form
expressions as integrals over the stress-energy distribution of the matter and
gravitational fields. The expressions of $I_L, \cdots, Z_L$ for general
sources are given by (5.15)--(5.20) in \cite{B98mult} and shall not be
reproduced here.\footnote{Below we give the source moments needed at the 3PN
order in a form already reduced to the case of compact binaries in circular
orbits.} Note that the above formalism can be applied only to PN sources,
which remain confined in their own near zone; the final expressions of the
source moments are valid only for sources that are semi-relativistic like
inspiralling compact binaries.

Consider the so-called ``canonical'' construction of the MPM metric in
harmonic coordinates, designated that way because it is based on Thorne's
\cite{Th80} canonical expression for the linearized approximation
$h_{\mathrm{can}\,1}^{\alpha\beta}$ [given by \eqref{hcan1} below]. The MPM
metric is parametrized by the canonical multipole moments $M_L$ and $S_L$ and
reads, to quadratic order,
\begin{equation}\label{hcan}
h_{\mathrm{can}}^{\alpha\beta} = G h_{\mathrm{can}\,1}^{\alpha\beta}[M_L,S_L]
+ G^2 h_{\mathrm{can}\,2}^{\alpha\beta}[M_L,S_L] + \mathcal{O}(G^3)\,,
\end{equation}
where the Newton constant $G$ is introduced as a convenient book-keeping
parameter for labeling the successive non-linear approximations. From
\eqref{dh}--\eqref{EE} the linearized approximation
$h_{\mathrm{can}\,1}^{\alpha\beta}$ obviously satisfies $\partial_\mu
h_{\mathrm{can}\,1}^{\alpha\mu}=0$ together with $\Box
h_{\mathrm{can}\,1}^{\alpha\beta}=0$. Following \cite{BD86,B87} we adopt the
following explicit retarded solution of these equations,
\begin{subequations}\label{hcan1}\begin{align}
h^{00}_{\mathrm{can}\,1} &= -\frac{4}{c^2}\sum_{\ell = 0}^{\infty}
\frac{(-)^\ell}{\ell !} \partial_L \left[ r^{-1} M_L (t-r/c)\right] \,,\\
h^{0i}_{\mathrm{can} 1} &= \frac{4}{c^3}\sum_{\ell = 1}^{\infty} \frac{(-)^\ell}{\ell
!}  \left\{ \partial_{L-1} \left[ r^{-1} M_{iL-1}^{(1)} (t-r/c)\right] +
\frac{\ell}{\ell+1} \varepsilon_{iab} \partial_{aL-1} \left[ r^{-1} S_{bL-1}
(t-r/c)\right]\right\} \,,\\ h^{ij}_{\mathrm{can}\,1} &=
-\frac{4}{c^4}\sum_{\ell = 2}^{\infty} \frac{(-)^\ell}{\ell !} \left\{ \partial_{L-2}
\left[ r^{-1} M_{ijL-2}^{(2)} (t-r/c)\right] + \frac{2\ell}{\ell+1}
\partial_{aL-2} \left[ r^{-1} \varepsilon_{ab(i} S_{j)bL-2}^{(1)}
(t-r/c)\right]\right\}\,,
\end{align}\end{subequations}
with $F^{(n)}(t)$ denoting $n$ time derivatives of $F(t)$. These expressions
represent the most general solution of the vacuum linearized field equations
modulo a change of gauge \cite{Th80}.

Next, the quadratically non-linear term $h_{\mathrm{can}\,2}^{\alpha\beta}$
--- and subsequently all non-linear terms $h_{\mathrm{can}\,n}^{\alpha\beta}$
--- is constructed by the following algorithm. We first define
\begin{equation}\label{u2}
u^{\alpha\beta}_{\mathrm{can}\,2} = \mathop{\mathrm{FP}}_{B=0}\,
\Box^{-1}_\mathrm{ret} \biggl[ \left(\frac{r}{r_0}\right)^B
N_2^{\alpha\beta}(h_{\mathrm{can}\,1}) \biggr]\,,
\end{equation}
where $\Box^{-1}_\mathrm{ret}$ represents the ordinary (flat) retarded
integral operator acting on the source
$N_2^{\alpha\beta}(h_{\mathrm{can}\,1})$ which is obtained by insertion of the
linearized metric \eqref{hcan1} into the quadratic source term given by
\eqref{N2}. The symbol $\mathop{\mathrm{FP}}_{B=0}$ refers to a specific
operation of taking the finite part when the complex number $B$ tends to
zero. Such a finite part involves the multiplication of the source term by a
regularization factor $(r/r_0)^B$, where $r_0$ represents an arbitrary
constant length scale (and $B\in\mathbb{C}$). The finite part is necessary for
dealing with multipolar expansions which are singular at the origin $r=0$
[like in \eqref{hcan1}]. It will not be further detailed here and we refer to
\cite{BD86,B95,B98mult} for full details. The point is that the object
\eqref{u2} obeys the d'Alembertian equation we want to solve, namely
\begin{equation}\label{boxu2}
\Box u^{\alpha\beta}_{\mathrm{can}\,2} =
N_2^{\alpha\beta}(h_{\mathrm{can}\,1})\,.
\end{equation}
However, such a solution is \textit{a priori} not divergenceless and so the
harmonic coordinate condition needs not to be satisfied. To obtain a solution
which is divergenceless we add to $u^{\alpha\beta}_{\mathrm{can}\,2}$ another
piece $v^{\alpha\beta}_{\mathrm{can}\,2}$ defined as follows. Computing the
divergence
$w^{\alpha}_{\mathrm{can}\,2}=\partial_{\mu}u^{\alpha\mu}_{\mathrm{can}\,2}$,
we readily find
\begin{equation}\label{du2}
w^{\alpha}_{\mathrm{can}\,2} = \mathop{\mathrm{FP}}_{B=0}\,
\Box^{-1}_\mathrm{ret} \biggl[ B\left(\frac{r}{r_0}\right)^B \frac{n_i}{r}
\,N_2^{\alpha i}(h_{\mathrm{can}\,1}) \biggr]\,,
\end{equation}
where we used the fact that the source term of \eqref{boxu2}, 
as an immediate consequence of \eqref{dN2},
is divergenceless, $\partial_{\beta}N^{\alpha\beta}_{\mathrm{can}\,2}=0$. 
Again, because the source term is
divergenceless, the divergence $w^{\alpha}_{\mathrm{can}\,2}$ must be a
(retarded) solution of the source-free d'Alembertian equation, $\Box
w^{\alpha}_{\mathrm{can}\,2}=0$. This can also be checked from the fact that
there is a factor $B$ explicit in the source of \eqref{du2} (appearing because
of the differentiation of the regularization factor $r^B$), and therefore the
finite part at $B=0$ is actually equal to the residue in the Laurent expansion
when $B\rightarrow 0$, and is necessarily a retarded solution of the
source-free equation \cite{BD86}.

Given any vector of the type $w^{\alpha}_{\mathrm{can}\,2}$, \textit{i.e.}
one which is of the form of a retarded solution of the d'Alembertian equation,
we can always find four sets of STF tensors $N_L$, $P_L$, $Q_L$ and $R_L$ such
that the following decomposition holds,
\begin{subequations}\label{w2}\begin{align}
w^0_{\mathrm{can}\,2} &= \sum_{\ell = 0}^{+\infty}\partial_L \left[r^{-1}
N_L(t-r/c)\right]\,,\\ w^i_{\mathrm{can}\,2} &= \sum_{\ell =
0}^{+\infty}\partial_{iL} \left[ r^{-1} P_L (t-r/c) \right] \nonumber\\&+
\sum_{\ell = 1}^{+\infty} \left\{ \partial_{L-1} \left[ r^{-1} Q_{iL-1}
(t-r/c) \right] + \varepsilon_{iab} \partial_{aL-1} \left[r^{-1} R_{bL-1}
(t-r/c) \right] \right\}\,.
\end{align}\end{subequations}
From this decomposition (which is unique) we define the object
$v^{\alpha\beta}_{\mathrm{can}\,2}$ by the formulas\footnote{We are adopting
here a modified version of the MPM algorithm (with respect to \cite{BD86}) as
proposed in \cite{B98quad}.}
\begin{subequations}\label{v2}\begin{align}
v^{00}_{\mathrm{can}\,2} &= - r^{-1} N^{(-1)} + \partial_a \left[ r^{-1}
\left(- N^{(-1)}_a+ Q^{(-2)}_a -3P_a\right) \right] \,,\\
v^{0i}_{\mathrm{can}\,2} &= r^{-1} \left( - Q^{(-1)}_i +3 P^{(1)}_i\right) -
\varepsilon_{iab} \partial_a \left[ r^{-1} R^{(-1)}_b \right] - \sum_{\ell =
2}^{+\infty}\partial_{L-1} \left[ r^{-1} N_{iL-1} \right] \,,\\
v^{ij}_{\mathrm{can}\,2} &= - \delta_{ij} r^{-1} P + \sum_{\ell = 2}^{+\infty}
\biggl\{ 2 \delta_{ij}\partial_{L-1} \left[ r^{-1} P_{L-1}\right] - 6
\partial_{L-2(i} \left[ r^{-1} P_{j)L-2}\right] \nonumber\\ & \quad +
\partial_{L-2} \left[ r^{-1} (N^{(1)}_{ijL-2} + 3 P^{(2)}_{ijL-2} - Q_{ijL-2})
\right] - 2 \partial_{aL-2}\left[ r^{-1} \varepsilon_{ab(i} R_{j)bL-2} \right]
\biggr\}\,.
\end{align}\end{subequations}
The superscript $(-p)$ denotes the time anti-derivatives (\textit{i.e.} time
integrals) of the moments. Such anti-derivatives yield some secular losses of
mass and momenta by gravitational radiation which have been checked to agree
with the corresponding gravitational radiation fluxes, see \textit{e.g.}
(4.12) in \cite{B98quad}. The formulas \eqref{v2} have been conceived in such
a way that the divergence of the so defined
$v^{\alpha\beta}_{\mathrm{can}\,2}$ cancels out the divergence of
$u^{\alpha\beta}_{\mathrm{can}\,2}$ which is
$w^{\alpha}_{\mathrm{can}\,2}$. In the following we shall denote by
$\mathcal{V}^{\alpha\beta}$ the operation for going from a vector such as
\eqref{w2} --- a retarded solution of the source-free wave equation --- to the
tensor \eqref{v2}. We therefore pose
\begin{equation}\label{v2p}
v^{\alpha\beta}_{\mathrm{can}\,2} =
\mathcal{V}^{\alpha\beta}\left[w_{\mathrm{can}\,2}\right]\,,
\end{equation}
and as mentioned before this tensor immediately satisfies $\Box
v^{\alpha\beta}_{\mathrm{can}\,2}=0$ (which is obvious) and also
\begin{equation}\label{dv2p}
\partial_\mu v^{\alpha\mu}_{\mathrm{can}\,2} =
-w^{\alpha}_{\mathrm{can}\,2}\,.
\end{equation}
This property can be directly checked from \eqref{v2} and \eqref{w2}. Finally,
it is clear from \eqref{boxu2} and \eqref{dv2p} that by posing
\begin{equation}\label{hcan2}
h^{\alpha\beta}_{\mathrm{can}\,2} = u^{\alpha\beta}_{\mathrm{can}\,2} +
v^{\alpha\beta}_{\mathrm{can}\,2}\,,
\end{equation}
we solve the Einstein vacuum field equations at quadratic order, namely
\begin{subequations}\label{EEquad}\begin{align}
\Box h^{\alpha\beta}_{\mathrm{can}\,2} &= N^{\alpha\beta}_{\mathrm{can}\,2}\,,\\
\partial_\mu h^{\alpha\mu}_{\mathrm{can}\,2} &= 0
\,.\end{align}\end{subequations} 
The MPM algorithm can be extended to any post-Minkowskian order $n$.

The structure of the quadratic metric $h^{\alpha\beta}_{\mathrm{can}\,2}$ so
constructed has been investigated in previous works \cite{BD92,B98quad}. It
consists of two types of terms: those which depend on the source moments at a
single instant, namely the current retarded time $t-r/c$, referred to as
\textit{instantaneous} terms, and the other ones which are sensitive to the
entire ``past history'' of the source, \textit{i.e.} which depend on all
previous times ($\tau \leq t-r/c$), and are referred to as the
\textit{hereditary} terms. The hereditary terms are themselves composed of
three types of contributions, the tail integrals --- made from interaction
between the mass of the source $M$ and the time-varying moments $M_L$ and
$S_L$ (having $\ell\geq 2$) ---, the memory integrals responsible for the
so-called ``non-linear memory'' or Christodoulou effect \cite{Chr91,WiW91,Th92}
(investigated within the present approach in \cite{BD92,B98quad}), and
semi-hereditary integrals which are in the form of simple anti-derivatives of
instantaneous terms and are associated with the secular variations of the
mass, linear momentum and angular momentum. The semi-hereditary integrals are
given by the time anti-derivatives present in the formula \eqref{v2}.

To obtain the radiative moments we expand the metric at future null infinity
in a radiative coordinate system $X^\mu=(c\,T,X^i)$, which is such that the
metric admits an expansion in simple powers of $1/R$ without the logarithms
which plague the harmonic coordinate system $x^\mu=(c\,t,x^i)$ \cite{B87}. Up
to quadratic order and for all multipole interactions we consider, we find
that it is sufficient to define for the radiative coordinates $X^i=x^i$ and
(denoting $T_R=T-R/c$)
\begin{equation}\label{TR}
T_R = t- \frac{r}{c} -\frac{2 G M}{c^3}\ln\left(\frac{r}{r_0}\right) +
\mathcal{O}(G^2)\,,
\end{equation}
where $r_0$ is the length scale introduced in \eqref{u2}. Expanding the metric
when $R\rightarrow\infty$ with $T_R=\mathrm{const}$, and applying the TT
projection we obtain the radiative moments $U_L$ and $V_L$ we are seeking by
comparing with their definition in \eqref{hijTT}. At linear order the
radiative moments agree with the $\ell$-th time derivatives of the canonical
moments, $M_L^{(\ell)}$ and $S_L^{(\ell)}$. At quadratic order we find that
tail and non-linear memory terms appear; these have already been investigated
in \cite{BD92,B98quad,B98tail}.\footnote{The semi-hereditary integrals
associated with secular gravitational radiation losses do not contribute to
the radiative moments.} Their general structure will also be given in
\eqref{eq:deltaUV} below. Finally, we have numerous instantaneous terms whose
determination necessitates the straightforward but long implementation of the
MPM algorithm \eqref{u2}--\eqref{hcan2}. This is the work required here: we
have implemented the MPM algorithm in a Mathematica program to obtain all the
instantaneous terms needed to control the 3PN waveform. The presentation of
the results is postponed to Section \ref{secVA}.

\section{Relation between the canonical and source moments}
 \label{secIV}

\subsection{General method}\label{secIVA}

We next need to connect the canonical moments $\{M_L,S_L\}$ to a convenient
choice of moments that are suitably defined to play the role of source
moments. As it turns out, the source moments are best represented by six
multipole moments $\{I_L,J_L,W_L,X_L,Y_L,Z_L\}$ admitting closed-form
expressions in the form of integrals over the source and the gravitational
field. To define them we consider a MPM construction which is more general
than the one given by \eqref{hcan}, namely (still up to quadratic order)
\begin{equation}\label{hgen}
h_{\mathrm{gen}}^{\alpha\beta} = G
h_{\mathrm{gen}\,1}^{\alpha\beta}[I_L,J_L,W_L,\cdots,Z_L] + G^2
h_{\mathrm{gen}\,2}^{\alpha\beta}[I_L,J_L,W_L,\cdots,Z_L] +
\mathcal{O}(G^3)\,,
\end{equation}
where the linearized metric $h_{\mathrm{gen}\,1}^{\alpha\beta}$ is defined by
the canonical expression $h_{\mathrm{can}\,1}^{\alpha\beta}$ explicitly given
in \eqref{hcan1} but parametrized by $\{I_L,J_L\}$ instead of $\{M_L,S_L\}$,
and augmented by a linearized \textit{gauge transformation} associated with
some vector $\varphi_1^{\alpha}$ parametrized by the remaining moments
$\{W_L,X_L,Y_L,Z_L\}$ which can thus rightly be called the gauge
moments. Thus,
\begin{equation}\label{hgen1}
h_{\mathrm{gen}\,1}^{\alpha\beta} = h_{\mathrm{can}\,1}^{\alpha\beta}[I_L,J_L]
+ \partial\varphi_1^{\alpha\beta}[W_L,X_L,Y_L,Z_L]\,,
\end{equation}
where for any vector $\varphi_1^{\alpha}$ we denote the gauge transformation
by
\begin{equation}\label{gaugetransf}
\partial\varphi_1^{\alpha\beta} =\partial^{\alpha}\varphi_1^{\beta}
+\partial^{\beta}\varphi_1^{\alpha}
-\eta^{\alpha\beta}\partial_{\mu}\varphi_1^{\mu}\,.
\end{equation}
Note that
$\partial_\mu\partial\varphi_1^{\alpha\mu}=\Box\varphi_1^{\alpha}$. The
expression of $\varphi_1^{\alpha}$ in terms of the gauge moments is
\begin{subequations}\label{phi1}\begin{align}
\varphi^0_1 =& {4\over c^3}\sum_{\ell\geq 0} {(-)^\ell\over \ell !} \partial_L
\left[ r^{-1} W_L (t-r/c)\right] \,,\\ \varphi^i_1 =& -{4\over
c^4}\sum_{\ell\geq 0} {(-)^\ell\over \ell !} \partial_{iL} \left[ r^{-1} X_L
(t-r/c)\right]\\ & -{4\over c^4}\sum_{\ell\geq 1} {(-)^\ell\over \ell !}
\left\{ \partial_{L-1} \left[ r^{-1} Y_{iL-1} (t-r/c)\right] + {\ell\over
\ell+1} \varepsilon_{iab} \partial_{aL-1} \left[ r^{-1} Z_{bL-1}
(t-r/c)\right]\right\}\,.
\end{align}\end{subequations}
The quadratic metric $h_{\mathrm{gen}\,2}^{\alpha\beta}$ will now be defined
by the same algorithm as for the canonical metric in Section \ref{secIII} but
starting from the general linearized metric \eqref{hgen1}. The result will be
another MPM metric (both the canonical and general metrics are legitimate to
describe the exterior field of any isolated matter source \cite{BD86}) and we
shall look for the relation between $\{M_L,S_L\}$ and
$\{I_L,J_L,W_L,X_L,Y_L,Z_L\}$ which is necessary in order that these two
metrics differ by a coordinate transformation (at quadratic order), and
therefore describe the same physical matter source.

To proceed, we have to define
\begin{equation}\label{ugen2}
u^{\alpha\beta}_{\mathrm{gen}\,2} = \mathop{\mathrm{FP}}_{B=0}\,
\Box^{-1}_\mathrm{ret} \biggl[ \left(\frac{r}{r_0}\right)^B
N_2^{\alpha\beta}(h_{\mathrm{gen}\,1}) \biggr]\,.
\end{equation}
The only difference with \eqref{u2} is that the quadratic source \eqref{N2} is
computed from $h_{\mathrm{gen}\,1}^{\alpha\beta}$ instead of
$h_{\mathrm{can}\,1}^{\alpha\beta}$. However, since the two linear metrics
$h_{\mathrm{gen}\,1}^{\alpha\beta}$ and $h_{\mathrm{can}\,1}^{\alpha\beta}$
differ by the gauge transformation \eqref{gaugetransf} the difference between
the corresponding sources must have a specific structure, and we find
\begin{equation}\label{diffN2}
N_2^{\alpha\beta}(h_{\mathrm{gen}\,1}) =
N_2^{\alpha\beta}(h_{\mathrm{can}\,1}) + \Box\Omega_2^{\alpha\beta} +
\partial\Delta_2^{\alpha\beta}\,.
\end{equation}
We employ the notation \eqref{gaugetransf} for the gauge term
$\partial\Delta_2^{\alpha\beta}$. The expressions of the tensor
$\Omega_2^{\alpha\beta}$ and vector $\Delta_2^{\alpha}$ are determined with
the help of \eqref{N2} and read
\begin{subequations}\label{OmDelta}\begin{align}
\Omega_2^{\alpha\beta} =&
-\partial_\mu\left(\varphi_1^\mu\left[h_{\mathrm{can}\,1}^{\alpha\beta} +
  \partial\varphi_1^{\alpha\beta}\right]\right) + 2
\partial_\mu\varphi_1^{(\alpha} h_{\mathrm{can}\,1}^{\beta)\mu}\nonumber\\ &+
\partial_\mu\varphi_1^{\alpha}\partial^\mu\varphi_1^{\beta}
+\frac{1}{2}\eta^{\alpha\beta}\Bigl[\partial_\mu\varphi_1^{\nu}\partial_\nu
  \varphi_1^\mu
  -\partial_\mu\varphi_1^\mu\partial_\nu\varphi_1^\nu\Bigr]\,,\label{Om2}\\
\Delta_2^{\alpha} =& - h_{\mathrm{can}\,1}^{\mu\nu}
\,\partial_{\mu}\partial_{\nu}\varphi_1^\alpha +
\partial_\mu\left(\varphi_1^\mu\Box\varphi_1^{\alpha}\right)\,.\label{Delta2}
\end{align}\end{subequations}
As a consequence of \eqref{OmDelta} we easily verify that
\begin{equation}\label{dOm2}
\partial_\mu\Omega_2^{\alpha\mu} + \Delta_2^{\alpha} = 0 \,.
\end{equation}
This relation is consistent with the fact that the source term
$N_2^{\alpha\beta}$ is divergenceless [because of \eqref{dN2}]. Hence we see
that the divergence of \eqref{diffN2} is automatically verified, where we use
the fact that
$\partial_\mu\partial\Delta_2^{\alpha\mu}=\Box\Delta_2^{\alpha}$.

Applying our specific finite part of the retarded integral operator on both
sides of \eqref{diffN2} we obtain the relation between
$u^{\alpha\beta}_{\mathrm{gen}\,2}$ defined by \eqref{ugen2} and the
corresponding $u^{\alpha\beta}_{\mathrm{can}\,2}$ defined by \eqref{u2} in the
canonical algorithm, namely
\begin{equation}\label{ugencan2}
u_{\mathrm{gen}\,2}^{\alpha\beta} = u_{\mathrm{can}\,2}^{\alpha\beta} +
\Omega_2^{\alpha\beta} + \partial\phi_2^{\alpha\beta} + X_2^{\alpha\beta} +
Y_2^{\alpha\beta}\,.
\end{equation}
The difference between the two prescriptions is made of various terms. The
terms $\Omega_2^{\alpha\beta}$ and $\partial\phi_2^{\alpha\beta}$ represent
what we would expect if the operation of taking the finite part of the
retarded integral would commute with partial derivatives. Here the gauge
transformation is associated with the gauge vector defined by the finite part
of the retarded integral of $\Delta_2^{\alpha}$, 
\begin{equation}\label{phi2}
\phi_2^{\alpha} = \mathop{\mathrm{FP}}_{B=0}\, \Box^{-1}_\mathrm{ret} \biggl[
\left(\frac{r}{r_0}\right)^B \Delta_2^{\alpha} \biggr]\,.
\end{equation}
The last two terms $X_2^{\alpha\beta}$ and $Y_2^{\alpha\beta}$ come from the
non commutation of the finite part of the retarded integral operator
$\mathop{\mathrm{FP}}_{B=0}\, \Box^{-1}_\mathrm{ret}(r/r_0)^B$ with the
differential operators $\Box$ and $\partial$ which are present in front of the
last two terms of \eqref{N2}, respectively. We have
\begin{subequations}\label{commXY}\begin{align}
X_2^{\alpha\beta} &= \mathop{\mathrm{FP}}_{B=0}\, \Box^{-1}_\mathrm{ret}
 \left[ \left(\frac{r}{r_0}\right)^B \Box \Omega_2^{\alpha\beta}\right] -
 \Omega_2^{\alpha\beta}\,,\\ Y_2^{\alpha\beta} &= \mathop{\mathrm{FP}}_{B=0}\,
 \Box^{-1}_\mathrm{ret} \left[ \left(\frac{r}{r_0}\right)^B
 \partial\Delta_2^{\alpha\beta}\right] - \partial\phi_2^{\alpha\beta}\,,
\end{align}\end{subequations}
which can also be seen more formally as the action of ``commutators'' namely
\begin{subequations}\label{commXY2}\begin{align}
X_2^{\alpha\beta} &= \Bigl[\mathrm{FP}\, \Box^{-1}_\mathrm{ret}, \,\Box\Bigr]
 \,\Omega_2^{\alpha\beta}\,,\\ Y_2^{\alpha\beta} &= \Bigl[\mathrm{FP}\,
 \Box^{-1}_\mathrm{ret}, \,\partial\Bigr]\,\Delta_2^{\alpha\beta}\,.
\end{align}\end{subequations}
Our notation for the commutators involved and for the partial derivative
$\partial$ should be clear. Here $\mathrm{FP}\, \Box^{-1}_\mathrm{ret}$ is a
short hand for $\mathop{\mathrm{FP}}_{B=0}\, \Box^{-1}_\mathrm{ret}(r/r_0)^B$
and we have used the fact that $\Box(\mathrm{FP}\,
\Box^{-1}_\mathrm{ret}f)=f$. It is evident that the non commutation of
$\mathrm{FP}\, \Box^{-1}_\mathrm{ret}$ with partial derivatives comes from the
presence of the regularization factor $r^B$. Thus $X_2^{\alpha\beta}$ and
$Y_2^{\alpha\beta}$ are built from the spatial differentiation of $r^B$, 
{\it i.e.} $\partial_i r^B=B\,n_i\,r^{B-1}$, and
 will involve an explicit factor $B$
in their sources. Their expressions read as
\begin{subequations}\label{XY}\begin{align}
X_2^{\alpha\beta} &= \mathop{\mathrm{FP}}_{B=0}\, \Box^{-1}_\mathrm{ret}
\left[ B \left(\frac{r}{r_0}\right)^B
\left(-\frac{B+1}{r^{2}}\,\Omega_2^{\alpha\beta} -
\frac{2}{r}\,\frac{\partial\Omega_2^{\alpha\beta}}{\partial
r}\right)\right]\,,\\ Y_2^{\alpha\beta} &= \mathop{\mathrm{FP}}_{B=0}\,
\Box^{-1}_\mathrm{ret} \left[ B \left(\frac{r}{r_0}\right)^B \frac{n_i}{r}
\left( - \delta^{i\alpha} \Delta_2^\beta - \delta^{i\beta} \Delta_2^\alpha +
\eta^{\alpha\beta}\Delta_2^i\right) \right]\,.
\end{align}\end{subequations}
In Section \ref{secIVB} we shall present a practical method to evaluate
$X_2^{\alpha\beta}$ and $Y_2^{\alpha\beta}$ at the lowest PN
order, given the general quadratic-type structure for the source terms
\eqref{Om2} and \eqref{Delta2}.

The first part of the MPM algorithm $u^{\alpha\beta}_{\mathrm{gen}\,2}$ has
been obtained in \eqref{ugencan2}, and we look now for the second part
$v^{\alpha\beta}_{\mathrm{gen}\,2}$. To this end we compute the divergence
$w^{\alpha}_{\mathrm{gen}\,2}=\partial_{\mu}u^{\alpha\mu}_{\mathrm{gen}\,2}$.
Using \eqref{ugencan2} and the property \eqref{dOm2} we readily find that
\begin{equation}\label{wgen2}
w^{\alpha}_{\mathrm{gen}\,2} = w^{\alpha}_{\mathrm{can}\,2} +
\partial_{\mu}U_2^{\alpha\mu}\,,
\end{equation}
where we pose for simplicity
\begin{equation}\label{U2XY}
U_2^{\alpha\beta} = X_2^{\alpha\beta}+Y_2^{\alpha\beta}\,.
\end{equation}
The structure \eqref{XY} of $X_2^{\alpha\beta}$ and $Y_2^{\alpha\beta}$
involving the retarded integral of a source term containing an explicit factor
$B$ implies that $U_2^{\alpha\beta}$ is necessarily a retarded solution of the
source-free d'Alembertian equation, $\Box U_2^{\alpha\beta} = 0$. Hence, there
must exist ten STF tensors $A_L, B_L, \cdots, L_L$ (functions of the retarded
time) parametrizing the ten components of $U_2^{\alpha\beta}$ in such a way
that
\begin{subequations}\label{U2comp}\begin{align}
U_2^{00} &= \sum_{\ell = 0}^{+\infty}\partial_L \left[r^{-1}
A_L(t-r/c)\right]\,,\\ U_2^{0i} &= \sum_{\ell = 0}^{+\infty}\partial_{iL}
\left[ r^{-1} B_L (t-r/c) \right] \nonumber\\&+ \sum_{\ell = 1}^{+\infty}
\left\{ \partial_{L-1} \left[ r^{-1} C_{iL-1} (t-r/c) \right] +
\varepsilon_{iab} \partial_{aL-1} \left[r^{-1} D_{bL-1} (t-r/c) \right]
\right\}\,,\\ U_2^{ij} &= \sum_{\ell = 0}^{+\infty}\left\{\partial_{ijL}
\left[ r^{-1} E_L (t-r/c) \right] + \delta_{ij}\partial_{L} \left[ r^{-1} F_L
(t-r/c) \right]\right\}\nonumber\\&+ \sum_{\ell = 1}^{+\infty} \left\{
\partial_{L-1(i} \left[ r^{-1} G_{j)L-1} (t-r/c) \right] + \varepsilon_{ab(i}
\partial_{j)aL-1} \left[r^{-1} H_{bL-1} (t-r/c) \right] \right\}\nonumber\\&+
\sum_{\ell = 2}^{+\infty}\left\{\partial_{L-2} \left[ r^{-1} K_{ijL-2} (t-r/c)
\right] + \partial_{aL-2} \left[ r^{-1} \varepsilon_{ab(i} L_{j)bL-2} (t-r/c)
\right]\right\} \,.\end{align}\end{subequations}
The divergence of this tensor, $W_2^\alpha=\partial_\mu U_2^{\alpha\mu}$, will
also be of that form and hence there will exist four STF tensors $N'_L, \cdots,
R'_L$ such that
\begin{subequations}\label{W2}\begin{align}
W_2^0 &= \sum_{\ell = 0}^{+\infty}\partial_L \left[r^{-1} N'_L(t-r/c)\right]\,,\\
W_2^i &= \sum_{\ell = 0}^{+\infty}\partial_{iL} \left[ r^{-1} P'_L (t-r/c)
\right] \nonumber\\&+ \sum_{\ell = 1}^{+\infty} \left\{ \partial_{L-1} \left[
r^{-1} Q'_{iL-1} (t-r/c) \right] + \varepsilon_{iab} \partial_{aL-1}
\left[r^{-1} R'_{bL-1} (t-r/c) \right] \right\}\,.
\end{align}
\end{subequations}
The four tensors $N'_L, \cdots, R'_L$ play exactly the same role as $N_L,
\cdots, R_L$ in \eqref{w2}, and we shall apply the same algorithm as the one
going from \eqref{w2} to \eqref{v2}. Thus, we define from the components of
$W_2^\alpha$ a new tensor $V_2^{\alpha\beta}$ by this algorithm, which was
denoted by $\mathcal{V}^{\alpha\beta}$ in \eqref{v2p}. Hence
\begin{equation}\label{algV2}
V_2^{\alpha\beta} = \mathcal{V}^{\alpha\beta}\left[W_2\right]\,,
\end{equation}
so  that in component form this tensor reads
\begin{subequations}\label{V2comp}\begin{align}
V^{00}_2 &= - r^{-1} N'^{(-1)} + \partial_a \left[ r^{-1} \left(- N'^{(-1)}_a+
Q'^{(-2)}_a -3P'_a\right) \right] \,,\\ V^{0i}_2 &= r^{-1} \left( - Q'^{(-1)}_i
+3 P'^{(1)}_i\right) - \varepsilon_{iab} \partial_a \left[ r^{-1} R'^{(-1)}_b
\right] - \sum_{\ell = 2}^{+\infty}\partial_{L-1} \left[ r^{-1} N'_{iL-1}
\right] \,,\\ V^{ij}_2 &= - \delta_{ij} r^{-1} P' + \sum_{\ell = 2}^{+\infty}
\biggl\{ 2 \delta_{ij}\partial_{L-1} \left[ r^{-1} P'_{L-1}\right] - 6
\partial_{L-2(i} \left[ r^{-1} P'_{j)L-2}\right] \nonumber\\ & \quad+
\partial_{L-2} \left[ r^{-1} (N'^{(1)}_{ijL-2} + 3 P'^{(2)}_{ijL-2} -
Q'_{ijL-2}) \right] - 2 \partial_{aL-2}\left[ r^{-1} \varepsilon_{ab(i}
R'_{j)bL-2} \right] \biggr\}\,.
\end{align}\end{subequations}
However, in the present case the tensors $N'_L, \cdots, R'_L$ can be directly
related to the ones parametrizing \eqref{U2comp}. By computing the divergence
$W_2^\alpha=\partial_\mu U_2^{\alpha\mu}$ we readily find
\begin{subequations}\label{NPQR}\begin{align}
N'_L &= A_L^{(1)} + B_L^{(2)} + C_L\,,\\ P'_L &= E_L^{(2)} + F_L + \frac{1}{2}G_L +
B_L^{(1)}\,,\\Q'_L &= \frac{1}{2}G_L^{(2)} + K_L + C_L^{(2)}\,,\\R'_L &=
\frac{1}{2}H_L^{(2)} + \frac{1}{2}L_L + D_L^{(1)}\,.
\end{align}\end{subequations}
Thus $V_2^{\alpha\beta}$ can be expressed directly in terms of $A_L, \cdots,
L_L$ by substituting \eqref{NPQR} into \eqref{V2comp}. In doing so we shall
discover that the time anti-derivatives present in \eqref{V2comp} become in
fact ``instantaneous'' because they are cancelled by some time derivatives
coming from \eqref{NPQR}. By construction of \eqref{V2comp} we have at once
$\Box V_2^{\alpha\beta}=0$ and $\partial_\mu
V_2^{\alpha\mu}=-W_2^{\alpha}$. Applying the MPM algorithm we therefore find
for the second part of the algorithm,
\begin{equation}\label{vgen2}
v_{\mathrm{gen}\,2}^{\alpha\beta} = v_{\mathrm{can}\,2}^{\alpha\beta} +
V_2^{\alpha\beta}\,.
\end{equation}
Gathering the results \eqref{ugencan2} and \eqref{vgen2} the complete
quadratic-order metric is obtained as
\begin{equation}\label{hgen2}
h_{\mathrm{gen}\,2}^{\alpha\beta} = h_{\mathrm{can}\,2}^{\alpha\beta} +
\Omega_2^{\alpha\beta} + \partial\phi_2^{\alpha\beta} + U_2^{\alpha\beta} +
V_2^{\alpha\beta}\,,
\end{equation}
and it satisfies the vacuum Einstein field equations in harmonic coordinates,
\textit{i.e.}
\begin{subequations}\label{EEgen}\begin{align}
\Box h_{\mathrm{gen}\,2}^{\alpha\beta} &= N_{\mathrm{gen}\,2}^{\alpha\beta}\,,\\
\partial_\mu h_{\mathrm{gen}\,2}^{\alpha\mu} &= 0\,.
\end{align}\end{subequations}

To find the relation between the source and canonical moments we notice that
the sum of the last two terms in \eqref{hgen2} is a solution of the linearized
vacuum equations, since it satisfies $\Box(U_2^{\alpha\beta} +
V_2^{\alpha\beta})=0$ and also $\partial_\mu(U_2^{\alpha\mu} +
V_2^{\alpha\mu})=0$. It must therefore be of the form of the general solution
$h_{\mathrm{gen}\,1}^{\alpha\beta}$ of these equations which has been given in
\eqref{hgen1}, \textit{i.e.}  there should exist some moments $\delta I_L$ and
$\delta J_L$ representing specific corrections to $I_L$ and $J_L$ (necessarily
at quadratic order) and some gauge vector $\psi^\alpha_2$ such that
\begin{equation}\label{U2V2}
U_2^{\alpha\beta} + V_2^{\alpha\beta} =
h_{\mathrm{can}\,1}^{\alpha\beta}[\delta I_L,\delta J_L] +
\partial\psi_2^{\alpha\beta}\,.
\end{equation}
Let us prove that the corrections we seek to the moments $I_L$ and $J_L$ that
are needed to reproduce the canonical moments are indeed provided by these
$\delta I_L$ and $\delta J_L$, \textit{i.e.}
\begin{subequations}\label{MS}\begin{align}
M_L &= I_L + G\,\delta I_L + \mathcal{O}(G^2)\,,\label{MSa}\\S_L &= J_L +
G\,\delta J_L + \mathcal{O}(G^2)\,.\label{MSb}
\end{align}\end{subequations}
To this end we have to check that the general metric
$h_{\mathrm{gen}}^{\alpha\beta}[I_L,J_L,W_L,\cdots,Z_L]$ constructed at
quadratic order in \eqref{hgen} is \textit{isometric} --- differs by a
coordinate transformation --- to the canonical metric
$h_{\mathrm{can}}^{\alpha\beta}[M_L,S_L]$ given by \eqref{hcan}. This
immediately follows from \eqref{hgen2} and \eqref{U2V2} which permits us
to recast the general metric \eqref{hgen} into the form
\begin{align}\label{hgencan}
h_{\mathrm{gen}}^{\alpha\beta}[I_L,J_L,\cdots] &= G\left[
h_{\mathrm{can}\,1}^{\alpha\beta}[M_L,S_L] +
\partial\varphi_1^{\alpha\beta}\right] \nonumber\\&+ G^2\left[
h_{\mathrm{can}\,2}^{\alpha\beta}[M_L,S_L] + \Omega_2^{\alpha\beta} +
\partial\varphi_2^{\alpha\beta}\right]+ \mathcal{O}(G^3)\,,
\end{align}
where we have posed $\varphi_2^{\alpha}=\phi_2^{\alpha}+\psi_2^{\alpha}$, and
where higher-order powers of $G$ are consistently neglected. From this result
we conclude that $h_{\mathrm{gen}}^{\alpha\beta}[I_L,J_L,\cdots]$ and
$h_{\mathrm{can}}^{\alpha\beta}[M_L,S_L]$ differ by the coordinate
transformation
\begin{equation}\label{coord}
x_{\mathrm{gen}}^{\alpha} = x_{\mathrm{can}}^{\alpha} + G\varphi_1^{\alpha} +
G^2\varphi_2^{\alpha} + \mathcal{O}(G^3)\,,
\end{equation}
as we have recognized that $\Omega_2^{\alpha\beta}$ represents precisely the
quadratic non-linear part of that coordinate transformation, \textit{i.e.}
the term which makes it to differ from a linearized gauge
transformation. Hence we have proved that the two sets of moments
$\{I_L,J_L,W_L,X_L,Y_L,Z_L\}$ and $\{M_L,S_L\}$ related by \eqref{MS} are
physically equivalent --- they describe the same physical matter source. Note
that the relations \eqref{MS} give the canonical moments as functionals of the
full set of source moments $\{I_L,J_L,W_L,X_L,Y_L,Z_L\}$. Consequently, just
two moments $M_L$ and $S_L$ are still sufficient to describe the external
field of any source \cite{BD86}.  Notice also that $M_L$ and $S_L$ are almost
equal to $I_L$ and $J_L$ in the sense that the corrections $\delta I_L$ and
$\delta J_L$ in \eqref{MS} will turn out to be very small in a PN expansion,
being of order 2.5PN \cite{B96}. This is of course the result of the fact that
the gauge moments $\{W_L,X_L,Y_L,Z_L\}$ do not play any physical role at the
linear approximation, where the coordinate transformation reduces to the gauge
transformation. However, since the theory is covariant with respect to
non-linear diffeomorphisms and not merely with respect to linear gauge
transformations, the moments $\{W_L,X_L,Y_L,Z_L\}$ do play a role at the
non-linear level.

\subsection{Practical implementation}\label{secIVB}

Finally let us sketch our practical method to compute the correction terms
$\delta I_L$ and $\delta J_L$. We remark first that they come \textit{via}
\eqref{U2V2} from the ten STF tensors $A_L, \cdots, L_L$ parametrizing
$U_2^{\alpha\beta}$ as given by \eqref{U2comp}. We can therefore express
$\delta I_L$ and $\delta J_L$ directly in terms of $A_L, \cdots, L_L$ by
following in details the steps \eqref{W2}--\eqref{NPQR}. The result is
\begin{subequations}\label{deltaIJ}\begin{align}
\delta I_L &= -c^2\frac{(-)^\ell\ell!}{4}\left[A_L + 4 B_L^{(1)} + 3 E_L^{(2)}
+ 3 F_L + G_L\right]\,,\\\delta J_L &= c^3\frac{(-)^\ell(\ell+1)!}{4 \ell}\left[
D_L + \frac{1}{2}H_L^{(1)}\right]\,.
\end{align}\end{subequations}
The next problem is to compute the tensors $A_L, \cdots, L_L$ in the PN
approximation. These are defined from the two objects $X_2^{\alpha\beta}$ and
$Y_2^{\alpha\beta}$ which are given in particular by their commutator form
\eqref{commXY2}. We thus need to compute the commutator between the operator
$\mathrm{FP}\, \Box^{-1}_\mathrm{ret}$ and derivative operators, when applied
either on the terms $\Omega_2^{\alpha\beta}$ or $\Delta_2^{\alpha\beta}$. The
relevant point for our purpose is that the general structure of
 these terms at the quadratic order is known.
 Namely $\Omega_2^{\alpha\beta}$ and
$\Delta_2^{\alpha\beta}$ are made of quadratic products of retarded multipolar
waves, \textit{i.e.} are given by sums of terms of the type
\begin{equation}\label{KPQ}
\mathcal{K}_{PQ} = \partial_{\langle P\rangle}\left[r^{-1}
F(t-r/c)\right]\partial_{\langle Q\rangle}\left[r^{-1} G(t-r/c)\right]\,,
\end{equation}
where the functions $F$ and $G$ stand for some time derivatives of moments in
the list $\{I_L,J_L,W_L,X_L,Y_L,Z_L\}$. It is convenient to suppress the
indices on these moments and to write only the ``active'' indices appearing in
the spatial multi-derivatives $\partial_P$ and $\partial_Q$, composed with the
multi-indices $P=a_1\cdots a_p$ and $Q=b_1\cdots b_q$ ($p$ and $q$ being the
number of partial derivatives in $\partial_P$ and $\partial_Q$). Furthermore
the multi-derivatives in \eqref{KPQ} are chosen to be STF (this can always be
assumed modulo a possible STF decomposition), hence the brackets
$\langle\rangle$ surrounding their indices. The problem is therefore reduced
to that of evaluating, in the PN approximation, the quantities\footnote{In the
case of $\mathcal{Y}^i_{PQ}$ we can restrict ourselves to a spatial derivative
$\partial^i$ because the time derivative $\partial_t$ commutes with the
operator $\mathrm{FP}\, \Box^{-1}_\mathrm{ret}$, thus $\mathcal{Y}^0_{PQ}=0$.}
\begin{subequations}\label{XYPQ}\begin{align}
\mathcal{X}_{PQ} &= \Bigl[\mathrm{FP}\, \Box^{-1}_\mathrm{ret},
  \,\Box\Bigr] \,\mathcal{K}_{PQ} \label{XPQdef}\,,\\ \mathcal{Y}^i_{PQ}
&= \Bigl[\mathrm{FP}\, \Box^{-1}_\mathrm{ret},
  \,\partial^i\Bigr]\,\mathcal{K}_{PQ}\label{YPQdef}\,.
\end{align}\end{subequations}
Indeed $X_2^{\alpha\beta}$ and $Y_2^{\alpha\beta}$ are given by some sums of
terms of the type $\mathcal{X}_{PQ}$ and $\mathcal{Y}^i_{PQ}$ respectively
(and multiplied by appropriate constant tensors involving Kronecker symbols to
perform the needed contractions).

The term $\mathcal{X}_{PQ}$ has in fact already been computed at the lowest PN
order in the Appendix of \cite{B96}. The result turned out to be quite simple,
namely
\begin{equation}\label{XPQ}
\mathcal{X}_{PQ} = \frac{1}{c}\partial_{\langle PQ\rangle}\Bigl[r^{-1}
\Bigl(\delta_{p,0}\,F^{(1)}\,G + \delta_{0,q}\,F\,G^{(1)}\Bigr)\Bigr] +
\mathcal{O}\left(\frac{1}{c^3}\right)\,.
\end{equation}
Here the functions are evaluated at retarded time $t-r/c$ (with $F^{(1)}$ and
$G^{(1)}$ denoting the time derivatives), and $\delta_{p,0}$ and
$\delta_{0,q}$ denote the usual Kronecker symbols. As we see in \eqref{XPQ}
the two STF multi-derivative operators $\partial_{\langle P\rangle}$ and
$\partial_{\langle Q\rangle}$ originally present in \eqref{KPQ} have merged
into a single STF derivative operator $\partial_{\langle PQ\rangle}$ with
$p+q$ indices. The formula \eqref{XPQ} constitutes a useful practical lemma
for doing computations at the lowest PN order. Because of the factor $1/c$ in
front of \eqref{XPQ} the PN ``parity'' of the result \eqref{XPQ} will be
opposite to that of the source term \eqref{KPQ}, which in practice will
typically be even. As a consequence we shall find that the PN order of
$X_2^{\alpha\beta}$ is dominantly ``odd'', starting in fact with 2.5PN.

As for the term $\mathcal{Y}^i_{PQ}$, it was not required in \cite{B96} but
will play a role here for the waveform at the 3PN order. We have worked out
the equivalent of \eqref{XPQ} for this term, and find, still at the lowest PN
order,
\begin{equation}\label{YPQ}
\mathcal{Y}^i_{PQ} = - \frac{p+q}{c(2p+2q+1)}
\left\{\delta_{p,0}\,\delta^i_{\langle b_q}\partial_{PQ-1\rangle}\Bigl[r^{-1}
\,F^{(1)}\,G\Bigr] + \delta_{0,q}\,\delta^i_{\langle
a_p}\partial_{P-1Q\rangle}\Bigl[r^{-1} \,F\,G^{(1)}\Bigr]\right\} +
\mathcal{O}\left(\frac{1}{c^3}\right)\,.
\end{equation}
Consistent with our notation we write $P-1=a_1\cdots a_{p-1}$ and
$Q-1=b_1\cdots b_{q-1}$. Again there is a factor $1/c$ and we shall find that
the corresponding $Y_2^{\alpha\beta}$ is dominantly ``odd'', starting at 2.5PN
order. Note that the new lemma \eqref{YPQ} is not independent from the
previous one \eqref{XPQ} and is actually more general than it. Indeed, by
computing the divergence of $\mathcal{Y}^i_{PQ}$ using its definition
\eqref{YPQdef}, we get
\begin{equation}\label{relYZ}
\partial_i\mathcal{Y}^i_{PQ} = \Bigl[\mathrm{FP}\,
  \Box^{-1}_\mathrm{ret}, \,\Box\Bigr] \,\mathcal{K}_{PQ} -
\Bigl[\mathrm{FP}\, \Box^{-1}_\mathrm{ret}, \,\partial^i\Bigr]
\,\partial_i\mathcal{K}_{PQ}\,,
\end{equation}
which can be used to check the consistency of the two formulas \eqref{XPQ} and
\eqref{YPQ}. The results needed at 3PN order for the relation between the
canonical and source moments as obtained by these means --- namely the
formulas \eqref{deltaIJ} and the lemmas \eqref{XPQ}--\eqref{YPQ} --- are
reported in Section \ref{secVB}.

\section{The moments for 3PN waveform}\label{secV}

Using the MPM algorithm of Section \ref{secIII} the radiative moments
$\{U_L,V_L\}$ are related to the canonical moments $\{M_L,S_L\}$, and
following Section \ref{secIV} the canonical moments are in turn expressed in
terms of the source moments $\{I_L,J_L,W_L,X_L,Y_L,Z_L\}$. In the current
Section we present the results (skipping some details) of the computation of
all the moments needed for controlling the FWF in the case of compact binary
systems up to 3PN order.

\subsection{The radiative moments for 3PN polarisations}\label{secVA}

To obtain the gravitational polarisations at 3PN order one must compute: the
mass radiative quadrupole $U_{ij}$ with 3PN accuracy; the current radiative
quadrupole $V_{ij}$ and mass radiative octupole $U_{ijk}$ with 2.5PN accuracy;
mass hexadecapole $U_{ijkl}$ and current octupole $V_{ijk}$ with 2PN
precision; $U_{ijklm}$ and $V_{ijkl}$ up to 1.5PN order; $U_{ijklmn}$,
$V_{ijklm}$ at 1PN; $U_{ijklmno}$, $V_{ijklmn}$ at 0.5PN; and finally
$U_{ijklmnop}$, $V_{ijklmno}$ to Newtonian order. The relations connecting
$U_L$ and $V_L$ to the canonical moments $M_L$ and $S_L$ are first obtained
following the MPM method of Section \ref{secIII}.\footnote{We have implemented
the MPM algorithm on the algebraic computing software Mathematica using the
powerful tensor package xTensor \cite{xtensor}.}

The quadratic contributions to the radiative mass (resp. current) moments are
found in the form of sums of terms $\delta_2 U_L(u)$ (resp. $\delta_2
V_L(u)/c$) whose general structure reads
\begin{equation} \label{eq:deltaUV}
\delta_2 U_L(u), \,\delta_2 V_L(u)/c =\frac{G}{c^{m-\ell+2}} \int_{-\infty}^u
ds \, \chi_{LK_1K_2}(u,s) \, A^{(p_1)}_{K_1}(s) \, A^{(p_2)}_{K_2}(s)\,,
\end{equation} 
The power of $1/c$ in front is chosen in such a way that $m$ represents the PN
order of our calculation of the waveform, \textit{i.e.} $m=6$ at the 3PN
order. The capital letter $A$ stands either for $M$ or $S$, meaning that we
are considering in \eqref{eq:deltaUV} interactions between canonical moments
of the type $M^{(p_1)}_{K_1}\, M^{(p_2)}_{K_2}$ or $M^{(p_1)}_{K_1}\,
S^{(p_2)}_{K_2}$ or $S^{(p_1)}_{K_1}\, S^{(p_2)}_{K_2}$, with the superscript
$(p)$ denoting time derivatives, and the multi-indices $K_{1}$ and $K_{2}$
having length $k_{1}$ and $k_{2}$ (\textit{e.g.}  $K_1=a_1\cdots
a_{k_1}$).\footnote{The reasonning we shall make can be easily generalized to
$n$-th non-linear terms $\delta_n U_L$, $\delta_n V_L/c$ involving $n$
canonical moments $A^{(p_1)}_{K_1}$, $A^{(p_2)}_{K_2}$, $\cdots$,
$A^{(p_n)}_{K_n}$.} The kernel $\chi_{LK_1K_2}$ has itself an algebraic
structure made of a sum of products of Kronecker and Levi-Civita symbols. Its
physical dimension depends on time only, and each of its three sets of
indices, $L$, $K_1$ and $K_2$, is symmetric and trace-free (STF). For
instantaneous terms, which are functions of the multipole moments
$A^{(p_1)}_{K_1}$, $A^{(p_2)}_{K_2}$ evaluated at the instant of emission
$u=t-r/c$, it is proportional to the Dirac function $\delta(s-u)$.

The above structure does not exist generically for an arbitrary pair of
multipole moments nor for any arbitrary value of $k_1$ and $k_2$. A closer
look will actually allow us to reduce the number of terms in the source we
shall focus on to a few ones, making our task much easier. As a product
$\varepsilon_{ijk} \varepsilon_{abc}$ can always be transformed into a linear
combination of $\delta_{ia'} \delta_{jb'} \delta_{kc'}$ with
$\{a',b',c'\}=\{a,b,c\}$, the number $\epsilon$ of Levi-Civita symbols in each
of the individual terms $\delta \delta ... \delta \varepsilon \varepsilon
... \varepsilon $ composing $\chi_{LK_1K_2}$ may be reduced to 0 or 1. The
symmetry of parity implies that this number is the same for all terms. Now, if
$\epsilon = 0$, the integer $\ell+k_1+k_2$ is even (equal to twice the number
of Kronecker symbols) and all indices in $L$ must contract with an index of
$K_1$ or $K_2$. Thus, we must necessarily have $k_1+k_2 \ge \ell$.  On the
other hand, if $\epsilon =1$, the Levi-Civita symbol carries one index from
each of the three STF sets, so that there remain $\ell-1$ free indices of type
$L$ carried by Kronecker symbols, as well as $k_1-1$ indices (resp. $k_2-1$)
of type $K_1$ (resp. $K_2$) involved in the contraction of some $\delta$'s
with the multipole moments. Then, the same arguments as before show that
$(\ell-1)+(k_1-1)+(k_2-1)$ must be even with $ (k_1-1)+(k_2-1) \ge \ell-1$.
The previous constraints can all be summarized by the single statement that
$k_1 + k_2 - \ell - \epsilon$ is always an \textit{even} positive integer.

The structure of the quadratic interactions may be further refined by noticing
that only the multipole moments that have dimensions compatible with
\eqref{eq:deltaUV} are allowed to enter $\delta_2 U,\,\delta_2 V/c$. Let us
pose for later convenience $[A^{(p_1)}_{K_1}]=[M] \,[L]^{a_1+k_1-p_1}
\,[V]^{\alpha_1+p_1}$ and $[A^{(p_2)}_{K_2}]=[M] \,[L]^{a_2+k_2-p_2}
\,[V]^{\alpha_2+p_2}$, where $[M]$, $[L]$ and $[V]$ denote the dimension of a
mass, a length and a velocity. Equating $[U_L]= [V_L/c]$ and $[ds
\,\chi_{LK_1K_2}]\,[G/c^{m-\ell+2} \,A^{(p_1)}_{K_1}\,A^{(p_2)}_{K_2}]$ on the
one hand, remembering on the other hand that $[\chi_{LK_1K_2}]$ is a certain
power $q\in \mathbb{Z}$ of the time dimension $[T]$, we find:
\begin{subequations}\label{selection}\begin{align}
&\sum_{i=1,2} (a_i + \alpha_i + k_i) = m-1\,,\\ 
&\sum_{i=1,2} (\alpha_i + p_i) = q+m+1 \,.
\end{align}\end{subequations}
Now, we know that $k_1 + k_2 -\ell - \epsilon \in \mathbb{N}$. Moreover, the
number $\epsilon=0,1$ of Levi-Civita symbols is itself governed by the parity
symmetry. More precisely, defining the integers $\tilde{\alpha}_1$,
$\tilde{\alpha}_2$ and $\tilde{\epsilon}$, associated with $A^{(p_1)}_{K_1}$,
$A^{(p_2)}_{K_2}$ and $\delta_2 U_L,\,\delta_2 V_L/c$ respectively, to be
equal to zero when the latter multipole moments are of mass type or to 1 when
they are of current type, the consistency of the transformation of both sides
of equation \eqref{eq:deltaUV} under parity imposes that $\tilde{\epsilon} =
\tilde{\alpha}_1 + \tilde{\alpha}_2+\epsilon$ [mod 2]. As a result, the
maximum multipolar order $k_1 + k_2 - \epsilon$ of the radiative moments
containing a quadratic interaction $A^{(p_1)}_{K_1}\,A^{(p_2)}_{K_2}$ is given
by\footnote{The remainder function means the usual division remainder:
$\text{remainder}[\frac{N}{2}]=0$ or $1$ depending on whether $N$ is an even
or odd integer.}
\begin{equation} \label{eq:lmax}
\ell_\text{max}(a_i,\alpha_i,\tilde{\alpha}_i,\tilde{\epsilon})=m-1-\sum_{i=1,2}
(a_i + \alpha_i) - \text{remainder}\biggl[\frac{1}{2}\Bigl(\sum_{i=1,2}
\tilde{\alpha}_i+\tilde{\epsilon}\Bigr)\biggr]\,.
\end{equation}
At the $\frac{m}{2}$PN approximation, such a contribution exists only if
$\ell_\text{max}\ge 2$, or equivalently $\sum_{i=1,2}(a_i + \alpha_i) +
\text{remainder}[\frac{1}{2}(\sum_{i=1,2}
\tilde{\alpha}_i+\tilde{\epsilon})]\le m-3$. Once this necessary condition is
fulfilled, the orders of multipolarity possibly affected by the piece of
non-linear correction \eqref{eq:deltaUV} are
$\ell_\text{max}(a_i,\alpha_i,\tilde{\alpha}_i,\tilde{\epsilon})$,
$\ell_\text{max}(a_i,\alpha_i,\tilde{\alpha}_i,\tilde{\epsilon})-2$, $\cdots$,
2 or 3, with $\tilde{\epsilon}=0$ (resp. $\tilde{\epsilon}=1$) for mass (resp.
current) radiative moments. The latter ``selection'' rules may be generalized
to interactions of any post-Minkowskian order $n$, in which case $m-1$ must be
replaced by $m+5-3n$ in the expression \eqref{eq:lmax} of $\ell_\text{max}$
while all summation ranges become $1\le i\le n$. With the selection rules
\eqref{selection}--\eqref{eq:lmax} we are able to know beforehand which
non-linear multipole interaction is needed to be computed in the radiative
moments $U_L$, $V_L$ at a given PN order.

The result concerning the 3PN mass quadrupole moment $U_{ij}$ is already known
\cite{BD92,B98quad,B98tail} and we simply report it here. Actually, at 3PN
order $U_{ij}$ involves a cubically non-linear term, composed of the so-called
tails of tails, whose computation necessitates an extension of the MPM
algorithm to cubic order $G^3$ \cite{B98tail}. We have
\begin{align}\label{U2}
U_{ij}(T_R) &= M^{(2)}_{ij} (T_R) + {2G M\over c^3} \int_{-\infty}^{T_R} d
\tau \left[ \ln \left({T_R-\tau\over 2\tau_0}\right)+{11\over12} \right]
M^{(4)}_{ij} (\tau) \nonumber \\ &+{G\over
c^5}\left\{-{2\over7}\int_{-\infty}^{T_R} d\tau M^{(3)}_{a\langle
i}(\tau)M^{(3)}_{j\rangle a}(\tau) \right.\nonumber \\ &\qquad~\left. + {1
\over7}M^{(5)}_{a\langle i}M_{j\rangle a} - {5 \over7} M^{(4)}_{a\langle
i}M^{(1)}_{j\rangle a} -{2 \over7} M^{(3)}_{a\langle i}M^{(2)}_{j\rangle a}
+{1 \over3}\varepsilon_{ab\langle i}M^{(4)}_{j\rangle a}S_{b}\right\}\nonumber
\\ &+ 2\left({G M\over c^3}\right)^2\int_{-\infty}^{T_R} d \tau \left[ \ln^2
\left({T_R-\tau\over2\tau_0}\right)+{57\over70} \ln\left({T_R-\tau\over
2\tau_0}\right)+{124627\over44100} \right] M^{(5)}_{ij} (\tau) \nonumber \\ &
+\,\, \mathcal{O}\left(\frac{1}{c^7}\right)\,.
\end{align}
Notice the tail integral at 1.5PN order, the tail-of-tail integral at 3PN
order, and the non-linear memory integral at 2.5PN. In the tail and
tail-of-tail integrals, $M$ represents the mass monopole moment or total mass
of the binary system. The constant $\tau_0$ in the tail integrals is given by
$\tau_0=r_0/c$, where $r_0$ is the arbitrary length scale originally
introduced in the MPM formalism through \eqref{u2}, and appearing also in the
relation between the radiative and harmonic coordinates as given by
\eqref{TR}.

The moments required at 2.5PN order are new with this paper (apart from the
tails) and involve some interactions between the mass quadrupole moment and
the mass octupole or current quadrupole moments. Which type of interactions is
determined by using the selection rules discussed above. These moments are
given by\footnote{In all formulas below the STF projection $\langle \rangle $
applies only to the ``free'' indices denoted $ijkl\cdots$ carried by the
moments themselves. Thus the dummy indices such as $abc\cdots$ are excluded
from the STF projection.}
\begin{subequations}\label{U3V2}\begin{align}
U_{ijk} (T_R) &= M^{(3)}_{ijk} (T_R) + {2G M\over c^3} \int_{-\infty}^{T_R}
d\tau\left[ \ln \left({T_R-\tau\over 2\tau_0}\right)+{97\over60} \right]
M^{(5)}_{ijk} (\tau)\nonumber \\ & +{G\over c^5}\left\{ \int_{-\infty}^{T_R}
d\tau \left[-{1\over3}M^{(3)}_{a\langle i} (\tau)M^{(4)}_{jk\rangle a}
(\tau)-{4\over5}\varepsilon_{ab\langle i} M^{(3)}_{ja} (\tau)S^{(3)}_{k\rangle
b} (\tau)\right]\right.\nonumber\\ & -{4\over3}M^{(3)}_{a\langle
i}M^{(3)}_{jk\rangle a}-{9\over4}M^{(4)}_{a\langle i}M^{(2)}_{jk\rangle a} +
{1\over4}M^{(2)}_{a\langle i}M^{(4)}_{jk\rangle a} -
{3\over4}M^{(5)}_{a\langle i}M^{(1)}_{jk\rangle a} +{1\over4}M^{(1)}_{a\langle
i}M^{(5)}_{jk\rangle a}\nonumber\\ &+ {1\over12}M^{(6)}_{a\langle
i}M_{jk\rangle a} +{1\over4}M_{a\langle i}M^{(6)}_{jk\rangle a} +
{1\over5}\varepsilon_{ab\langle i}\left[-12S^{(2)}_{ja}M^{(3)}_{k\rangle
b}-8M^{(2)}_{ja}S^{(3)}_{k\rangle b} -3S^{(1)}_{ja}M^{(4)}_{k\rangle
b}\right.\nonumber\\ &\left.\left.  -27M^{(1)}_{ja}S^{(4)}_{k\rangle
b}-S_{ja}M^{(5)}_{k\rangle b}-9M_{ja}S^{(5)}_{k\rangle b}
-{9\over4}S_{a}M^{(5)}_{jk\rangle b}\right] +{12\over5}S_{\langle
i}S^{(4)}_{jk\rangle}\right\}\nonumber\\
&+\,\mathcal{O}\left(\frac{1}{c^6}\right) \label{U3}\,,\\ V_{ij} (T_R) &=
S^{(2)}_{ij} (T_R) + {2G M\over c^3} \int_{-\infty}^{T_R} d \tau \left[ \ln
\left({T_R-\tau\over 2\tau_0}\right)+{7\over6} \right] S^{(4)}_{ij}
(\tau)\nonumber\\ &+ {G\over7\,c^{5}}\left\{4S^{(2)}_{a\langle
i}M^{(3)}_{j\rangle a}+8M^{(2)}_{a\langle i}S^{(3)}_{j\rangle a}
+17S^{(1)}_{a\langle i}M^{(4)}_{j\rangle a}-3M^{(1)}_{a\langle
i}S^{(4)}_{j\rangle a}+9S_{a\langle i}M^{(5)}_{j\rangle a}\right.\nonumber\\
&-3M_{a\langle i}S^{(5)}_{j\rangle
a}-{1\over4}S_{a}M^{(5)}_{ija}-7\varepsilon_{ab\langle
i}S_{a}S^{(4)}_{j\rangle b} +{1\over2}\varepsilon_{ac\langle
i}\left[3M^{(3)}_{ab}M^{(3)}_{j\rangle bc} +{353\over24}M^{(2)}_{j\rangle
bc}M^{(4)}_{ab}\right.\nonumber\\
&\left.\left.-{5\over12}M^{(2)}_{ab}M^{(4)}_{j\rangle
bc}+{113\over8}M^{(1)}_{j\rangle bc}M^{(5)}_{ab}
-{3\over8}M^{(1)}_{ab}M^{(5)}_{j\rangle bc}+{15\over4}M_{j\rangle
bc}M^{(6)}_{ab} +{3\over8}M_{ab}M^{(6)}_{j\rangle
bc}\right]\right\}\nonumber\\
&+\,\mathcal{O}\left(\frac{1}{c^6}\right)\,.\label{V2}
\end{align}\end{subequations}
At 2PN order we have the standard tails and some previously known interactions
of the mass quadrupole with itself \cite{B98quad}, namely
\begin{subequations}\label{U4V3}\begin{align}
U_{ijkl}(T_R) &= M^{(4)}_{ijkl} (T_R) + {G\over c^3} \left\{ 2 M
\int_{-\infty}^{T_R} d \tau \left[ \ln \left({T_R-\tau\over
2\tau_0}\right)+{59\over30} \right] M^{(6)}_{ijkl}(\tau) \right.\nonumber\\
&\quad\left. +{2\over5}\int_{-\infty}^{T_R} d\tau M^{(3)}_{\langle
ij}(\tau)M^{(3)}_{kl\rangle }(\tau) -{21\over5}M^{(5)}_{\langle ij}M_{kl\rangle }- {63
\over5}M^{(4)}_{\langle ij}M^{(1)}_{kl\rangle }- {102\over5}M^{(3)}_{\langle
ij}M^{(2)}_{kl\rangle }\right\}\nonumber \\ &+
\,\mathcal{O}\left(\frac{1}{c^5}\right)\label{U4}\,,\\ V_{ijk} (T_R) &=
S^{(3)}_{ijk} (T_R) + {G\over c^3} \left\{ 2 M \int_{-\infty}^{T_R} d \tau
\left[ \ln \left({T_R-\tau\over 2\tau_0}\right)+{5\over3} \right]
S^{(5)}_{ijk} (\tau) \right.\nonumber \\
&\quad\left.+{1\over10}\varepsilon_{ab\langle i}M^{(5)}_{ja}M_{k\rangle b}-
{1\over2}\varepsilon_{ab\langle i}M^{(4)}_{ja}M^{(1)}_{k\rangle b} - 2
S_{\langle i}M^{(4)}_{jk\rangle } \right\} \nonumber\\ & +
\,\mathcal{O}\left(\frac{1}{c^5}\right)\,.\label{V3}
\end{align}\end{subequations}
At 1.5PN we again have some non-linear interactions (new with this paper)
involving the mass octupole and current quadrupole and given by
\begin{subequations}\label{U5V4}\begin{align}U_{ijklm}(T_R) &=
M^{(5)}_{ijklm}(T_R) + {G\over c^3}\left\{2 M \int_{-\infty}^{T_R} d \tau
\left[ \ln \left({T_R-\tau\over2\tau_0}\right)+ {232\over105} \right]
M^{(7)}_{ijklm} (\tau)\right.\nonumber\\ &+{20\over21}\int_{-\infty}^{T_R} d
\tau M^{(3)}_{\langle ij} (\tau) M^{(4)}_{klm\rangle} (\tau)
-{710\over21}M^{(3)}_{\langle
ij}M^{(3)}_{klm\rangle}-{265\over7}M^{(2)}_{\langle ijk}M^{(4)}_{lm\rangle}
-{120\over7}M^{(2)}_{\langle ij}M^{(4)}_{klm\rangle}\nonumber\\
&\left.-{155\over7}M^{(1)}_{\langle
ijk}M^{(5)}_{lm\rangle}-{41\over7}M^{(1)}_{\langle ij}M^{(5)}_{klm\rangle}
-{34\over7}M_{\langle ijk}M^{(6)}_{lm\rangle}-{15\over7}M_{\langle
ij}M^{(6)}_{klm\rangle}\right\}\nonumber\\&
+\mathcal{O}\left(\frac{1}{c^4}\right)\label{U5}\,,\\ V_{ijkl}(T_R) &=
S^{(4)}_{ijkl}(T_R) + {G\over c^3}\left\{2 M \int_{-\infty}^{T_R} d \tau \left[
\ln \left({T_R-\tau\over2\tau_0}\right)+{119\over60} \right]S^{(6)}_{ijkl}
(\tau)\right.\nonumber\\ &\left.-{35\over3}S^{(2)}_{\langle
ij}M^{(3)}_{kl\rangle}-{25\over3}M^{(2)}_{\langle ij}S^{(3)}_{kl\rangle}
-{65\over6}S^{(1)}_{\langle ij}M^{(4)}_{kl\rangle}-{25\over6}M^{(1)}_{\langle
ij}S^{(4)}_{kl\rangle} -{19\over6}S_{\langle
ij}M^{(5)}_{kl\rangle}\right.\nonumber\\ &-{11\over6}M_{\langle
ij}S^{(5)}_{kl\rangle}-{11\over12}S_{\langle i}M^{(5)}_{jkl\rangle}
+{1\over6}\varepsilon_{ab\langle i}\left[-5M^{(3)}_{ja}M^{(3)}_{kl\rangle b}
-{11\over2}M^{(4)}_{ja}M^{(2)}_{kl\rangle b}
-{5\over2}M^{(2)}_{ja}M^{(4)}_{kl\rangle b}\right.\nonumber\\
&\left.\left.-{1\over2}M^{(5)}_{ja}M^{(1)}_{kl\rangle b}
+{37\over10}M^{(1)}_{ja}M^{(5)}_{kl\rangle b}
+{3\over10}M^{(6)}_{ja}M_{kl\rangle b}+{1\over2}M_{ja}M^{(6)}_{kl\rangle
b}\right]\right\} \nonumber\\&
+\mathcal{O}\left(\frac{1}{c^4}\right)\,.\label{V4}
\end{align}\end{subequations}
For all the other moments that are required, it is sufficient to assume the
agreement between the radiative and canonical moments,
\begin{subequations}\label{ULVL}
\begin{align}
U_L (T_R) &= M^{(\ell)}_L(T_R) + \mathcal{O}\left(\frac{1}{c^3}\right)\,,\\ V_L
(T_R) &= S^{(\ell)}_L(T_R) + \mathcal{O}\left(\frac{1}{c^3}\right)\,.
\end{align}\end{subequations}

\subsection{The canonical moments for 3PN polarisations}\label{secVB}

Following the investigation of Section \ref{secIV} we now give the canonical
moments in terms of source-rooted multipole moments. It turns out that the
difference between these two types of moments --- which is due to the presence
of the gauge moments defined by \eqref{phi1} --- arises only at the small
2.5PN order. The consequence is that we have to worry about this difference
only for the 3PN canonical mass quadrupole moment $M_{ij}$, the 2.5PN mass
octopole moment $M_{ijk}$, and the 2.5PN current quadrupole moment
$S_{ij}$. For the mass quadrupole moment, the requisite correction has already
been used in \cite{ABIQ04} and is given by\footnote{The equation (11.7a) in
\cite{BIJ02} contains a sign error with respect to the original result
\cite{B96} (with no consequence for any of the results in \cite{BIJ02}). The
correct sign is reproduced here.}
\begin{equation}\label{M2}
M_{ij}= I_{ij}+\frac{4G}{c^5} \left[W^{(2)}I_{ij}-W^{(1)}I_{ij}^{(1)}\right] +
\mathcal{O}\left(\frac{1}{c^7}\right)\,,
\end{equation}
where $I_{ij}$ denotes the source mass quadrupole, and where $W$ is the
monopole corresponding to the gauge moments $W_L$ (\textit{i.e.} $W$ is the
moment having $\ell=0$). At the PN order we are working, $W$ is needed only at
Newtonian order and will be provided in Section \ref{secVC}. Notice that the
remainder in \eqref{M2} is at order 3.5PN --- consistently with the accuracy
we aim here. The expression \eqref{M2} is valid in a mass-centred frame
defined by the vanishing of the mass dipole moment: $I_i=0$. Note that a
formula generalizing \eqref{M2} to all PN orders (and all multipole
interactions) is not possible at present and needs to be investigated anew for
specific cases. Thus it is convenient in the present approach to use
systematically the source moments $\{I_L,J_L,W_L,X_L,Y_L,Z_L\}$ as the
fundamental variables describing the source.

Similarly, the other moments $M_{ijk}$ and $S_{ij}$ will admit some correction
terms starting at the 2.5PN order. We have computed these new corrections,
together with recomputed and confirmed those in \eqref{M2}, by following the
method of Section \ref{secIV}, \textit{i.e.} evaluating the STF tensors $A_L,
\cdots, L_L$ in \eqref{U2comp} by means of the two practical lemmas
\eqref{XPQ} and \eqref{YPQ}, then plugging these tensors into
\eqref{deltaIJ}. We also performed an independent calculation by implementing
the general MPM algorithm of Section \ref{secIII} starting directly with the
general linearized metric \eqref{hgen1} parametrized by the source moments,
instead of the canonical metric \eqref{hcan1} parametrized by the canonical
moments.

In this second approach, which fully confirmed the previous results, we need
to know beforehand the relevant multipole interactions and we used the same
selection rules \eqref{selection}--\eqref{eq:lmax} as before. The only
difference is that the quadratic interaction we consider,
$A^{(p_1)}_{K_1}\times A^{(p_2)}_{K_2}$, are between any two source multipole
moments composed of the main moments $\{I_L,J_L\}$ and the gauge moments
$\{W_L,X_L,Y_L,Z_L\}$, \textit{i.e.} the letter $A$ symbolizes now any of the
$I$, $J$, $W$, $X$, $Y$ or $Z$. By applying the selection rules
\eqref{selection}--\eqref{eq:lmax} at 3PN order, \textit{i.e.} for $m=6$, it
is straightforward to check that (i) no gauge multipole moment can enter cubic
interactions up to our approximation level, and (ii) all quadratic
interactions involving at least one gauge moment have to be instantaneous,
meaning that $q=-1$ for them. We can in fact determine all possible
contributions by inspection. Their full list is given in table
\ref{tab:gauge_moments}.
\begin{table}[here]\caption{ \label{tab:gauge_moments} 
Non-linear corrections in $M_{ijk}$ and $S_{ij}$ involving at least one gauge
multipole moment at 3PN order. The first entry indicates, for each
interaction, which radiative moment it belongs to, whereas the second entry
tells us how many time derivatives are involved. STF symbols are omitted.}
\vspace{0.3cm}\begin{tabular}{l c c| c c c c| c} \hline number of
$\partial_t$ & \multicolumn{2}{c|}{1 ($\ge p$)} & \multicolumn{4}{c|}{2 ($\ge
p$)} & \multicolumn{1}{c}{3 ($\ge p$)}\\
 \hline
in $M_{ijk}$ & $I
\!\!\times\!\! Y_{ijk}^{(1)}$ & $I_{ij}^{(1-p)} \!\!\times\!\! Y_{k}^{(p)}$ &
$I_{ijk}^{(2-p)} \!\!\times\!\! W^{(p)}$ & $I \!\!\times\!\! W_{ijk}^{(2)}$ &
$I_{ij}^{(2-p)} \!\!\times\!\! W_k^{(p)}$ &-- & --\\ \hline in $S_{ij}$ & $J_i
\!\!\times\!\!  Y_j^{(1)}$ & --& $I \!\!\times\!\! Z_{ij}^{(2)}$ &
$\varepsilon_{iab} I_{aj}^{(2-p)} \!\!\times\!\! Y_b^{(p)}$ & $J_i
\!\!\times\!\! W_j^{(2)}$ & $J_{ij}^{(2-p)} \!\!\times\!\! W^{(p)}$ &
$\varepsilon_{iab} I_{aj}^{(3-p)} \!\!\times\!\! W_b^{(p)}$ \\ \hline
\end{tabular} 
\end{table}

Further rules of selection might be used to discard some candidates, but all
the contributions to $U_{ijk}$ and $V_{ij}$ that are presented here have been
computed explicitly. By retaining only the interactions that involve the pairs
of multipole moments composing the elements of table
\ref{tab:gauge_moments},\footnote{\textit{I.e.} the interactions
$I\!\!\times\!\! Y_{ijk}$, $I_i\!\!\times\!\! Y_j$, $I\!\!\times\!\! W_{ijk}$,
$I_{ij}\!\!\times\!\!W_k$, $J_i\!\!\times\!\!Y_j$, $I\!\!\times\!\!Z_{ij}$,
$J_i\!\!\times\!\!W_j$ and $J_{ij}\!\!\times\!\!W$.}  the source used in our
algorithms could indeed be reduced to a finite, reasonably small number of
terms. However the detailed calculation of some of these interactions turns
out to yield zero; this is the case for instance of the interaction $I
\!\!\times\!\!  Z_{ij}^{(2)}$ which does not contribute to $S_{ij}$.

Finally our explicit results for $M_{ijk}$ and $S_{ij}$ are
\begin{subequations}\label{M3S2}\begin{align}
M_{ijk} &= I_{ijk} + {4G\over
c^5}\left[W^{(2)}I_{ijk}-W^{(1)}I_{ijk}^{(1)}+3\,I_{\langle ij}Y_{k\rangle
}^{(1)}\right] + \mathcal{O}\left(\frac{1}{c^6}\right)\label{M3} \,,\\ S_{ij} &=
J_{ij} +{2G\over c^5}\left[\varepsilon_{ab\langle i}\left(-I_{j\rangle
b}^{(3)}W_{a}-2I_{j\rangle b}Y_{a}^{(2)} +I_{j\rangle
b}^{(1)}Y_{a}^{(1)}\right)+3J_{\langle i}Y_{j\rangle
}^{(1)}-2J_{ij}^{(1)}W^{(1)}\right] \nonumber\\& +
\mathcal{O}\left(\frac{1}{c^6}\right)\,,\label{S2}
\end{align}\end{subequations}
where $W_i$ and $Y_i$ are the dipole moments corresponding to the moments
$W_L$ and $Y_L$. The remainders in \eqref{M3S2} are consistent with our
approximation 3PN for the FWF. Besides the mass quadrupole moment \eqref{M2},
and mass octopole and current quadrupole moments \eqref{M3S2}, we can state
that, with the required 3PN precision, all the other moments $M_L$ agree with
their corresponding $I_L$, and similarly the $S_L$ agree with $J_L$, namely 
\begin{subequations}\label{MLSL}
\begin{align}
M_L &= I_L + \mathcal{O}\left(\frac{1}{c^5}\right)\,,\\ S_L &= J_L +
\mathcal{O}\left(\frac{1}{c^5}\right)\,.
\end{align}\end{subequations}

\subsection{The source moments for 3PN polarisations}\label{secVC}

We have finally succeeded in parametrizing the FWF entirely in terms of the
source moments $\{I_L,J_L,W_L,X_L,Y_L,Z_L\}$ up to 3PN order. The interest of
this construction lies in the fact that the source moments are known for
general PN matter systems. They were obtained by matching the external MPM
field of the source to the internal PN field valid in the source's near zone
\cite{B95,B96,B98mult}. The source moments have been worked out in the case of
compact binary systems with increasing PN precision
\cite{BDI95,BIJ02,BI04mult,ABIQ04}. Here we list all the required $I_L$'s and
$J_L$'s (and also the few needed gauge moments) for non-spinning compact
objects and for circular orbits. We do not enter the details because the
derivation of these moments follows exactly the same techniques as in
\cite{BIJ02,BI04mult}.

The only moment needed at the 3PN order is the mass quadrupole moment
$I_{ij}$, first computed for circular orbits in \cite{BIJ02} and subsequently
extended to general orbits in \cite{BI04mult}. We write it as
\begin{equation}\label{I3PN}
I_{ij} = \nu\,m\,\left(A\,x_{\langle
ij\rangle}+B\,{r^3\over G m} \,v_{\langle ij\rangle} +
C\,\sqrt{{r^3\over G m}} \,x_{\langle
i}v_{j\rangle}\right)+\mathcal{O}\left(\frac{1}{c^7}\right)\,.
\end{equation} 
The relative position and velocity of the two bodies in harmonic coordinates
are denoted by $x^i =y_1^i - y_2^i$ and $v^i = dx^i/dt = v_1^i - v_2^i$
(spatial indices are lowered and raised with the Kronecker metric so that
$x_i=x^i$ and $v_i=v^i$). The distance between the two particles in harmonic
coordinates is denoted $r=|\mathbf{x}|$. The two masses are $m_1$ and $m_2$,
the total mass is $m=m_1+m_2$ (not to be confused with the mass monopole
moment $M$), the symmetric mass ratio $\nu=m_1m_2/m^2$ satisfies $0<\nu\leq
1/4$, and the mass difference ratio is $\Delta=(m_1-m_2)/m$ which reads also
$\Delta=\pm\sqrt{1-4\nu}$ (according to the sign of $m_1- m_2$). To express
the coefficients $A$, $B$ and $C$ in \eqref{I3PN} as PN series we introduce
the small post-Newtonian parameter
\begin{equation}\label{gamma}
\gamma = {Gm\over rc^2}\,.
\end{equation} 
With these notations we have (in the frame of the `center-of-mass' and for
circular orbits)
\begin{subequations}\label{ABC}\begin{align}
A &= 1 + \gamma \left(-{1\over 42}-{13\over 14}\nu \right) + \gamma^2
\left(-{461\over 1512} -{18395\over 1512}\nu - {241\over 1512} \nu^2\right)
\nonumber \\ &~~+ \gamma^3 \left({395899\over 13200}-{428\over 105}\ln
\left({r\over r_0}\right) +\left[{3304319\over 166320} - {44\over 3}\ln
\left({r\over {r'}_0}\right)\right]\nu \right.\nonumber\\ &~~\qquad +
\left. {162539\over 16632} \nu^2 + {2351\over 33264}\nu^3 \right) \,,\\ B &=
\gamma \left({11\over 21}-{11\over 7}\nu\right) + \gamma^2 \left({1607\over
378}-{1681\over 378} \nu +{229\over 378}\nu^2\right) \nonumber \\ &~~+
\gamma^3 \left(-{357761\over 19800}+{428\over 105} \ln \left({r\over r_0}
\right) - {92339\over 5544} \nu + {35759\over 924} \nu^2 + {457\over 5544}
\nu^3 \right)\,,\\C &= {48\over 7} \,\gamma^{5/2} \,\nu\,.
\end{align}\end{subequations}
The coefficients $A$ and $B$ correspond to conservative PN orders (which are
even), while the coefficient $C$ involves a single term at the odd 2.5PN order
due to radiation reaction.

Notice the appearance of logarithms in both $A$ and $B$ at the 3PN
order. These logarithms have two distinct origins, depending on whether they
are scaled with the constant $r_0$ associated with the finite part
prescription in \eqref{u2}, or with an alternative constant denoted
${r'}_0$. The logarithms with $r_0$ will combine later with other
contributions due to tails and tails-of-tails, and the constant $r_0$ will be
absorbed into some unobservable shift of the binary's orbital phase, as can
already be seen from the fact that $r_0$ is associated with the difference of
origin of time between harmonic and radiative coordinates, see \eqref{TR}. 

The other constant ${r'}_0$ is defined by
$m\ln{r'}_0=m_1\ln{r'}_1+m_2\ln{r'}_2$, where ${r'}_1$ and ${r'}_2$ are two
\textit{regularization constants} appearing in a Hadamard self-field
regularization scheme for the 3PN equations of motion of point masses in
harmonic coordinates \cite{BF00,BFeom}. The constant ${r'}_0$ is therefore
present in the 3PN equations of motion and we shall thus also meet this
constant in the 3PN orbital frequency given by \eqref{omega3PN} below. The
regularization constant ${r'}_0$ is unobservable, since it can be removed by a
coordinate transformation at 3PN order --- ${r'}_0$ can rightly be called a
\textit{gauge constant}. In practice this means that ${r'}_0$ will cancel out
when using the 3PN equations of motion to compute the time derivatives of the
3PN quadrupole moment, as will be explicitly verified in Section
\ref{secVI}.\footnote{Note also that the 3PN quadrupole moment
\cite{BIJ02,BI04mult} depended originally on three constants $\xi$, $\kappa$,
$\zeta$ (called ambiguity parameters) reflecting some incompleteness of the
Hadamard self-field regularization. These constants have been computed by
means of the powerful dimensional regularization \cite{BDE04,BDEI04}, and we
have replaced the result, which was $\xi=-\frac{9871}{9240}$, $\kappa=0$ and
$\zeta=-\frac{7}{33}$, back into \eqref{ABC}.}

The list of required moments continues with the 2.5PN order at which we need
the mass octupole and current quadrupole given by (with
$\Delta=\frac{m_1-m_2}{m}$)
\begin{subequations}\begin{align}
I_{ijk} &= -\nu\,m\,\Delta\,\left\{ {x}_{\langle ijk\rangle} \left[1 -\gamma
\nu- \gamma^2 \left({139\over 330}+{11923\over 660}\nu +{29\over
110}\nu^2\right) \right]\right.\nonumber\\ &~~+{r^2\over c^2}\,x_{\langle
i}v_{jk\rangle} \left[1 - 2\nu - \gamma \left(-{1066\over
165}+\left.{1433\over 330}\nu -{21\over 55} \nu^2\right)
\right]\right.\nonumber\\ &~~\left.+{196\over 15} \,{r\over
c}\,\gamma^2\,\nu\,x_{\langle ij}v_{k\rangle} \right\}
+\mathcal{O}\left(\frac{1}{c^6}\right)\label{I3}\,,\\ J_{ij} &= -\nu\,m\,
\Delta\,\left\{ \varepsilon_{ab\langle i} x_{j\rangle a}v_b \left[1 +\gamma
\left({67\over 28}-{2\over 7}\nu \right)+\gamma^2\left({13\over 9} -{4651\over
252}\nu -{1\over 168}\nu^2 \right)\right]\right.\nonumber\\
&~~\left.-{188\over 35} \,{r\over c}\,\gamma^2\,\nu\,\varepsilon_{ab\langle
i} v_{j\rangle a}x_b \right\} +\mathcal{O}\left(\frac{1}{c^6}\right)\,.\label{J2}
\end{align}\end{subequations}
At 2PN order we require:
\begin{subequations}\begin{align}
I_{ijkl} &= \nu \,m\,\left\{ x_{\langle ijkl\rangle}\left[1 - 3\nu + \gamma
\left({3\over 110} - {25\over 22}\nu + {69\over
22}\nu^2\right)\right.\right.\nonumber\\
&~~+\left.\left.\gamma^2\left(-{126901\over200200}-{58101\over2600}\nu
+{204153\over2860}\nu^2+{1149\over1144}\nu^3\right) \right]\right.\nonumber\\
&~~\left.+{r^2\over c^2}\,x_{\langle ij}v_{kl\rangle}\left[{78\over55} ( 1 -
5\nu + 5\nu^2 )\right.\right. \nonumber \\
&~~+\left.\left.\gamma\,\left({30583\over3575}-{107039\over3575}\nu
+{8792\over715}\nu^2-{639\over715}\nu^3\right) \right]\right. \nonumber \\
&~~\left.+{71\over715}\,{r^4\over c^4}\,v_{\langle
ijkl\rangle}\left(1-7\nu+14\nu^2-7\nu^3\right)\right\}
+\mathcal{O}\left(\frac{1}{c^5}\right)\label{I4}\,,\\ J_{ijk} &= \nu\,m
\,\left\{ \varepsilon_{ab\langle i} x_{jk\rangle a} v_b \left[1 - 3\nu +
\gamma \left({181\over 90} - {109\over 18}\nu + {13\over
18}\nu^2\right)\right. \right.\nonumber \\
&~~+\left.\left. \gamma^2\left({1469\over3960}-{5681\over264}\nu
+{48403\over660}\nu^2-{559\over3960}\nu^3\right)\right]\right.\nonumber \\
&~~+\left.{r^2\over c^2}\,\varepsilon_{ab\langle i}x_a v_{jk\rangle
b}\left[{7\over45}\left(1 - 5\nu +
5\nu^2\right)+\gamma\left({1621\over990}-{4879\over990}\nu+{1084\over495}\nu^2
-{259\over990}\nu^3\right)\right]\right\} \nonumber \\
&~~+\mathcal{O}\left(\frac{1}{c^5}\right)\,.\label{J3}
\end{align}\end{subequations}
At 1.5PN order:
\begin{subequations}\begin{align}
I_{ijklm}&=-\nu\,m\,\Delta\, \left\{ x_{\langle
ijklm\rangle}\left[1-2\nu+\gamma\left({2\over 39}-{47\over 39}\nu+{28\over
13}\nu^2 \right)\right]\right.\nonumber \\ &~~+\left. \frac{70}{39}\,{r^2
\over c^2} x_{\langle ijk}v_{lm\rangle}\left(1-4\nu+3\nu^2\right)\right\}
+\mathcal{O}\left(\frac{1}{c^4}\right)\label{I5}\,,\\ J_{ijkl} &= -\nu\,m \,
\Delta\,\left\{ \varepsilon_{ab\langle i}x_{jkl\rangle a} v_b \left[1-2\nu
+\gamma \left({20\over 11}-{155\over 44}\nu+{5\over 11}\nu^2\right)
\right]\right.\nonumber\\&~~\left.  +\frac{4}{11}\,{r^2\over
c^2}\,\varepsilon_{ab\langle i}x_{ja} v_{kl\rangle
b}\left(1-4\nu+3\nu^2\right)\right\}
+\mathcal{O}\left(\frac{1}{c^4}\right)\,.\label{J4}
\end{align}\end{subequations}
At 1PN order:
\begin{subequations}\begin{align}
I_{ijklmn}&=\nu\,m\,\left\{ x_{\langle ijklmn\rangle}\left[1-5 \nu+5 \nu^2
+\gamma\,\left({1\over14}-{3\over2}\nu+6\nu^2-{11\over2}\nu^3\right)\right]\right.\nonumber
\\ &~~+\left.{15\over7}\,{r^2\over c^2}\, x_{\langle
ijkl}v_{mn\rangle}\left(1-7\nu+14\nu^2-7\nu^3\right) \right\} +
\mathcal{O}\left(\frac{1}{c^4}\right)\label{I6}\,,\\ J_{ijklm} &= \nu\,m \,
\left\{ \varepsilon_{ab\langle i}x_{jklm\rangle a}v_b\left[ 1-5\nu + 5\nu^2 +
\gamma\left({1549\over910}-{1081\over130}\nu+{107\over13}\nu^2
-{29\over26}\nu^3\right)\right]\right. \nonumber \\
&~~\left.+{54\over91}\,{r^2\over c^2}\varepsilon_{ab\langle
i}x_{jka}v_{lm\rangle b} \left(1-7\nu+14\nu^2-7\nu^3\right) \right\} +
\mathcal{O}\left(\frac{1}{c^4}\right)\,.\label{J5}
\end{align}\end{subequations}
At 0.5PN order:
\begin{subequations}\begin{align}
I_{ijklmno}&= -\nu\,m\,\Delta\, (1-4 \nu + 3 \nu^2)\,x_{\langle
ijklmno\rangle} + \mathcal{O}\left(\frac{1}{c^2}\right)\label{I7}\,,\\ J_{ijklmn}
&= -\nu\,m \,\Delta\,(1-4\nu + 3\nu^2)\,\varepsilon_{ab\langle i}
x_{jklmn\rangle a}v_{b} + \mathcal{O}\left(\frac{1}{c^2}\right)\,.\label{J6}
\end{align}\end{subequations}
At Newtonian order:
\begin{subequations}\begin{align}
I_{ijklmnop}&=\nu\,m\,\left(1-7\nu+14\nu^2-7\nu^3\right)\,x_{\langle
ijklmnop\rangle} + \mathcal{O}\left(\frac{1}{c^2}\right)\label{I8}\,,\\
J_{ijklmno}&= \nu\,m \,\left(1-7\nu+14\nu^2
-7\nu^3\right)\,\varepsilon_{ab\langle i}x_{jklmno\rangle a}v_{b} +
\mathcal{O}\left(\frac{1}{c^2}\right)\,.\label{J7}
\end{align}\end{subequations}
The 2.5PN correction terms in $I_{ijk}$ and $J_{ij}$, the 2PN terms in
$I_{ijkl}$ and $J_{ijk}$, and the 1PN terms in $I_{ijklm}$ and $J_{ijkl}$ are
new with this paper. The higher-order Newtonian moments $I_{ijklmno}$ and
$J_{ijklmn}$ were also not needed before, but Newtonian moments are trivial
and are given for general $\ell$ by
\begin{subequations}\begin{align}
I_L &= \nu\,m\,s_\ell(\nu)\,x_{\langle L\rangle}+
\mathcal{O}\left(\frac{1}{c^2}\right)\label{Il}\,,\\ J_{L-1} &=
\nu\,m\,s_\ell(\nu)\,\varepsilon_{ab\langle i_{\ell-1}} x_{L-2\rangle a}v_b+
\mathcal{O}\left(\frac{1}{c^2}\right)\,,\label{Jl}
\end{align}\end{subequations}
in which we pose
\begin{equation}\label{sell}
s_\ell(\nu)=X_2^{\ell-1}+(-)^\ell X_1^{\ell-1}\,.
\end{equation}
Here we define $X_1=\frac{m_1}{m}=\frac{1}{2}(1+\Delta)$ and
$X_2=\frac{m_2}{m}=\frac{1}{2}(1-\Delta)$ with
$\Delta=\frac{m_1-m_2}{m}=\pm\sqrt{1-4\nu}$, so that $X_1+X_2=1$ and $X_1
X_2=\nu$.\footnote{Note that the coefficient $s_\ell(\nu)$ is equal to the
product $\tilde{m} f_k(\nu)$ in the notation of \cite{K07}. The equivalence of
the two expressions follows from the Waring formulas \cite{MacMahon} for
$X_1^n+X_2^n$ and $X_1^n-X_2^n$. We find (where $\left({n\atop p}\right)$ is
the usual binomial coefficient)
\begin{eqnarray*}
s_{2k}(\nu)&=&\sum_{p=0}^{k-1}(-)^p\frac{2k-1}{2k-1-p}\left({2k-1-p\atop
  p}\right)\nu^p\,,\\s_{2k+1}(\nu)&=&-\Delta\sum_{p=0}^{k-1}(-)^p
  \left({2k-1-p\atop   
  p}\right)\nu^p\,.
\end{eqnarray*}
}

In addition we shall need the mass monopole $I$ agreeing with its canonical
counterpart $M$ which parametrizes the various tail terms in Section
\ref{secVA}. Since the tails arise at 1.5PN order we need $M$ only at the
1.5PN relative order. It is given by
\begin{equation}\label{M0}
I = M = m \left(1 - \frac{\nu}{2}\,\gamma\right) +
\mathcal{O}\left(\frac{1}{c^4}\right)\,.
\end{equation}
We require also the current dipole moment or angular momentum $J_i$ (agreeing
with its canonical counterpart $S_i$) since it appears in some non-linear
terms, for instance in \eqref{U2}. It is needed only at Newtonian order,
\begin{equation}\label{J1}
J_i = S_i = \nu\,m \, \varepsilon_{iab}x_av_b +
\mathcal{O}\left(\frac{1}{c^2}\right)\,.
\end{equation}

Finally, we have to provide the few gauge moments that enter the relations
between canonical and source moments found in \eqref{M2} and
\eqref{M3S2}. They are readily computed from the general expressions of all
the gauge moments $\{W_L,X_L,Y_L,Z_L\}$ given in (5.15)--(5.20) of
\cite{B98mult}. The calculation is quite simple because these moments, namely
the monopolar moment $W$ and the two dipole moments $W_i$ and $Y_i$, are
Newtonian. For circular orbits we find
\begin{subequations}\label{WY}\begin{align}
W &= \mathcal{O}\left(\frac{1}{c^2}\right)\label{W0}\,,\\ W_i &= {1\over 10} \,
\nu\,m \,\Delta\, r^2\,v^i +
\mathcal{O}\left(\frac{1}{c^2}\right)\label{W1}\,,\\Y_i &= {1\over 5} \,
\frac{G\,m^2\,\nu}{r} \,\Delta\,x^i +
\mathcal{O}\left(\frac{1}{c^2}\right)\,.\label{Y1}
\end{align}\end{subequations}
We are done with all the source multipole moments needed to control the 3PN
accurate FWF generated by compact binary sources in quasi-circular orbits.

\section{Time derivatives of the source multipole moments}
\label{secVI}

For the purpose of computing the time derivatives of the source moments we
require the 3PN accurate equations of motion of compact binary sources. Like
in the computation of the moments we have to take into account both the
conservative effects at 1PN, 2PN and 3PN orders, and the effect of radiation
reaction at 2.5PN order.

We consider non-spinning objects so the motion takes place in a fixed plane,
say the x-y plane. The relative position $\mathbf{x} = \mathbf{y}_1 -
\mathbf{y}_2$, velocity $\mathbf{v} = d\mathbf{x}/dt$, and acceleration
$\mathbf{a} = d\mathbf{v}/dt$ are given by
\begin{subequations}\label{xva}\begin{align}
\mathbf{x} &= r \,\mathbf{n} \,,\label{xx}\\ \mathbf{v} &= \dot r \,\mathbf{n}
+ r \,\omega \,\bm{\lambda}\,,\label{vv}\\ \mathbf{a} &= (\ddot{r} - r
\,\omega^2) \,\mathbf{n} + (r \,\dot{\omega} + 2 \dot{r} \,\omega)
\,\bm{\lambda}\,.\label{aa}
\end{align}\end{subequations}
For a while the time derivative will be denoted using an over dot. Here
$\bm{\lambda} = \hat{\mathbf{z}}\times\mathbf{n}$ is perpendicular to the unit
vector $\hat{\mathbf{z}}$ along the z-direction orthogonal to the orbital
plane, and to the binary's separation direction $\mathbf{n}$. The orbital
frequency $\omega$ is related in the usual way to the orbital phase $\phi$ by
$\omega = \dot \phi$.

Through 3PN order, it is possible to model the motion of the binary as a
\textit{quasi-circular} orbit decaying by the effect of radiation reaction at
the 2.5PN order. This effect is computed by balancing the change in the
orbital energy with the total energy flux radiated by the gravitational
waves. At 2.5PN order this yields (see \textit{e.g.} \cite{KBI07})
\begin{subequations}\label{romdot}\begin{align}
\dot{r} &= - \frac{64}{5} \sqrt{\frac{G m}{r}}~\nu\,\gamma^{5/2} +
\mathcal{O}\left(\frac{1}{c^7}\right)\label{rdot}\,,\\ \dot{\omega} &=
\frac{96}{5} \,\frac{G m}{r^3}\,\nu\,\gamma^{5/2}+
\mathcal{O}\left(\frac{1}{c^7}\right)\label{omdot}\,,
\end{align}\end{subequations}
where $\gamma$ is given by \eqref{gamma}. By substituting those expressions
into \eqref{xva},\footnote{We notice that $\ddot{r}=\mathcal{O}(c^{-10})$ is
of the order of the \textit{square} of radiation-reaction effects and is
therefore zero with this approximation.} we obtain
the expressions for the inspiral velocity and acceleration,
\begin{subequations}\label{va3PN}\begin{align}
\mathbf{v} &= r \,\omega \,\bm{\lambda} - \frac{64}{5} \sqrt{\frac{G
m}{r}}~\nu\,\gamma^{5/2}\,\mathbf{n} +
\mathcal{O}\left(\frac{1}{c^7}\right)\label{v3PN}\,,\\ \mathbf{a} &= -\omega^2
\,\mathbf{x} - \frac{32}{5}\,\sqrt{\frac{G
m}{r^3}}\,\,\nu\,\gamma^{5/2}\,\mathbf{v} +
\mathcal{O}\left(\frac{1}{c^7}\right)\label{a3PN}\,.
\end{align}\end{subequations}

A central result of PN calculations of the equations of motion is the
expression of the orbital frequency $\omega$ in terms of the binary's
separation $r$ up to 3PN order. This result has been obtained in harmonic
coordinates in \cite{BF00,BFeom,BDE04} and independently in
\cite{IFA01,itoh1,itoh2}, and in ADM coordinates in
\cite{JaraS98,JS99,DJSdim}. In the present work $r$ is given in harmonic
coordinates and the expression of the 3PN orbital frequency is
\begin{align}\label{omega3PN}
\omega^2 &= {G m\over r^3}\biggl\{ 1+\gamma\Bigl(-3+\nu\Bigr) + \gamma^2
\left(6+\frac{41}{4}\nu +\nu^2\right) \\ &~~+\gamma^3
\left(-10+\left[-\frac{75707}{840}+\frac{41}{64}\pi^2
+22\ln\left(\frac{r}{r'_0}\right) \right]\nu +\frac{19}{2}\nu^2+\nu^3\right) +
\mathcal{O}\left(\frac{1}{c^8}\right) \biggr\}\nonumber\,.
\end{align}
Note that the logarithm at 3PN order involves the same constant ${r'}_0$ as in
the source quadrupole moment \eqref{I3PN}--\eqref{ABC}. This logarithm comes
from a Hadamard self-field regularization scheme and its appearance is
specific to harmonic coordinates.

As often convenient we shall use in place of the parameter $\gamma$ given by
\eqref{gamma} an alternative parameter $x$ directly linked to the orbital
frequency \eqref{omega3PN}, namely
\begin{equation}\label{x}
x = \left({G\,m\,\omega \over c^3}\right)^{2/3}\,.
\end{equation}
The interest in this parameter stems from its invariant meaning in a large
class of coordinate systems including the harmonic and ADM coordinate
systems. At 3PN order it is given in terms of $x$ by
\begin{align}\label{gammax}
\gamma &= x \biggl\{ 1+x \left(1-\frac{\nu}{3}\right) + x^2
 \left(1-\frac{65}{12}\nu\right) \,\\ &~~+ x^3
 \left(1+\left[-\frac{2203}{2520}-\frac{41}{192}\pi^2
 -\frac{22}{3}\ln\left(\frac{r}{r'_0}\right) \right]\nu
 +\frac{229}{36}\nu^2+\frac{\nu^3}{81}\right) +
 \mathcal{O}\left(\frac{1}{c^8}\right) \biggr\}\nonumber\,.
\end{align}
Combining \eqref{omega3PN} with \eqref{gammax} we find that the velocity
squared $v^2=r^2\omega^2+\dot{r}^2=r^2\omega^2+\mathcal{O}(c^{-10})$ is
related to $x$ by
\begin{align}\label{vv2}
\left(\frac{v}{c}\right)^2 &= x \biggl\{ 1+x \left(-2+\frac{2}{3}\nu\right) +
x^2 \left(1+\frac{53}{6}\nu+\frac{\nu^2}{3}\right) \,\\ &~~+ x^3
 \left(\left[-\frac{36227}{1260}+\frac{41}{96}\pi^2
 +\frac{44}{3}\ln\left(\frac{r}{r'_0}\right) \right]\nu
 -\frac{29}{9}\nu^2+\frac{10}{81}\nu^3\right) +
 \mathcal{O}\left(\frac{1}{c^8}\right) \biggr\}\nonumber\,.
\end{align}

During the computation of the time derivatives of the source moments, each
time an acceleration is produced the result is consistently \textit{order
reduced}, \textit{i.e.} the acceleration is replaced with \eqref{a3PN} at the
right PN order. Such an order reduction will generate in particular some 2.5PN
radiation-reaction terms which are to be taken into account in the 3PN
waveform. This occurs when computing the time derivatives of the moments
$I_{ij}$, $I_{ijk}$ and $J_{ij}$ that appear in the FWF at Newtonian and 0.5PN
orders. On the other hand, when computing the polarization states following
\eqref{hpc} we shall meet some scalar products of the polarization vectors
$\mathbf{P}$ and $\mathbf{Q}$ with the relative velocity $\mathbf{v}$. If
those scalar products occur at Newtonian and 0.5PN orders (\textit{i.e.} in
multipolar pieces corresponding to the moments $I_{ij}$, $I_{ijk}$ and
$J_{ij}$) we shall have to take into account the 2.5PN radiation-reaction term
coming from the expression of $\mathbf{v}$ given by \eqref{v3PN}.\footnote{Not
considering the radiation-reaction contribution in $\mathbf{v}$ given by
\eqref{v3PN} has been the source of an error in \cite{ABIQ04} which has been
pointed out and corrected in \cite{KBI07}.} However it was shown in
\cite{KBI07} that the radiation-reaction terms in the FWF at the 2.5PN order
can be absorbed into a modification of the orbital phase, where they appear to
constitute in fact a very small phase modulation, comparable with unknown
contributions in the phase being at least of order 5PN --- negligible here
since the phase is known only to 3.5PN order. In the present paper, we have
chosen\footnote{As usual there are many different ways of presenting PN
results at a given order of approximation, and choosing one or another is
often a matter of convenience.} to include all the radiation-reaction terms
coming from both \eqref{v3PN} and \eqref{a3PN}, and to present them as 2.5PN
and 3PN amplitude corrections in our final results which will be presented in
\eqref{Hplus}--\eqref{Hcross} and \eqref{hlm} below.

Let us next check that the Hadamard self-field regularization constant
${r'}_0$ appearing both in the 3PN orbital frequency \eqref{omega3PN} and in
the 3PN quadrupole moment \eqref{ABC},\footnote{The other moments are given at
2.5PN order at most; they do not depend on ${r'}_0$ since the appearance of
regularization constants is a feature of the 3PN approximation.} is actually a
gauge constant. To this end we simply verify that ${r'}_0$ will be eliminated
when expressing the FWF in terms of the gauge invariant parameter
\eqref{x}. From \eqref{ABC} we see that the dependence on ${r'}_0$ of the 3PN
quadrupole moment is
\begin{equation}\label{Ir0p}
I_{ij} = \nu\,m \left[ 1 - {44\over 3}\gamma^3\,\nu\,\ln \left({r\over
{r'}_0}\right)\right]\,x_{\langle ij\rangle} + \cdots +
\mathcal{O}\left(\frac{1}{c^7}\right)\,.
\end{equation}
We indicate by dots all the terms that are independent of ${r'}_0$ (for
convenience we also show the Newtonian term). Now the FWF depends on the
second time derivative of the quadrupole moment. For circular orbits this
reads [coming back to the superscript notation $(n)$ for time derivatives]
\begin{equation}\label{Ir0p2}
I_{ij}^{(2)} = 2 \nu\,m \left[ 1 - {44\over 3}\gamma^3\,\nu\,\ln \left({r\over
{r'}_0}\right) \right]\Bigl(v_{\langle ij\rangle}+x_{\langle
i}a_{j\rangle}\Bigr) + \cdots + \mathcal{O}\left(\frac{1}{c^7}\right)\,.
\end{equation}
Replacing $v_i$ and $a_i$ by their values \eqref{va3PN} we get with the
required approximation (still being interested only in the fate of the
constant ${r'}_0$)
\begin{equation}\label{Ir0p2'}
I_{ij}^{(2)} = 2 \nu\,m\,v^2\left[ 1 - {44\over 3}\gamma^3\,\nu\,\ln
\left({r\over {r'}_0}\right) \right]\Bigl(\lambda_{\langle
ij\rangle}-n_{\langle ij\rangle}\Bigr) + \cdots +
\mathcal{O}\left(\frac{1}{c^7}\right)\,.
\end{equation}
The squared velocity $v^2=r^2\omega^2+\mathcal{O}(c^{-10})$ appears in
factor. It is now clear that replacing $v^2$ by its expression in terms of the
parameter $x$ following \eqref{vv2}, we produce another logarithmic term
containing ${r'}_0$, namely
\begin{equation}\label{v2log}
v^2 = c^2\,x \left[ 1 + {44\over 3}x^3\,\nu\,\ln \left({r\over {r'}_0}\right)
\right] + \cdots + \mathcal{O}\left(\frac{1}{c^7}\right)\,,
\end{equation}
which will cancel out the dependence of the quadrupole moment on ${r'}_0$ at
3PN order (using the fact that $\gamma$ can be replaced by $x$ in a small 3PN
term). Thus, finally,
\begin{equation}\label{Ir0p2x}
I_{ij}^{(2)} = 2 \nu\,m\,c^2\,x\Bigl(\lambda_{\langle ij\rangle}-n_{\langle
ij\rangle}\Bigr) + \cdots + \mathcal{O}\left(\frac{1}{c^7}\right)\,,
\end{equation}
is independent on ${r'}_0$, which means that this constant cannot affect any
physical result at the 3PN order.

\section{Computation of the tail and memory integrals}
\label{secVII}

The results of Sections \ref{secV}--\ref{secVI} yield the complete control of
the \textit{instantaneous} part of the FWF. We now tackle the computation of
the \textit{hereditary} part, which is composed of tails (and tails-of-tails
and squared-tails) and non-linear memory terms. The hereditary integrals have
been explicitly provided in Section \ref{secVA} as contributions to the
various radiative moments $U_L$ and $V_L$ given by
\eqref{U2}--\eqref{U5V4}. Our computation will basically be a straightforward
extension of the computation performed at 2.5PN order in Section 4 of
\cite{ABIQ04}. Since we employ exactly the same techniques, we skip most of
the details and rely on \cite{ABIQ04} for justification of the method and
proofs.

We first consider the non-linear memory terms. Up to 3PN order we have the
2.5PN memory integrals in the radiative mass quadrupole moment $U_{ij}$ given
by \eqref{U2} and the radiative mass hexadecapole moment $U_{ijkl}$ given by
\eqref{U4} --- these are the memory terms contributing to the FWF at 2.5PN
order \cite{ABIQ04} --- and, in addition, we have the memory integral in the
mass octupole moment $U_{ijk}$ given by \eqref{U3} and the one in $U_{ijklm}$
given by \eqref{U5} --- these contribute specifically at 3PN
order.\footnote{Recall that the non-linear memory terms occur only in the
  \textit{mass-type} radiative multipole moments $U_L$.} Like in \cite{ABIQ04}
we obtain the corresponding integrands (\textit{i.e.} the terms under the
integral sign) and compute directly their contributions to the two wave
polarizations $h_+$ and $h_\times$. Indeed it is convenient to perform the
relevant contractions of the integrands with the polarization vectors
$\mathbf{P}$ and $\mathbf{Q}$ (see Section \ref{secVIII} for the conventions
we adopt) so as to only deal with \textit{scalar} quantities.

We find that the memory integrals in $h_+$ and $h_\times$ are composed of two
types of terms. First there is a term, only present in the plus polarization
$h_+$, which does not depend on the orbital phase and can thus be viewed as a
\textit{zero-frequency} (DC) term. Actually, because of the steady inspiral,
this term is a steadily varying function of time, with an amplitude increasing
like some power law of the time remaining till the coalescence. Strictly
speaking, this term is to be regarded as \textit{the} memory contribution
because it does depend on the behaviour of the system in the remote past, and
therefore must be computed using some model for the evolution of the binary
system in the past. In the present paper we find that the only zero-frequency
term up to 3PN order is the one which appeared already at 2.5PN order and was
evaluated in \cite{ABIQ04} --- interestingly there are no other terms of this
type at the 3PN order. Because of the cumulative effect of integration over
the whole past we know that this term, though originating from 2.5PN order,
finally contributes in the FWF at the Newtonian level
\cite{Chr91,WiW91,Th92}. In practice the computation of this DC term reduces
(in the circular orbit case) to the evaluation of the single elementary
integral
\begin{equation}\label{intI}
\mathrm{I}(T_R) = \frac{(G\,m)^{p-1}}{c^{2p-3}} \int_{-\infty}^{T_R}
\frac{d\tau}{r^p(\tau)}\,.
\end{equation}
Here $r(\tau)$ denotes the binary's separation at any time $\tau\leq T_R$
(where $T_R=T-R/c$ is the current time). The coefficient in front of
\eqref{intI} is chosen for convenience to make the integral dimensionless. The
integral \eqref{intI} is easily computed using a simplified model of binary
evolution in the past in which the orbit is assumed to remain circular apart
from the gradual inspiral at any time. In this model the binary separation
evolves like $r(\tau)\propto (T_c-\tau)^{1/4}$ where $T_c$ denotes the instant
of coalescence (see \cite{ABIQ04} for more details). In the remote past we
thus have $r(\tau)\sim (-\tau)^{1/4}$ so the integral \eqref{intI} converges
when $p>4$ (actually we shall only need the case $p=5$ like in
\cite{ABIQ04}). The result reads
\begin{equation}\label{intIres}
\mathrm{I}(T_R) = \frac{5}{64(p-4)}\,\frac{x^{p-4}(T_R)}{\nu}\,,
\end{equation}
where $x(T_R)$ denotes the \textit{current} value (\textit{i.e.} at the
current retarded time $T_R$) of the parameter $x$ defined by
\eqref{x}. Witness the memory effect: the end result \eqref{intIres} is of
order $x^{p-4}=\mathcal{O}(c^{-2p+8})$ which is a factor $c^5$ \textit{larger}
than the original formal PN order $\mathcal{O}(c^{-2p+3})$ as shown in
\eqref{intI}. Hence, although the memory term is formally of order 2.5PN, its
actual contribution to the waveform is comparable to a Newtonian term. As
mentioned 
above we do not find memory (zero-frequency) contributions originating from
the next 3PN order, and therefore finally no DC term at 0.5PN order.

Second there are other terms, present in both polarizations, which depend on
the orbital phase, and oscillate like some harmonics of the orbital phase (say
$n\,\phi$). Such phase-dependent, oscillating terms do not exhibit the memory
effect, essentially because the oscillations, due to the sequence of orbital
cycles in the entire life of the binary system, more or less compensate each
other. As a result these terms, in contrast with
\eqref{intI}--\eqref{intIres}, keep on their formal PN order. We recover the
2.5PN terms investigated in \cite{ABIQ04} and in addition we obtain several
other terms at 3PN order. The latter are computed by a slight generalization
of the method followed in \cite{ABIQ04}: instead of (4.18) in \cite{ABIQ04} we
need to consider the integral
\begin{equation}\label{intJ}
\mathrm{J}(T_R) = \frac{(G\,m)^{p-1}}{c^{2p-3}} \int_{-\infty}^{T_R}
d\tau\,\frac{e^{i \,n\,\phi(\tau)}}{r^p(\tau)}\,,
\end{equation}
where $\phi(\tau)$ is the orbital phase at any time, where $n$ and $p$ range
over integer or half-integer values (\textit{e.g.} $n=1,3,5$ and $p=11/2$ at
3PN order), and where the coefficient is chosen to make the integral
dimensionless. Following the steps (4.18)--(4.23) in \cite{ABIQ04} we compute
this integral using our model of binary's past evolution, and in the adiabatic
limit, which means that the \textit{current} value of the adiabatic parameter
$\xi$ associated with the binary inspiral is considered to be small and of PN
order $\xi(T_R)=\mathcal{O}(c^{-5})$. We then find
\begin{equation}\label{intJres}
\mathrm{J}(T_R) = x^{p-\frac{3}{2}}(T_R)\,\frac{e^{i \,n\,\phi(T_R)}}{i \,n}\left[1 +
\mathcal{O}\left(\frac{1}{c^5}\right)\right]\,.
\end{equation}
This result (valid only if $n\not= 0$) permits to handle all the
phase-dependent oscillating terms coming from the memory integrals.

We next turn to the computation of the tails and tails-of-tails present in the
radiative moments \eqref{U2}--\eqref{U5V4}. Again we closely follow the
previous investigation \cite{ABIQ04} on which we refer for more details. The
computation of tails reduces to the evaluation of an elementary integral
involving a logarithmic kernel,
\begin{equation}\label{intK}
\mathrm{K}(T_R) = \frac{(G\,m)^{p-1}}{c^{2p-3}} \int_{-\infty}^{T_R}
d\tau\,\frac{e^{i \,n\,\phi(\tau)}}{r^p(\tau)}
\ln\left(\frac{T_R-\tau}{T_c-T_R}\right)\,, 
\end{equation}
in which the logarithm has been scaled with the constant time $T_c-T_R$,
instead of the previous normalization by $2\tau_0$, where $T_c$ is the instant
of coalescence in the model of \cite{ABIQ04}. Such scaling can always be done
at the price of adding another term proportional to some integral of the type
$\mathrm{J}(T_R)$ computed previously. Following the derivation of this
integral in \cite{ABIQ04}, we find that, at dominant order in the adiabatic
approximation,
\begin{equation}\label{intKres}
\mathrm{K}(T_R) = x^{p-\frac{3}{2}}(T_R)\,\frac{e^{i
\,n\,\phi(T_R)}}{i\,n}\left[\frac{\pi}{2 i} -
\ln\left(\frac{n}{\xi(T_R)}\right) - C + \mathcal{O}\left(\frac{\ln
c}{c^5}\right)\right]\,.
\end{equation}
Here $C=0.577\cdots$ is the Euler constant, and $\xi(T_R)$ denotes the current
value of the adiabatic parameter associated with the inspiral, which is
defined by $\xi(T_R)=[(T_c-T_R)\omega(T_R)]^{-1}$ in the model of
\cite{ABIQ04}. The adiabatic parameter is related to the PN parameter $x$ by
\begin{equation}\label{abiabxi}
\xi(T_R)=\frac{256\nu}{5}\,x^{5/2}(T_R)\,.
\end{equation}
The squared-tails are computed using the same integral
\eqref{intK}--\eqref{intKres}. Concerning the tails-of-tails we simply have to
consider an integral involving a logarithm squared,
\begin{equation}\label{intL}
\mathrm{L}(T_R) = \frac{(G\,m)^{p-1}}{c^{2p-3}} \int_{-\infty}^{T_R}
d\tau\,\frac{e^{i \,n\,\phi(\tau)}}{r^p(\tau)}
\ln^2\left(\frac{T_R-\tau}{T_c-T_R}\right)\,,
\end{equation}
which is computed using the same technique with the result
\begin{equation}\label{intLres}
\mathrm{L}(T_R) = x^{p-\frac{3}{2}}(T_R)\,\frac{e^{i \,n\,\phi(T_R)}}{i
\,n}\left[\frac{\pi^2}{6} + \left(C + \ln\left(\frac{n}{\xi(T_R)}\right)+
\frac{i \pi}{2}\right)^2 + \mathcal{O}\left(\frac{\ln c}{c^5}\right)\right]\,.
\end{equation}
We are done with the computation of all tails and tails-of-tails in the 3PN
waveform.

For completeness let us give also the two technical formulas which enable ones
to arrive at the results \eqref{intKres} and \eqref{intLres}. Posing
$y=(T_R-\tau)/(T_c-T_R)$ and $\lambda=n/\xi$, and working at the leading order
in the adiabatic limit $\xi\rightarrow 0$ or equivalently when
$\lambda\rightarrow +\infty$, the formulas express that, for any positive or
negative $\lambda$ (see \textit{e.g.}  \cite{GZ} p. 573 and 574),
\begin{subequations}\label{formules}\begin{align}
\int_0^1 dy\,\ln y \, e^{-i\lambda y} &=
\frac{1}{\lambda}\left[-\frac{\pi}{2}\mathrm{sign}(\lambda)+i\bigl(\ln
|\lambda| + C \bigr)\right]
+\mathcal{O}\left(\frac{1}{\lambda^2}\right)\,,\label{formtail}\\ \int_0^1 d y
\,\ln^2 y \,e^{-i\,\lambda\,y} &=
\frac{i}{\lambda}\,\left(-\frac{\pi^2}{6}+\left[-\frac{\pi}{2}\mathrm{sign}(\lambda)+i(\ln
|\lambda|+C)\right]^2\right)
+\mathcal{O}\left(\frac{1}{\lambda^3}\right)\,.\label{formtail2}
\end{align}\end{subequations}
Notice that we are only interested in the \textit{recent past} contribution to
the integrals \eqref{formules}, corresponding to the interval $0\leq y\leq 1$
equivalent to the time interval $2T_R-T_c\leq \tau\leq T_R$. The reason is
that the \textit{remote past} contribution, given by $1<y<+\infty$ or
equivalently $-\infty<\tau<2T_R-T_c$, is small in the adiabatic limit. This is
a characteristic feature of tails: they die out very rapidly, therefore they
depend essentially on the recent past evolution of the matter source
\cite{BD92,BS93}. In the case at hand this technically means that the
remote-past contributions to the integrals are of order
\begin{subequations}\label{remotepast}\begin{align}
\int_1^{+\infty} dy\,\ln y\,e^{-i \,\lambda\,y} &=
  \mathcal{O}\left(\frac{1}{\lambda^2}\right)\,,\\ \int_1^{+\infty} dy\,\ln^2
  y\,e^{-i \,\lambda\,y} &= \mathcal{O}\left(\frac{1}{\lambda^3}\right)\,,
\end{align}\end{subequations}
as can easily be verified by using integration by parts.

\section{3PN polarization waveforms for data analysis}\label{secVIII}

We specify our conventions for the orbital phase and polarization vectors
defining the polarization waveforms \eqref{hpc} in the case of quasi-circular
binary systems of non-spinning compact objects. If the orbital plane is chosen
to be the x-y plane (like in Section \ref{secVI}), with the orbital phase
$\phi$ measuring the direction of the unit vector $\mathbf{n} = \mathbf{x}/r$
along the relative separation vector, then
\begin{equation}\label{n}
\mathbf{n} = \hat{\mathbf{x}}\,\cos{\phi} + \hat{\mathbf{y}}\,\sin{\phi}\,,
\end{equation}
where $\hat{\mathbf{x}}$ and $\hat{\mathbf{y}}$ are the unit directions along
x and y. Following \cite{BIWW96,ABIQ04} we choose the polarization vector
$\mathbf{P}$ to lie along the x-axis and the observer to be in the y-z plane
with
\begin{equation}\label{N}
\mathbf{N} = s_i\,\hat{\mathbf{y}} + c_i\,\hat{\mathbf{z}}\,,
\end{equation}
where we pose $c_i=\cos{i}$ and $s_i=\sin{i}$, with $i$ being the orbit's
inclination angle ($0\leq i\leq\pi$). With this choice $\mathbf{P}$ lies along
the intersection of the orbital plane with the plane of the sky in the
direction of the \textit{ascending node} $\mathcal{N}$, \textit{i.e.} that
point at which the bodies cross the plane of the sky moving toward the
observer. The orbital phase $\phi$ is the angle between the ascending node
$\mathcal{N}$ and the direction of body one (say). The rotating orthonormal
triad $(\mathbf{n},{\bm \lambda},\hat{\mathbf{z}})$ describing the motion of
the binary [see \eqref{xva}] is then related to the fixed polarization triad
$(\mathbf{N},\mathbf{P},\mathbf{Q})$ by
\begin{subequations}\label{nlz}\begin{align}
\mathbf{n} &= \mathbf{P}\,\cos{\phi} + \bigl(c_i \,\mathbf{Q} + s_i
\,\mathbf{N}\bigr)\,\sin{\phi} \,,\\ {\bm \lambda} &= - \mathbf{P}\,\sin{\phi} +
\bigl(c_i \,\mathbf{Q} + s_i \,\mathbf{N}\bigr)\,\cos{\phi} \,,\\
\hat{\mathbf{z}} &= - s_i \,\mathbf{Q} + c_i \,\mathbf{N}\,.
\end{align}\end{subequations}

As in previous works \cite{BIWW96,ABIQ04} we shall present the wave
polarizations \eqref{hpc} as expansion series in powers of the gauge-invariant
PN parameter $x$ defined by \eqref{x}. With a convenient overall factorization
we write them as
\begin{align}\label{hpcPN}
\left(\begin{array}{l}h_+\\[0.5cm]h_\times
\end{array}\right) &= \frac{2\,G \,m \,\nu
  \,x}{c^2\,R}\,\left(\begin{array}{l}H_+ \\[0.5cm]H_\times
\end{array}\right) +
\mathcal{O}\left(\frac{1}{R^2}\right)\,,
\end{align}
with the following PN expansion series
\begin{equation}\label{HpcPN}
H_{+,\times} = \sum^{+\infty}_{n=0} x^{n/2}\,H_{+,\times}^{(n/2)}\,.
\end{equation}
The PN coefficients $H_{+,\times}^{(n/2)}$ will be given as functions of the
orbital phase $\phi$, and will also be polynomials in the symmetric mass ratio
$\nu$ and depend on the inclination angle $i$. In addition they will involve,
at high PN order, the logarithm of $x$ as we shall discuss below.

Following \cite{BIWW96,ABIQ04} it is convenient to perform a change of phase
variable, from the actual orbital phase $\phi$ satisfying $\dot{\phi}=\omega$,
to some new variable denoted $\psi$. Recall that the orbital phase $\phi$
evolves by gravitational radiation reaction and its expression as a function
of time is known from previous work \cite{BIJ02,BFIJ02,BDEI04} up to 3.5PN
order. We then pose\footnote{A similar phase variable is also introduced in
black-hole perturbation theory \cite{Sasa94,TSasa94,TTS96}.}
\begin{equation}\label{psi}
\psi=\phi - {2G M \omega \over c^3} \ln\left({\omega\over \omega_0}\right)\,,
\end{equation}
where $M$ is the binary's total mass given by \eqref{M0}, and where $\omega_0$
denotes the constant
\begin{equation}\label{omega0}
\omega_0 = \frac{e^{\frac{11}{12}-C}}{4\tau_0}\,.
\end{equation}
Here $\tau_0=r_0/c$ is the normalization of logarithms in the tail integrals
of the radiative moments \eqref{U2}--\eqref{U5V4}; $r_0$ is the constant
included in the definition of the finite part in \eqref{u2}. Like $\tau_0$ the
constant $\omega_0$ is arbitrary, because it is linked to the difference of
origins of time in the far zone and in the near zone, see \eqref{TR}. For
instance we can choose $\omega_0=\pi f_\mathrm{seismic}$ where
$f_\mathrm{seismic}$ is the entry frequency of some ground-based
interferometric detector. Using \eqref{M0} and the notation \eqref{x} the new
phase variable reads
\begin{equation}\label{psix}
\psi=\phi- 3 x^{3/2}\left[1 - \frac{\nu}{2}x\right]\ln\left({x\over
x_0}\right)
\,,\end{equation}
where $x_0=(\frac{G m \omega_0}{c^3})^{2/3}$.\footnote{We have $\ln
x_0=\frac{11}{18}-\frac{2}{3}C-\frac{4}{3}\ln 2+\frac{2}{3}\ln\left(\frac{G
m}{c^2 r_0}\right)$ in agreement with the equation (68) of \cite{K07}.} Our
modified phase variable \eqref{psi}--\eqref{psix} will be valid up to 3PN
order but in fact it turns out to be the same as at the previous 2.5PN order
\cite{ABIQ04}.

The logarithmic term in $\psi$ corresponds to some spreading of the different
frequency components of the wave along the line of sight from the source to
the far-away detector, and expresses physically the tail effect as a small
delay in the arrival time of gravitational waves. However, practically
speaking, the main interest of this term is to minimize the occurence of
logarithms in the FWF. Indeed we notice that the logarithmic term in
\eqref{psi}, although of formal PN order $\mathcal{O}(c^{-3})$, represents in
fact a very small modulation of the orbital phase: compared with the dominant
phase evolution whose order is that of the inverse of radiation reaction,
\textit{i.e.} $\phi=\mathcal{O}(\xi^{-1})=\mathcal{O}(c^5)$, this term is of
order $\mathcal{O}(c^{-8})$ namely 4PN in the phase evolution, which can be
regarded as negligible to the present accuracy. Thus the logarithms associated
with the phase modulation in \eqref{psi} will be ``eliminated'' from the FWF
at 3PN order. This does not mean that we should ignore them but that the
formulation in terms of the small phase modulation \eqref{psi} is quite
natural (for the data analysis it is probably better to keep the logarithm as
it stands in the definition of the phase variable $\psi$). However all the
logarithms will not be ``removed'' by this process, and we shall find that
some ``true'' logarithms remain starting at the 3PN order. Such logarithms
cannot be absorbed into some small modulation of the orbital phase, so 3PN
will remain as the true order of magnitude of these logarithms in the FWF.

With those conventions and notation we find for the plus
polarization\footnote{We also requote the previous 2.5PN results
\cite{ABIQ04} taking into account the published Erratum \cite{ABIQ04} and the
correcting term associated with radiation reaction and pointed out in
\cite{KBI07}.}
\begin{subequations}\label{Hplus}\begin{align}
H^{(0)}_+&=-(1+\,c_i^2) \cos 2\psi-{1\over96}\,\,s_i^2\,(17+\,c_i^2)\,,\label{HplusN}\\
H^{(0.5)}_+&=-\,s_i \,\Delta\left[\cos\psi
\left({5\over8}+{1\over8}\,c_i^2\right)-\cos
3\psi\left({9\over8}+{9\over8}\,c_i^2\right)\right]\,,\\
H^{(1)}_+&=\cos2\psi\left[{19\over6} + {3\over2}\,c_i^2-{ 1\over3}\,c_i^4
+\nu\left(-{19\over6}+{11\over6}\,c_i^2 +\,c_i^4\right)\right]\nonumber\\
&-\cos4\psi\left[{4\over3}\,s_i^2(1+\,c_i^2)(1-3\nu)\right]\,,\\ H^{(1.5)}_+
&=\,s_i\,\Delta\cos\psi\left[{19\over64} +{5\over16}\,c_i^2-{1\over192}\,c_i^4
+\nu\left(-{49\over96}
+{1\over8}\,c_i^2+{1\over96}\,c_i^4\right)\right]\nonumber\\
&+\cos2\psi\left[-2\pi(1+\,c_i^2)\right]\nonumber\\
&+\,s_i\,\Delta\cos3\psi\left[ -{657\over128}
-{45\over16}\,c_i^2+{81\over128}\,c_i^4 \right.\nonumber\\
&~~+\left.\nu\left({225\over64}-{9\over8}\,c_i^2
-{81\over64}\,c_i^4\right)\right]\nonumber\\ &+\,s_i\,\Delta
\cos5\psi\left[{625\over 384}\,s_i^2(1+\,c_i^2)(1 -2\nu)\right]\,,\\ H^{(2)}_+
&=\,\pi \,s_i\,\Delta\cos\psi\left[-{5\over8}
-{1\over8}\,c_i^2\right]\nonumber\\ &+\cos2\psi\left[
{11\over60}+{33\over10}\,c_i^2+{29\over24}\,c_i^4 -{1\over24}\,c_i^6
\right.\nonumber\\ &~~+\nu\left({353\over36}-3\,c_i^2-{251\over72}\,c_i^4
+{5\over24}\,c_i^6\right)\nonumber\\
&~~+\left.\nu^2\left(-{49\over12}+{9\over2}\,c_i^2-{7\over24}\,c_i^4
-{5\over24}\,c_i^6\right)\right]\nonumber\\ &+\pi \,s_i\,\Delta
\cos3\psi\left[{27\over8}(1+\,c_i^2)\right]\nonumber\\ &+{2\over15}\,s_i^2\,\cos4\psi\left[
59 + 35\,c_i^2-8\,c_i^4 -{5\over3}\,\nu\left(131 +59\,c_i^2 -24\,c_i^4\right) 
\right.\nonumber\\
&~~+\left.5\,\nu^2\left(21-3\,c_i^2-8\,c_i^4\right)\right]\nonumber\\
&+\cos6\psi\left[-{81\over40}\,s_i^4(1+\,c_i^2)\left(1-5\nu
+5\nu^2\right)\right]\nonumber\\ &+\,s_i\,\Delta
\sin\psi\left[{11\over40}+{5\ln2\over
4}+\,c_i^2\left({7\over40}+{\ln2\over4}\right)\right]\nonumber\\
&+\,s_i\,\Delta \sin3\psi\left[\left(-{189\over
40}+{27\over4}\ln(3/2)\right)(1+\,c_i^2)\right]\,,\\ H^{(2.5)}_+ &= s_i \,\Delta
\cos\psi\left[{1771\over 5120}-{1667\over5120}\,c_i^2+{217\over
9216}\,c_i^4-{1\over 9216}\,c_i^6\right.\nonumber\\
&~~+\nu\left({681\over256}+{13\over768}\,c_i^2-{35\over
768}\,c_i^4+{1\over2304}\,c_i^6\right)\nonumber\\
&~~+\left.\nu^2\left(-{3451\over9216}+{673\over3072}\,c_i^2-
{5\over9216}\,c_i^4-{1\over3072}\,c_i^6\right)\right]\nonumber\\
&+{\pi}\cos2\psi\left[{19\over3}+3\,c_i^2-{2\over3}\,c_i^4+\nu
\left(-{16\over3}+{14\over3}\,c_i^2+2\,c_i^4\right)\right]\nonumber\\
&+s_i\,\Delta \cos3\psi\left[{3537\over1024}-{22977\over5120}\,c_i^2-{15309
\over5120}\,c_i^4+{729\over5120}\,c_i^6\right.\nonumber\\
&~~+\nu\left(-{23829\over1280}+{5529\over1280}\,c_i^2+{7749\over1280}
\,c_i^4-{729\over1280}\,c_i^6\right)\nonumber\\
&~~+\left.\nu^2\left({29127\over5120}-{27267\over5120}\,c_i^2-{1647
\over5120}\,c_i^4+{2187\over 5120}\,c_i^6\right)\right]\nonumber\\
&+\cos4\psi\left[-{16\pi\over3}\,(1+\,c_i^2)\,s_i^2(1-3\nu)\right]\nonumber\\
&+s_i \,\Delta \cos5\psi\left[-{108125\over9216}+{40625\over9216}\,c_i^2
+{83125\over9216}\,c_i^4-{15625\over 9216}\,c_i^6\right.\nonumber\\
&~~+\nu\left({8125\over
256}-{40625\over2304}\,c_i^2-{48125\over2304}\,c_i^4+{15625\over
2304}\,c_i^6\right)\nonumber\\
&~~+\left.\nu^2\left(-{119375\over9216}+{40625\over 3072}\,c_i^2+{44375\over
9216}\,c_i^4-{15625\over 3072}\,c_i^6\right)\right]\nonumber\\ &+\Delta
\cos7\psi\left[{117649\over46080}\,s_i^5(1+\,c_i^2)(1-4\nu+3\nu^2)
\right]\nonumber\\
&+\sin2\psi\left[-{9\over5}+{14\over5}\,c_i^2+{7\over5}\,c_i^4+\nu
\left(32+{56\over5}\,c_i^2-{28\over5}\,c_i^4\right)\right]\nonumber\\
&+s_i^2(1+\,c_i^2)\sin4\psi\left[{56\over5}-{32\ln2\,\over3}
+\nu\left(-{1193\over30}+32\ln2 \right)\right]\,,\\ H^{(3)}_+ &=\pi\,\Delta
\,s_i\cos\psi\left[{19\over64}+{5\over16}\,c_i^2-{1\over192}\,c_i^4
+\nu\left(-{19\over96}+{3\over16}\,c_i^2+{1\over96}\,c_i^4\right)\right]
\nonumber\\
&+\,\cos2\psi\left[-{465497\over11025}+\left({856
\,C\over105}-{2\,\pi^2\over3}+{428\over105}\,\ln(16\,x)\right)(1+c_i^2)\right.
\nonumber\\
&~~\left.-{3561541\over88200}\,c_i^2
-{943\over720}\,c_i^4+{169\over720}\,c_i^6-{1\over360}\,c_i^8\right.\nonumber\\
&~~+\nu\left({2209\over360}-{41
\pi^2\over96}(1+c_i^2)+{2039\over180}\,c_i^2+{3311\over720}\,c_i^4
-{853\over720}\,c_i^6+{7\over360}\,c_i^8\right)\nonumber\\
&~~+\nu^2\left({12871\over540}-{1583\over60}\,c_i^2-{145\over108}\,c_i^4
+{56\over45}\,c_i^6-{7\over180}\,c_i^8\right)\nonumber\\
&~~\left.
+\nu^3\left(-{3277\over810}+{19661\over3240}\,c_i^2-{281\over144}\,c_i^4
-{73\over720}\,c_i^6+{7\over360}\,c_i^8\right)\right]\nonumber\\
&+\,\pi\,\Delta
\,s_i\cos3\psi\left[-{1971\over128}-{135\over16}\,c_i^2+{243\over128}\,c_i^4
+\nu\left({567\over64}-{81\over16}\,c_i^2-{243\over64}\,c_i^4\right)\right]
\nonumber\\
&+\,s_i^2\cos4\psi\left[-{2189\over210}+{1123\over210}\,c_i^2
+{56\over9}\,c_i^4-{16\over45}\,c_i^6\right.\nonumber\\
&~~+\nu\left({6271\over90}-{1969\over90}\,c_i^2-{1432\over45}\,c_i^4
+{112\over45}\,c_i^6\right)\nonumber\\
&~~+\nu^2\left(-{3007\over27}+{3493\over135}\,c_i^2+{1568\over45}\,c_i^4
-{224\over45}\,c_i^6\right)\nonumber\\
&~~\left.+\nu^3\left({161\over6}-{1921\over90}\,c_i^2-{184\over45}\,c_i^4
+{112\over45}\,c_i^6\right)\right]\nonumber\\ &+\,\Delta
\cos5\psi\left[{3125\,\pi \over384}\,s_i^3(1+c_i^2)(1-2\nu)\right]\nonumber\\
&+\,s_i^4\cos6\psi\left[{1377\over80}+{891\over80}\,c_i^2
-{729\over280}\,c_i^4\right.\nonumber\\
&~~+\nu\left(-{7857\over80}-{891\over16}\,c_i^2+{729\over40}\,c_i^4\right)
\nonumber\\
&~~+\nu^2\left({567\over4}+{567\over10}\,c_i^2-{729\over20}\,c_i^4\right)
\nonumber\\
&~~\left.+\nu^3\left(-{729\over16}-{243\over80}\,c_i^2
+{729\over40}\,c_i^4\right)\right]\nonumber\\
&+\,\cos8\psi\left[-{1024\over315}\,s_i^6\,(1+c_i^2)(1-7\nu+14\nu^2
-7\nu^3)\right]\nonumber\\ &+\,\Delta
\,s_i\sin\psi\left[-{2159\over40320}-{19\,\ln
2\over32}+\left(-{95\over224}-{5\,\ln
2\over8}\right)\,c_i^2+\left({181\over13440}+{\ln
2\over96}\right)\,c_i^4\right.\nonumber\\
&~~\left.+\nu\biggl({1369\over160}+{19\,\ln
2\over48}+\left(-{41\over48}-{3\,\ln
2\over8}\right)\,c_i^2+\left(-{313\over480}-{\ln
2\over48}\right)\,c_i^4\biggr)\right]\nonumber\\
&+\,\sin2\psi\left[-{428\,\pi\over105}\,(1+c_i^2)\right]\nonumber\\ &+\,\Delta
\,s_i\sin3\psi\left[{205119\over8960}-{1971\over64}\,\ln(3/2)
+\left({1917\over224}-{135\over8}\,\ln(3/2)\right)\,c_i^2\right.\nonumber\\
&~~~~+\left(-{43983\over8960}+{243\over64}\ln(3/2)\right)\,c_i^4\nonumber\\
&~~+\nu\biggl(-{54869\over960}+{567\over32}\,\ln(3/2)+\left(-{923\over80}
-{81\over8}\,\ln(3/2)\right)\,c_i^2\nonumber\\
&~~~~\left.+\left({41851\over2880}-{243\over32}\,\ln(3/2)\right)\,
c_i^4\biggr)\right]\nonumber\\ &+\,\Delta
\,s_i^3(1+c_i^2)\sin5\psi\left[-{113125\over5376}+{3125\over192}\,\ln(5/2)
+\nu\left({17639\over320}-{3125\over96}\,\ln(5/2)\right)\right]\,.
\end{align}\end{subequations}
For the cross polarizations we obtain
\begin{subequations}\label{Hcross}\begin{align}
H^{(0)}_\times&=-2c_i \sin2\psi\,,\\ H^{(0.5)}_\times&= \,s_ic_i\,\Delta
\left[-{3\over4}\sin\psi+{9\over4}\sin3\psi\right]\,,\\ H^{(1)}_\times&= c_i
\sin2\psi\left[{17\over3}-{4\over3}\,c_i^2
+\nu\left(-{13\over3}+4\,c_i^2\right)\right]\nonumber\\&+c_i
\,s_i^2\sin4\psi\left[-{8\over3}(1-3\nu)\right]\,,\\
H^{(1.5)}_\times&=\,s_ic_i \,\Delta \sin\psi\left[{21\over32}
-{5\over96}\,c_i^2 +\nu\left(-{23\over48}
+{5\over48}\,c_i^2\right)\right]\nonumber\\ &-4\pi\,c_i \sin2\psi\nonumber\\
&+\,s_ic_i\,\Delta \,\sin3\psi\left[-{603\over64} +{135\over64}\,c_i^2
+\nu\left({171\over32} -{135\over32}\,c_i^2\right)\right]\nonumber\\
&+\,s_ic_i\,\Delta \,\sin5\psi\left[{625\over192}(1-2\nu)\,s_i^2\right]\,,\\
H^{(2)}_\times&=\,s_ic_i\,\Delta\cos\psi\left[-{9\over20}
-{3\over2}\ln2\right]\nonumber\\&+\,s_ic_i\,\Delta
\cos3\psi\left[{189\over20}-{27\over2}\ln(3/2)\right]\nonumber\\
&-\,s_ic_i\,\Delta\left[{3\,\pi\over4}\right] \sin\psi\nonumber\\
&+c_i\sin2\psi\left[{17\over15}+{113\over30}\,c_i^2 -{1\over4}\,c_i^4
\right.\nonumber\\
&~~+\nu\left({143\over9}-{245\over18}\,c_i^2+{5\over4}\,c_i^4\right)\nonumber\\
&~~+\left.\nu^2\left(-{14\over3}+{35\over6}\,c_i^2
-{5\over4}c_i^4\right)\right]\nonumber\\ &+\,s_ic_i\,\Delta
\sin3\psi\left[{27\pi\over4}\right]\nonumber\\
&+{4\over15}\,c_i\,s_i^2\,\sin4\psi\left[55-12\,c_i^2
-{5\over3}\,\nu\left(119-36\,c_i^2\right)+5\,\nu^2\left(17-12\,c_i^2\right)
\right]\nonumber\\
&+c_i\sin6\psi\left[-{81\over20}\,s_i^4(1-5\nu+5\nu^2)\right]\,,\\
H^{(2.5)}_\times &={6\over5}\,s_i^2\,c_i\,\nu\nonumber\\
&+\,c_i\cos2\psi\left[2-{22\over5}\,c_i^2+\nu\left(-{282\over5}+{94\over5}
\,c_i^2\right)\right]\nonumber\\ &+\,c_i\,s_i^2
\cos4\psi\left[-{112\over5}+{64\over3}\ln2+\nu\left({1193\over15}-64\ln2
\right)\right]\nonumber\\ &+\,s_i\,c_i\,\Delta \sin\psi\left[-{913\over
7680}+{1891\over11520} \,c_i^2-{7\over4608}\,c_i^4\right.\nonumber\\
&~~+\left.\nu\left({1165\over384}-{235\over576}\,c_i^2
+{7\over1152}\,c_i^4\right)\right.\nonumber\\
&~~+\left.\nu^2\left(-{1301\over4608}+{301\over
2304}\,c_i^2-{7\over1536}\,c_i^4\right)\right]\nonumber\\ &+\pi
\,c_i\sin2\psi\left[{34\over3}-{8\over3}\,c_i^2+\nu\left(-{20\over3}
+8\,c_i^2\right)\right]\nonumber\\ &+\,s_i\,c_i\,\Delta
\sin3\psi\left[{12501\over2560}-{12069\over1280}\,c_i^2+{1701\over2560}
\,c_i^4\right.\nonumber\\
&~~+\nu\left(-{19581\over640}+{7821\over320}\,c_i^2-{1701\over640}
\,c_i^4\right)\nonumber\\ &~~+\left.\nu^2\left({18903\over 2560}-{11403\over
1280}\,c_i^2+{5103\over2560}\,c_i^4\right)\right]\nonumber\\
&+\,s_i^2\,c_i\sin4\psi\left[-{32\pi \over3}(1-3\nu)\right]\nonumber\\
&+\Delta \,s_i\,c_i\sin5\psi\left[-{101875\over4608}+{6875\over256}\,c_i^2-
{21875\over4608}\,c_i^4\right.\nonumber\\
&~~+\left.\nu\left({66875\over1152}-{44375\over576}\,c_i^2+{21875\over1152}
\,c_i^4\right)\right.\nonumber\\
&~~+\left.\nu^2\left(-{100625\over4608}+{83125\over2304}\,c_i^2-{21875\over1536}
\,c_i^4\right)\right]\nonumber\\ &+\Delta
\,s_i^5\,c_i\sin7\psi\left[{117649\over23040}
\left(1-4\nu+3\nu^2\right)\right]\,,\\
H^{(3)}_\times &= \Delta
\,s_i\,c_i\cos\psi\left[{11617\over20160}+{21\over16}\,\ln
2+\left(-{251\over2240}-{5\over48}\,\ln 2\right)\,c_i^2\right.\nonumber\\
&~~\left.+\nu\biggl(-{2419\over240}-{5\over24}\,\ln
2+\left({727\over240}+{5\over24}\,\ln
2\right)\,c_i^2\biggr)\right]\nonumber\\
&+\,c_i\cos2\psi\left[{856\,\pi\over105}\right]\nonumber\\ &+\,\Delta
\,s_i\,c_i\cos3\psi\left[-{36801\over896}+{1809\over32}\,\ln(3/2)
+\left({65097\over4480}-{405\over32}\,\ln(3/2)\right)\,c_i^2\right.\nonumber\\
&~~\left.+\nu\biggl({28445\over288}-{405\over16}\,\ln(3/2)
+\left(-{7137\over160}+{405\over16}\,\ln(3/2)\right)\,c_i^2\biggr)\right]
\nonumber\\ &+\,\Delta
\,s_i^3\,c_i\cos5\psi\left[{113125\over2688}-{3125\over96}\,\ln(5/2)
+\nu\left(-{17639\over160}+{3125\over48}\,\ln(5/2)\right)\right]\nonumber\\
&+\,\pi\,\Delta \,s_i\,c_i\sin\psi\left[{21\over32}-{5\over96}\,c_i^2
+\nu\left(-{5\over48}+{5\over48}\,c_i^2\right)\right]\nonumber\\
&+\,c_i\sin2\psi\left[-{3620761\over44100}+{1712\,C\over105}
-{4\,\pi^2\over3}+{856\over105}\,\ln(16\,x)\right.\nonumber\\
&~~~~-{3413\over1260}\,c_i^2
+{2909\over2520}\,c_i^4-{1\over45}\,c_i^6\nonumber\\
&~~+\nu\left({743\over90}-{41\,\pi^2\over48}+{3391\over180}\,c_i^2
-{2287\over360}\,c_i^4+{7\over45}\,c_i^6\right)\nonumber\\
&~~+\nu^2\left({7919\over270}-{5426\over135}\,c_i^2
+{382\over45}\,c_i^4-{14\over45}\,c_i^6\right)\nonumber\\
&~~\left.+\nu^3\left(-{6457\over1620}+{1109\over180}\,c_i^2-{281\over120}\,c_i^4
+{7\over45}\,c_i^6\right)\right]\nonumber\\ &+\,\pi\,\Delta
\,s_i\,c_i\sin3\psi\left[-{1809\over64}+{405\over64}\,c_i^2
+\nu\left({405\over32}-{405\over32}\,c_i^2\right)\right]\nonumber\\
&+\,s_i^2\,c_i\sin4\psi\left[-{1781\over105}+{1208\over63}\,c_i^2
-{64\over45}\,c_i^4\right.\nonumber\\
&~~+\nu\left({5207\over45}-{536\over5}\,c_i^2
+{448\over45}\,c_i^4\right)\nonumber\\
&~~+\nu^2\left(-{24838\over135}+{2224\over15}\,c_i^2
-{896\over45}\,c_i^4\right)\nonumber\\
&~~\left.+\nu^3\left({1703\over45}-{1976\over45}\,c_i^2
+{448\over45}\,c_i^4\right)\right]\nonumber\\ &+\,\Delta
\sin5\psi\left[{3125\,\pi\over192}\,s_i^3\,c_i(1-2\nu)\right]\nonumber\\
&+\,s_i^4\,c_i\sin6\psi\left[{9153\over280}-{243\over35}\,c_i^2
+\nu\left(-{7371\over40}+{243\over5}\,c_i^2\right)\right.\nonumber\\
&~~\left.+\nu^2\left({1296\over5}-{486\over5}\,c_i^2\right)
+\nu^3\left(-{3159\over40}+{243\over5}\,c_i^2\right)\right]\nonumber\\
&+\,\sin8\psi\left[-{2048\over315}\,s_i^6\,c_i(1-7\nu+14\nu^2-7\nu^3)\right]\,.
\end{align}\end{subequations}
Notice the obvious fact that the polarization waveforms remain invariant when
we rotate by $\pi$ the separation direction between the particles and
simultaneously exchange the labels of the two particles, \textit{i.e.} when we
apply the transformation $(\psi,\Delta)\rightarrow(\psi+\pi,-\Delta)$.
Moreover, due to the parity invariance, $H_+$ is unchanged after the
replacement $i\rightarrow \pi - i$, while $H_\times$ being the projection of
$h_{ij}^\mathrm{TT}$ on a tensorial product of two vectors of inverse parity
types, is changed into its opposite.

We have performed two important tests on these expressions. First of all we
have verified that the perturbative limit $\nu\rightarrow 0$ of the
polarization waveforms \eqref{Hplus}--\eqref{Hcross} is in full agreement up
to 3PN order with the result of black-hole perturbation theory as reported in
the Appendix B of \cite{TSasa94}.\footnote{In \cite{ABIQ04} a misprint was
spotted in the Appendix B of \cite{TSasa94}: the sign of the harmonic
coefficient $\zeta_{7,3}^\times$ should be changed, so that one should read
$\zeta_{7,3}^\times=+
\frac{729}{10250240}\cos(\theta)(167+...)\sin(\theta)(v^5 \cos(3 \psi)-...)$.}
Our second test is the verification that the wave polarizations
\eqref{Hplus}--\eqref{Hcross} give back the correct energy flux at 3PN
order. The asymptotic flux is given in terms of the polarizations by
\begin{equation}\label{FGW}
\mathcal{F}^\mathrm{GW} = \lim_{R\rightarrow+\infty}\,\frac{R^2\,c^3}{4 G}\int
\frac{d\Omega}{4\pi}\,\Bigl[\bigl(\dot{h}_+\bigr)^2 +
\bigl(\dot{h}_\times\bigr)^2 \Bigr] \,,
\end{equation}
where $d\Omega$ is the solid angle element associated with the direction of
propagation $\mathbf{N}$. We have $d\Omega=\sin\Theta\,d\Theta\,d\Phi$ where
$(\Theta,\Phi)$ are the angles defining $\mathbf{N}$, following the notation
of Section \ref{secII}. To obtain the polarizations corresponding to this
general convention for $\mathbf{N}$ we have to make some simple replacements
in \eqref{Hplus}--\eqref{Hcross} for $i$ and $\psi$. As is clear from the
geometry of the problem we must replace
$(i,\psi)\rightarrow(\Theta,\psi+\pi/2-\Phi)$. The time derivative of the
polarizations is computed in the adiabatic approximation, using
$\dot{\phi}=\omega$ and $\dot{\omega}$ given by \eqref{omdot}. Of course one
must take into account the difference between $\phi$ and the variable $\psi$
used in \eqref{Hplus}--\eqref{Hcross}. Finally, the angular integration in
\eqref{FGW} is readily performed and the result is in perfect agreement with
the 3PN energy flux given by (12.9) of \cite{BIJ02}.\footnote{The ambiguity
parameters therein are now known to be $\lambda=-\frac{1987}{3080}$
\cite{DJSdim,BDE04} and $\theta=\xi+2\kappa+\zeta=-\frac{11831}{9240}$
\cite{BDEI04}.}

As already mentioned there are some ``true'' logarithms which remain in the
FWF at the 3PN order --- \textit{i.e.} after it has been expressed with the
help of the PN parameter $x$ and the phase variable $\psi$. Inspection of
\eqref{Hplus}--\eqref{Hcross} shows that these logarithms have the effect of
correcting the Newtonian polarizations in the following way:
\begin{equation}\label{hpclog}
\left(\begin{array}{l}H_+\\[0.5cm]H_\times
\end{array}\right) = \left(\begin{array}{c}-(1+\,c_i^2) \cos 2\psi\\[0.5cm]-2c_i
\sin2\psi\end{array}\right)\left( 1 - \frac{428}{105} \,x^3\,\ln(16\,x)\right)
+ \cdots + \mathcal{O}\left(\frac{1}{c^7}\right)\,,
\end{equation}
where the dots represent the terms independent of logarithms. In our previous
computation of the 3PN flux using \eqref{FGW} we have already checked that
these logarithms are consistent with similar logarithms occuring at 3PN order
in the flux. Indeed we easily see that they correspond in the 3PN flux to the
terms
\begin{equation}\label{FGWlog}
\mathcal{F}^\mathrm{GW} = \frac{32 c^5}{5 G}\nu^2 x^5\left[1 - \frac{856}{105}
\,x^3\,\ln(16\,x) + \cdots + \mathcal{O}\left(\frac{1}{c^7}\right)\right]\,,
\end{equation}
already known from (12.9) in \cite{BIJ02}. Technically the logarithm in
\eqref{hpclog} or \eqref{FGWlog} is due to the tails-of-tails at 3PN
order. Notice that this logarithm survives in the test-mass limit
$\nu\rightarrow 0$ and is therefore also seen to appear in linear black hole
perturbations \cite{Sasa94,TSasa94,TTS96}.

\section{3PN spherical harmonic modes for numerical relativity}
\label{secIX}

The spin-weighted spherical harmonic modes of the polarization waveforms at
3PN order can now be obtained from using the angular integration
\eqref{decomp}. An alternative route would be to use the relations
\eqref{inv}--\eqref{UV} giving the modes directly in terms of separate
contributions of the radiative moments $U_L$ and $V_L$. In the present paper
the two routes are equivalent because all the radiative moments are
``uniformly'' given with the approximation that is necessary and sufficient to
control the 3PN waveform. 

In this respect one should be careful about what we mean by controlling the
modes up to 3PN order. We mean --- having in mind the standard PN practice ---
that the accuracy of the modes is exactly the one which is needed to obtain
the 3PN waveform. Thus the dominant mode $h^{22}$ will have full 3PN accuracy,
but higher-order modes, which start at some higher PN order, will have a lower
\textit{relative} PN accuracy. For instance we shall see that the mode
$h^{44}$ starts at 1PN order thus it will be given only with 2PN relative
accuracy.

The angular integration in \eqref{decomp} is over the angles
$(\Theta,\Phi)$. Like in our previous computation of the flux \eqref{FGW}, it
should be performed after substituting
$(i,\psi)\rightarrow(\Theta,\psi+\pi/2-\Phi)$ 
in the wave polarizations. Denoting $h = h_+ - i h_\times$ the integral we
consider is thus
\begin{equation}\label{decomp2}
h^{\ell m} = \int d\Omega \,h(\Theta,\psi+\pi/2-\Phi) \,
\overline{Y}^{\,\ell m}_{-2} (\Theta,\Phi)\,.
\end{equation}
Changing $\Phi$ into $\psi+\pi/2-\psi'$ and $\Theta$ into $i'=\arccos c_i'$, and
using the known dependence of the spherical harmonics on the azimuthal angle
$\Phi$ [see \eqref{harm}], we obtain
\begin{equation}\label{decomp3}
h^{\ell m} = (-i)^m\,e^{-i m \psi}\,\int_0^{2\pi}\!\! d\psi' \int_{-1}^1 dc_i'
\, h(i',\psi')\,Y^{\,\ell m}_{-2} (i',\psi')\,,
\end{equation}
exhibiting the azimuthal factor $e^{-i m \psi}$ appropriate for each mode. Let
us factorize out in all the modes an overall coefficient including $e^{-i m
\psi}$, and such that the dominant mode with $(\ell,m)=(2,2)$ starts with one
(by pure convention) at the Newtonian order. Remembering also our previous
factorization in \eqref{hpcPN} we pose
\begin{subequations}\label{Hhat}\begin{align}
h^{\ell m} &= \frac{2 G \,m \,\nu \,x}{R \,c^2} \,H^{\ell m}\,,\\H^{\ell m} &=
\sqrt{\frac{16\pi}{5}}\,\hat{H}^{\ell m}\,e^{-i m \psi}\,,
\end{align}\end{subequations}
and list all the results in terms of $\hat{H}^{\ell m}$,\footnote{The modes
having $m<0$ are easily deduced using
$\hat{H}^{\ell,-m}=(-)^\ell\overline{\hat{H}}^{\ell m}$.}
\begin{subequations} \label{hlm}\begin{align}
 \hat{H}^{22}&=1+x \left(-\frac{107}{42}+\frac{55 \nu }{42}\right)+2 \pi
x^{3/2}+x^2 \left(-\frac{2173}{1512}-\frac{1069 \nu }{216}+\frac{2047 \nu
^2}{1512}\right) \nonumber \\ &+x^{5/2} \left(-\frac{107 \pi }{21}-24 i \nu
+\frac{34 \pi \nu }{21}\right)+x^3 \bigg(\frac{27027409}{646800}-\frac{856
C}{105}+\frac{428 i \pi }{105}+\frac{2 \pi ^2}{3}\nonumber \\ &
+\left(-\frac{278185}{33264}+\frac{41 \pi^2}{96}\right) \nu -\frac{20261 \nu
^2}{2772}+\frac{114635 \nu^3}{99792}-\frac{428}{105} \ln (16 x)\bigg) +
\mathcal{O}\left(\frac{1}{c^7}\right)\label{h22}\,,\\ \hat{H}^{21}
&=\frac{1}{3} i \,\Delta \bigg[x^{1/2}+x^{3/2} \left(-\frac{17}{28}+\frac{5
\nu }{7}\right)+x^2 \left(\pi +i \left(-\frac{1}{2}-2 \ln 2\right)\right)
\nonumber \\ & +x^{5/2} \left(-\frac{43}{126}-\frac{509 \nu }{126}+\frac{79
\nu^2}{168}\right)+x^3 \bigg(-\frac{17 \pi }{28}+\frac{3 \pi \nu }{14}
\nonumber \\ &+i \left(\frac{17}{56}+\nu \left(-\frac{353}{28}-\frac{3 \ln
2}{7}\right)+\frac{17 \ln 2}{14}\right)\bigg)\bigg] +
\mathcal{O}\left(\frac{1}{c^7}\right)\,,\\ \hat{H}^{20}& =-\frac{5}{14
\sqrt{6}} + \mathcal{O}\left(\frac{1}{c^7}\right)\,,\\ \hat{H}^{33}
&=-\frac{3}{4} i \sqrt{\frac{15}{14}} \,\Delta \bigg[x^{1/2}+x^{3/2} (-4+2 \nu
)+x^2 \left(3 \pi +i \left(-\frac{21}{5}+6 \ln \left(3/2\right)\right)\right)
\nonumber \\ &+x^{5/2} \left(\frac{123}{110}-\frac{1838 \nu }{165}+\frac{887
\nu ^2}{330}\right)+x^3 \bigg(-12 \pi +\frac{9 \pi \nu }{2} \nonumber \\ &+i
\left(\frac{84}{5}-24 \ln \left(3/2\right)+\nu \left(-\frac{48103}{1215}+9 \ln
\left(3/2\right)\right)\right)\bigg)\bigg] +
\mathcal{O}\left(\frac{1}{c^7}\right)\,,\\ \hat{H}^{32} &=\frac{1}{3}
\sqrt{\frac{5}{7}} \bigg[x (1-3 \nu )+x^2 \left(-\frac{193}{90}+\frac{145 \nu
}{18}-\frac{73 \nu ^2}{18}\right)+x^{5/2} \left(2 \pi -6 \pi \nu +i
\left(-3+\frac{66 \nu }{5}\right)\right) \nonumber \\ &+x^3
\left(-\frac{1451}{3960}-\frac{17387 \nu }{3960}+\frac{5557
\nu^2}{220}-\frac{5341 \nu^3}{1320}\right)\bigg] +
\mathcal{O}\left(\frac{1}{c^7}\right)\,,\\ \hat{H}^{31} &=\frac{i \,\Delta}{12
\sqrt{14}} \bigg[x^{1/2}+x^{3/2} \left(-\frac{8}{3}-\frac{2 \nu
}{3}\right)+x^2 \left(\pi +i \left(-\frac{7}{5}-2 \ln 2\right)\right)
\nonumber \\ &+x^{5/2} \left(\frac{607}{198}-\frac{136 \nu }{99}-\frac{247
\nu^2}{198}\right)+x^3 \bigg(-\frac{8 \pi }{3}-\frac{7 \pi \nu }{6} \nonumber
\\ &+i \left(\frac{56}{15}+\frac{16 \ln 2}{3}+\nu \left(-\frac{1}{15}+\frac{7
\ln 2}{3}\right)\right)\bigg)\bigg] +
\mathcal{O}\left(\frac{1}{c^7}\right)\,,\\ \hat{H}^{30} &=-\frac{2}{5} i
\sqrt{\frac{6}{7}} x^{5/2} \nu + \mathcal{O}\left(\frac{1}{c^7}\right)\,,\\
\hat{H}^{44} &=-\frac{8}{9} \sqrt{\frac{5}{7}} \bigg[x (1-3 \nu )+x^2
\left(-\frac{593}{110}+\frac{1273 \nu }{66}-\frac{175 \nu^2}{22}\right)
\nonumber \\ &+x^{5/2} \left(4 \pi -12 \pi \nu +i \left(-\frac{42}{5}+\nu
\left(\frac{1193}{40}-24 \ln 2\right)+8 \ln 2\right)\right) \nonumber \\ &+x^3
\left(\frac{1068671}{200200}-\frac{1088119 \nu }{28600}+\frac{146879 \nu
^2}{2340}-\frac{226097 \nu^3}{17160}\right)\bigg] +
\mathcal{O}\left(\frac{1}{c^7}\right)\,,\\ \hat{H}^{43} &=-\frac{9 i
\,\Delta}{4 \sqrt{70}} \bigg[x^{3/2} (1-2 \nu )+x^{5/2}
\left(-\frac{39}{11}+\frac{1267 \nu }{132}-\frac{131 \nu^2}{33}\right)
\nonumber \\ &+x^3 \left(3 \pi -6 \pi \nu +i \left(-\frac{32}{5}+\nu
\left(\frac{16301}{810}-12 \ln \left(3/2\right)\right)+6 \ln
\left(3/2\right)\right)\right)\bigg] +
\mathcal{O}\left(\frac{1}{c^7}\right)\,,\\ \hat{H}^{42} &=\frac{1}{63}
\sqrt{5} \bigg[x (1-3 \nu )+x^2 \left(-\frac{437}{110}+\frac{805 \nu
}{66}-\frac{19 \nu^2}{22}\right)+x^{5/2} \bigg(2 \pi -6 \pi \nu \nonumber \\ &
+i \left(-\frac{21}{5}+\frac{84 \nu }{5}\right)\bigg) +x^3
\left(\frac{1038039}{200200}-\frac{606751 \nu }{28600}+\frac{400453 \nu
^2}{25740}+\frac{25783 \nu^3}{17160}\right)\bigg] +
\mathcal{O}\left(\frac{1}{c^7}\right)\,,\\ \hat{H}^{41} &=\frac{i \,\Delta}{84
\sqrt{10}} \bigg[x^{3/2} (1-2 \nu )+x^{5/2} \left(-\frac{101}{33}+\frac{337
\nu }{44}-\frac{83 \nu^2}{33}\right) \nonumber \\ & +x^3 \left(\pi -2 \pi \nu
+i \left(-\frac{32}{15}-2 \ln 2+\nu \left(\frac{1661}{30}+4 \ln
2\right)\right)\right)\bigg] + \mathcal{O}\left(\frac{1}{c^7}\right)\,,\\
\hat{H}^{40} &=-\frac{1}{504 \sqrt{2}} +
\mathcal{O}\left(\frac{1}{c^7}\right)\,,\\ \hat{H}^{55} &=\frac{625 i
\,\Delta}{96 \sqrt{66}} \bigg[x^{3/2} (1-2 \nu )+x^{5/2}
\left(-\frac{263}{39}+\frac{688 \nu }{39}-\frac{256 \nu^2}{39}\right)
\nonumber \\ &+x^3 \left(5 \pi -10 \pi \nu +i \left(-\frac{181}{14}+\nu
\left(\frac{105834}{3125}-20 \ln \left(5/2\right)\right)+10 \ln
\left(5/2\right)\right)\right)\bigg] +
\mathcal{O}\left(\frac{1}{c^7}\right)\,,\\ \hat{H}^{54} &=-\frac{32}{9
\sqrt{165}} \bigg[x^2 \left(1-5 \nu +5 \nu^2\right)+x^3
\left(-\frac{4451}{910}+\frac{3619 \nu }{130}-\frac{521 \nu ^2}{13}+\frac{339
\nu^3}{26}\right)\bigg] + \mathcal{O}\left(\frac{1}{c^7}\right)\,,\\
\hat{H}^{53} &=-\frac{9}{32} i \sqrt{\frac{3}{110}} \,\Delta \bigg[x^{3/2}
(1-2 \nu )+x^{5/2} \left(-\frac{69}{13}+\frac{464 \nu }{39}-\frac{88 \nu
^2}{39}\right)\nonumber \\ & +x^3 \left(3 \pi -6 \pi \nu +i
\left(-\frac{543}{70}+\nu \left(\frac{83702}{3645}-12 \ln
\left(3/2\right)\right)+6 \ln \left(3/2\right)\right)\right)\bigg] +
\mathcal{O}\left(\frac{1}{c^7}\right)\,,\\ \hat{H}^{52} &=\frac{2}{27
\sqrt{55}} \bigg[x^2 \left(1-5 \nu +5 \nu^2\right)+x^3
\left(-\frac{3911}{910}+\frac{3079 \nu }{130}-\frac{413 \nu ^2}{13}+\frac{231
\nu^3}{26}\right)\bigg] + \mathcal{O}\left(\frac{1}{c^7}\right)\,,\\
\hat{H}^{51} &=\frac{i \,\Delta}{288 \sqrt{385}} \bigg[x^{3/2} (1-2 \nu
)+x^{5/2} \left(-\frac{179}{39}+\frac{352 \nu }{39}-\frac{4 \nu
^2}{39}\right)\nonumber \\ & +x^3 \left(\pi -2 \pi \nu +i
\left(-\frac{181}{70}-2 \ln 2+\nu \left(\frac{626}{5}+4 \ln
2\right)\right)\right)\bigg] + \mathcal{O}\left(\frac{1}{c^7}\right)\,,\\
\hat{H}^{50} &= \mathcal{O}\left(\frac{1}{c^7}\right)\,,\\ \hat{H}^{66}
&=\frac{54}{5 \sqrt{143}} \bigg[x^2 \left(1-5 \nu +5 \nu^2\right)+x^3
\left(-\frac{113}{14}+\frac{91 \nu }{2}-64 \nu^2+\frac{39 \nu
^3}{2}\right)\bigg] + \mathcal{O}\left(\frac{1}{c^7}\right)\,,\\ \hat{H}^{65}
&=\frac{3125 i\,x^{5/2} \,\Delta}{504 \sqrt{429}} \bigg[1-4 \nu +3 \nu^2
\bigg] + \mathcal{O}\left(\frac{1}{c^7}\right)\,,\\ \hat{H}^{64}
&=-\frac{128}{495} \sqrt{\frac{2}{39}} \bigg[x^2 \left(1-5 \nu +5
\nu^2\right)+x^3 \left(-\frac{93}{14}+\frac{71 \nu }{2}-44 \nu^2+\frac{19 \nu
^3}{2}\right)\bigg] + \mathcal{O}\left(\frac{1}{c^7}\right)\,,\\ \hat{H}^{63}
&=-\frac{81 i\,x^{5/2} \,\Delta}{616 \sqrt{65}} \bigg[1-4 \nu +3 \nu^2 \bigg]
+ \mathcal{O}\left(\frac{1}{c^7}\right)\,,\\ \hat{H}^{62} &=\frac{2}{297
\sqrt{65}} \bigg[x^2 \left(1-5 \nu +5 \nu^2\right)+x^3
\left(-\frac{81}{14}+\frac{59 \nu }{2}-32 \nu^2+\frac{7 \nu
^3}{2}\right)\bigg] + \mathcal{O}\left(\frac{1}{c^7}\right)\,,\\ \hat{H}^{61}
&=\frac{i\,x^{5/2} \,\Delta}{8316 \sqrt{26}} \bigg[1-4 \nu +3 \nu^2\bigg] +
\mathcal{O}\left(\frac{1}{c^7}\right)\,,\\ \hat{H}^{60} &=
\mathcal{O}\left(\frac{1}{c^7}\right)\,,\\ \hat{H}^{77} &=-\frac{16807
i\,x^{5/2} \,\Delta}{1440} \sqrt{\frac{7}{858}} \bigg[1-4 \nu +3 \nu^2\bigg] +
\mathcal{O}\left(\frac{1}{c^7}\right)\,,\\ \hat{H}^{76} &=\frac{81}{35}
\sqrt{\frac{3}{143}} x^3 \left(1-7 \nu +14 \nu^2-7 \nu^3\right) +
\mathcal{O}\left(\frac{1}{c^7}\right)\,,\\ \hat{H}^{75} &=\frac{15625
i\,x^{5/2} \,\Delta}{26208 \sqrt{66}} \bigg[1-4 \nu +3 \nu^2\bigg] +
\mathcal{O}\left(\frac{1}{c^7}\right)\,,\\ \hat{H}^{74} &=-\frac{128
x^3}{1365}\sqrt{\frac{2}{33}} \left(1-7 \nu +14 \nu^2-7 \nu^3\right) +
\mathcal{O}\left(\frac{1}{c^7}\right)\,,\\ \hat{H}^{73} &=-\frac{243
i\,x^{5/2} \,\Delta}{160160}\sqrt{\frac{3}{2}} \bigg[1-4 \nu +3 \nu^2\bigg]+
\mathcal{O}\left(\frac{1}{c^7}\right)\,,\\ \hat{H}^{72} &=\frac{x^3 \left(1-7
\nu +14 \nu^2-7 \nu^3\right)}{3003 \sqrt{3}} +
\mathcal{O}\left(\frac{1}{c^7}\right)\,,\\ \hat{H}^{71} &=\frac{i\,x^{5/2}
\,\Delta}{864864 \sqrt{2}} \bigg[1-4 \nu +3 \nu^2\bigg] +
\mathcal{O}\left(\frac{1}{c^7}\right)\,,\\ \hat{H}^{70} &=
\mathcal{O}\left(\frac{1}{c^7}\right)\,,\\ \hat{H}^{88} &=-\frac{16384}{63}
\sqrt{\frac{2}{85085}} x^3 \left(1-7 \nu +14 \nu^2-7 \nu^3\right) +
\mathcal{O}\left(\frac{1}{c^7}\right)\,,\\ \hat{H}^{87} &=
\mathcal{O}\left(\frac{1}{c^7}\right)\,,\\ \hat{H}^{86} &=\frac{243}{35}
\sqrt{\frac{3}{17017}} x^3 \left(1-7 \nu +14 \nu^2-7 \nu^3\right) +
\mathcal{O}\left(\frac{1}{c^7}\right)\,,\\ \hat{H}^{85} &=
\mathcal{O}\left(\frac{1}{c^7}\right)\,,\\ \hat{H}^{84} &=-\frac{128
}{4095}\sqrt{\frac{2}{187}} x^3 \left(1-7 \nu +14 \nu^2-7 \nu ^3\right) +
\mathcal{O}\left(\frac{1}{c^7}\right)\,,\\ \hat{H}^{83} &=
\mathcal{O}\left(\frac{1}{c^7}\right)\,,\\ \hat{H}^{82} &=\frac{x^3}{9009
\sqrt{85}} \left(1-7 \nu +14 \nu^2-7 \nu^3\right) +
\mathcal{O}\left(\frac{1}{c^7}\right)\,,\\ \hat{H}^{81} &=
\mathcal{O}\left(\frac{1}{c^7}\right)\,,\\ \hat{H}^{80} &=
\mathcal{O}\left(\frac{1}{c^7}\right) \,,\end{align}
\end{subequations}
while all the higher-order modes fall into the PN remainder and are
negligible. However, we shall give here for the reader's convenience their
leading order expressions for non zero $m$ (see the derivation in
\cite{K07}). For $\ell + m$ even we find:
\begin{align}\label{lmeven}
\hat{H}^{\ell m} =& \frac{(-)^{(\ell-m+2)/2}}{2^{\ell+1} (\frac{\ell+m}{2})!
(\frac{\ell-m}{2})!(2\ell-1)!!}
\left(\frac{5(\ell+1)(\ell+2)(\ell+m)!(\ell-m)!}{\ell
(\ell-1)(2\ell+1)}\right)^{1/2} s_\ell(\nu) \,(i m)^\ell \,x^{\ell/2-1}
\nonumber\\ &+ \mathcal{O}\left(\frac{1}{c^{\ell-2}}\right)\,,
\end{align}
where we recall that the function $s_\ell(\nu)$ is defined in
\eqref{sell}. For $\ell + m$ odd we have:
\begin{align}\label{lmodd}
\hat{H}^{\ell m} =& \frac{(-)^{(\ell-m-1)/2}}{2^{\ell-1} (\frac{\ell+m-1}{2})!
(\frac{\ell-m-1}{2})! (2\ell+1)!!}
\left(\frac{5(\ell+2)(2\ell+1)(\ell+m)!(\ell-m)!}{\ell (\ell-1) (\ell+1)
}\right)^{1/2} \nonumber\\ &\times s_{\ell+1}(\nu) \,i \,(i m)^\ell
\,x^{(\ell-1)/2}+ \mathcal{O}\left(\frac{1}{c^{\ell-2}}\right)\,.
\end{align}
When $m=0$, $\hat{H}^{\ell m}$ may not vanish due to DC contributions of the
memory integrals. We already know that such an effect arises at Newtonian
order [see \eqref{HplusN}], hence the non zero values of $\hat{H}^{20}$ and
$\hat{H}^{40}$.

We find that the result for $\hat{H}^{22}$ at 3PN order given by \eqref{h22}
is in complete agreement with the result of Kidder \cite{K07}. The only
difference is our use of the particular phase variable \eqref{psix} which
permits to remove most of the logarithmic terms, showing that they are
actually negligible modulations of the orbital phase. For the other harmonics
we find agreement with the results of \cite{K07} up to 2.5PN order, but the
results have here been completed by all the 3PN contributions.

\acknowledgments A part of the results of this paper was obtained by means of
the tensor package xTensor developed for Mathematica by J. Mart\'in-Garc\'ia
\cite{xtensor}. We thank him for the technical support he provided as we were
elaborating the code. LB, GF and BRI are grateful to the Indo-French
collaboration IFCPAR for its support during the initial stages of this work.

\appendix

\section{List of Symbols}\label{appA}

\begin{tabbing}

$\mathbf{a}$~~~~~~~~~~~~~~\= relative acceleration of binary masses in harmonic
coordinates\\
$\alpha_L^{\ell m}$\>STF tensor connecting the usual spherical
harmonics basis $Y^{\ell m}$ \\ \>~~~to the STF tensors basis $N_{\langle
L\rangle}$\\ 
$c_i$\>$\cos i$\\ 
$\Delta$\> mass difference ratio;~$\Delta=(m_1-m_2)/m$\\
$\mathcal{F}^\mathrm{GW}$\> total gravitational wave energy flux\\
FWF\> full gravitational waveform at 3PN order\\
$\gamma$\> PN parameter;~$\gamma = {G\,m\over r\,c^2}$\\
$h_{ij}^\mathrm{TT}$\> transverse traceless (TT) projection of metric
deviation; Eq. \eqref{hijTT}\\
$h_{+,\times}$\>``plus'' and ``cross'' polarisation states of the FWF;
Eqs. \eqref{hpc}\\
$H_{+,\times}$\> same as $h_{+,\times}$ modulo an overall factor;
Eqs. \eqref{HpcPN}, \eqref{Hplus} and \eqref{Hcross}\\
$h^{\ell m}$\> spin-weighted spherical harmonic modes of the FWF;
Eq. \eqref{decomp}\\
$H^{\ell m}$, $\hat{H}^{\ell\,m}$\> same as $h^{\ell m}$ modulo overall
factors; Eqs. \eqref{Hhat}\\
$I_L$\> mass-type source multipole moment STF with $\ell$ multipolar spatial
indices; \\ \>~~~ is given for $\ell=2,3,4,5,6,7,8$ by
Eqs.~\eqref{I3PN}, \eqref{I3},\\ \>~~~\eqref{I4}, \eqref{I5}, \eqref{I6},
\eqref{I7}, \eqref{I8} respectively\\
$i$\> inclination angle of the binary orbit\\ 
$J_L$\> current-type source multipole moment STF with $\ell$ multipolar spatial
indices; \\ \>~~~ is given for $\ell=2,3,4,5,6,7$ by
Eqs. \eqref{J2}, \eqref{J3}, \eqref{J4},\\ \>~~~ \eqref{J5}, \eqref{J6},
\eqref{J7} respectively \\
$\ell$\> multipolar order\\
$\bm{\lambda}$\> unit vector in the orbital plane;~$\bm{\lambda} =
\hat{\mathbf{z}}\times\mathbf{n}$\\
$m_1\,,m_2$\> individual masses of binary components\\ $m$\> total mass of the
binary; $m=m_1 + m_2$\\
$M$\> source mass-type monopole moment; Eq. \eqref{M0}\\
$M_L$\> canonical mass-type STF moment with $\ell$ multipolar spatial indices,
related to\\ \>~~~ source moment $I_L$ by Eqs. \eqref{MSa} and \eqref{M2},
\eqref{M3}\\
$\mathbf{n}$\> binary's separation direction, from $m_2$ to $m_1$\\
$\nu$\> symmetric mass-ratio; $\nu={m_1\,m_2\over m^2}$\\
$\mathbf{N}$\> direction of propagation of gravitational wave\\
$\mathbf{P}$\>unit vector along the direction of the ascending node
$\mathcal{N}$\\
$\mathcal{P}^\mathrm{TT}_{ijkl}$\> TT projection operator; Eq. \eqref{hijTT}\\
$\Phi$\> azimuthal angle of $\mathbf{N}$ in spherical polar
coordinates\\ 
$\phi(t)$\> orbital phase of the binary's relative orbit, the angle between
$\mathbf{n}$ and $\mathbf{P}$,\\ \>~~~ increasing in the direction of
$\bm{\lambda}$ \\
$\psi(t)$\> effective orbital phase of the binary's relative orbit, as
modified by \\ \>~~~ 4PN log term; Eq. \eqref{psix}\\
$\mathbf{Q}$\> unit polarization vector in the plane of the sky; $\mathbf{Q}
=\mathbf{N}\times \mathbf{P} $ \\
RWF\> restricted post-Newtonian gravitational waveform\\
$r$\> relative separation of binary masses in harmonic coordinates\\
$R$\> distance to the source in radiative coordinates\\
STF\> symmetric-trace-free projection\\
$s_i$\>$\sin i$\\
$S_L$\> canonical current-type STF moment with $\ell$ multipolar spatial
indices, related to\\ \>~~~ source moment $J_L$ by Eqs. \eqref{MSb} and
\eqref{S2}\\
$\Theta$\> polar angle of $\mathbf{N}$ in
spherical polar coordinates\\ 
TT\> transverse traceless projection\\ 
$U_L$\> mass-type radiative multipole moment STF with $\ell$ multipolar
spatial indices; \\ \>~~~ is given for $\ell=2,3,4,5$ by Eqs.~\eqref{U2},
\eqref{U3}, \eqref{U4}, \eqref{U5} respectively\\
$U^{\ell m}$\> radiative mass moment (non-STF form) corresponding to $h^{\ell
m}$; Eq.\eqref{UV}\\
$\mathbf{v}$\> relative velocity of binary masses in harmonic coordinates\\
$V_L$\> mass-type radiative multipole moment STF with $\ell$ multipolar
spatial indices;\\ \>~~~ is given for $\ell=2,3,4$ by Eqs.~\eqref{V2},
\eqref{V3}, \eqref{V4} respectively\\
$V^{\ell m}$\> radiative current moment (non-STF form) corresponding to
$h^{\ell m}$; Eq. \eqref{UV}\\
$\omega$\> angular velocity of the relative orbit in harmonic coordinates;
Eq. \eqref{omega3PN}\\
$W_L$\> gauge moment, entering the relation between canonical and source
moments; \\ \>~~~ Eqs. \eqref{M2}, \eqref{M3S2}\\
$X_L$\> gauge moment\\
$x$\>gauge invariant PN expansion parameter;~Eq. \eqref{x}\\
$x^\mu$\> harmonic coordinate
system in the near-zone\\ 
$X^\mu$\> radiative-type coordinate system in the
far-zone\\
$Y^{\ell m}_{-2}(\Theta,\Phi)$\> spin-weighted spherical harmonics of weight $-2$;
Eq. \eqref{harm}\\
$Y^{\ell m}(\Theta,\Phi)$\> standard spherical harmonics\\ 
$Y_L$\> gauge moment\\
$Z_L$\> gauge moment\\
$\hat{\mathbf{z}}$\> unit vector normal to the binary orbital
plane\\
\end{tabbing}

\bibliography{}

\begin{thebibliography}{70}
\expandafter\ifx\csname natexlab\endcsname\relax\def\natexlab#1{#1}\fi
\expandafter\ifx\csname bibnamefont\endcsname\relax
  \def\bibnamefont#1{#1}\fi
\expandafter\ifx\csname bibfnamefont\endcsname\relax
  \def\bibfnamefont#1{#1}\fi
\expandafter\ifx\csname citenamefont\endcsname\relax
  \def\citenamefont#1{#1}\fi
\expandafter\ifx\csname url\endcsname\relax
  \def\url#1{\texttt{#1}}\fi
\expandafter\ifx\csname urlprefix\endcsname\relax\def\urlprefix{URL }\fi
\providecommand{\bibinfo}[2]{#2}
\providecommand{\eprint}[2][]{\url{#2}}

\bibitem[{lig()}]{ligo}
\bibinfo{howpublished}{\url{http://www.ligo.caltech.edu}}.

\bibitem[{vir()}]{virgo}
\bibinfo{howpublished}{\url{http://www.virgo.infn.it}}.

\bibitem[{lis()}]{lisa}
\bibinfo{howpublished}{\url{http://lisa.jpl.nasa.gov}}.

\bibitem[{\citenamefont{Blanchet}(2002)}]{Bliving}
\bibinfo{author}{\bibfnamefont{L.}~\bibnamefont{Blanchet}},
  \bibinfo{journal}{Living Rev. Rel.} \textbf{\bibinfo{volume}{5}},
  \bibinfo{pages}{3} (\bibinfo{year}{2002}), \eprint{gr-qc/0202016}.

\bibitem[{\citenamefont{Pretorius}(2005)}]{Pretorius05}
\bibinfo{author}{\bibfnamefont{F.}~\bibnamefont{Pretorius}},
  \bibinfo{journal}{Phys. Rev. Lett.} \textbf{\bibinfo{volume}{95}},
  \bibinfo{pages}{121101} (\bibinfo{year}{2005}), \eprint{gr-qc/0507014}.

\bibitem[{\citenamefont{Baker et~al.}(2006)\citenamefont{Baker, Centrella,
  Choi, Koppitz, and van Meter}}]{Baker}
\bibinfo{author}{\bibfnamefont{J.~G.} \bibnamefont{Baker}},
  \bibinfo{author}{\bibfnamefont{J.}~\bibnamefont{Centrella}},
  \bibinfo{author}{\bibfnamefont{D.-I.} \bibnamefont{Choi}},
  \bibinfo{author}{\bibfnamefont{M.}~\bibnamefont{Koppitz}}, \bibnamefont{and}
  \bibinfo{author}{\bibfnamefont{J.}~\bibnamefont{van Meter}},
  \bibinfo{journal}{Phys. Rev. Lett.} \textbf{\bibinfo{volume}{96}},
  \bibinfo{pages}{111102} (\bibinfo{year}{2006}), \eprint{gr-qc/0511103}.

\bibitem[{\citenamefont{Campanelli et~al.}(2006)\citenamefont{Campanelli,
  Lousto, Marronetti, and Zlochower}}]{Campanelli}
\bibinfo{author}{\bibfnamefont{M.}~\bibnamefont{Campanelli}},
  \bibinfo{author}{\bibfnamefont{C.~O.} \bibnamefont{Lousto}},
  \bibinfo{author}{\bibfnamefont{P.}~\bibnamefont{Marronetti}},
  \bibnamefont{and}
  \bibinfo{author}{\bibfnamefont{Y.}~\bibnamefont{Zlochower}},
  \bibinfo{journal}{Phys. Rev. Lett.} \textbf{\bibinfo{volume}{96}},
  \bibinfo{pages}{111101} (\bibinfo{year}{2006}), \eprint{gr-qc/0511048}.

\bibitem[{\citenamefont{Baker et~al.}(2007)\citenamefont{Baker, Campanelli,
  Pretorius, and Zlochower}}]{BCPZ}
\bibinfo{author}{\bibfnamefont{J.}~\bibnamefont{Baker}},
  \bibinfo{author}{\bibfnamefont{M.}~\bibnamefont{Campanelli}},
  \bibinfo{author}{\bibfnamefont{F.}~\bibnamefont{Pretorius}},
  \bibnamefont{and}
  \bibinfo{author}{\bibfnamefont{Y.}~\bibnamefont{Zlochower}},
  \bibinfo{journal}{Class. Quant. Grav} \textbf{\bibinfo{volume}{24}},
  \bibinfo{pages}{S25} (\bibinfo{year}{2007}).

\bibitem[{\citenamefont{Buonanno et~al.}(2007)\citenamefont{Buonanno, Cook, and
  Pretorius}}]{BCP07}
\bibinfo{author}{\bibfnamefont{A.}~\bibnamefont{Buonanno}},
  \bibinfo{author}{\bibfnamefont{G.~B.} \bibnamefont{Cook}}, \bibnamefont{and}
  \bibinfo{author}{\bibfnamefont{F.}~\bibnamefont{Pretorius}},
  \bibinfo{journal}{Phys. Rev. D} \textbf{\bibinfo{volume}{75}},
  \bibinfo{pages}{124018} (\bibinfo{year}{2007}), \eprint{gr-qc/0610122}.

\bibitem[{\citenamefont{Berti et~al.}(2007)\citenamefont{Berti, Cardoso,
  Gonzalez, Sperhake, Hannam, Husa, and Bruegmann}}]{Berti}
\bibinfo{author}{\bibfnamefont{E.}~\bibnamefont{Berti}},
  \bibinfo{author}{\bibfnamefont{V.}~\bibnamefont{Cardoso}},
  \bibinfo{author}{\bibfnamefont{J.~A.} \bibnamefont{Gonzalez}},
  \bibinfo{author}{\bibfnamefont{U.}~\bibnamefont{Sperhake}},
  \bibinfo{author}{\bibfnamefont{M.}~\bibnamefont{Hannam}},
  \bibinfo{author}{\bibfnamefont{S.}~\bibnamefont{Husa}}, \bibnamefont{and}
  \bibinfo{author}{\bibfnamefont{B.}~\bibnamefont{Bruegmann}},
  \bibinfo{journal}{Phys. Rev. D} \textbf{\bibinfo{volume}{76}},
  \bibinfo{pages}{064034} (\bibinfo{year}{2007}).

\bibitem[{\citenamefont{Hannam et~al.}(2008)\citenamefont{Hannam, Husa,
  Sperhake, Brugmann, and Gonzalez}}]{Jena}
\bibinfo{author}{\bibfnamefont{M.}~\bibnamefont{Hannam}},
  \bibinfo{author}{\bibfnamefont{S.}~\bibnamefont{Husa}},
  \bibinfo{author}{\bibfnamefont{U.}~\bibnamefont{Sperhake}},
  \bibinfo{author}{\bibfnamefont{B.}~\bibnamefont{Brugmann}}, \bibnamefont{and}
  \bibinfo{author}{\bibfnamefont{J.~A.} \bibnamefont{Gonzalez}},
  \bibinfo{journal}{Phys. Rev. D} \textbf{\bibinfo{volume}{77}},
  \bibinfo{pages}{044020} (\bibinfo{year}{2008}), \eprint{arXiv:0706.1305}.

\bibitem[{\citenamefont{Boyle et~al.}(2007)}]{Boyle}
\bibinfo{author}{\bibfnamefont{M.}~\bibnamefont{Boyle}} \bibnamefont{et~al.},
  \bibinfo{journal}{Phys. Rev. D} \textbf{\bibinfo{volume}{76}},
  \bibinfo{pages}{124038} (\bibinfo{year}{2007}), \eprint{arXiv:0710.0158}.

\bibitem[{\citenamefont{Blanchet
  et~al.}(2002{\natexlab{a}})\citenamefont{Blanchet, Iyer, and Joguet}}]{BIJ02}
\bibinfo{author}{\bibfnamefont{L.}~\bibnamefont{Blanchet}},
  \bibinfo{author}{\bibfnamefont{B.~R.} \bibnamefont{Iyer}}, \bibnamefont{and}
  \bibinfo{author}{\bibfnamefont{B.}~\bibnamefont{Joguet}},
  \bibinfo{journal}{Phys. Rev. D} \textbf{\bibinfo{volume}{65}},
  \bibinfo{pages}{064005} (\bibinfo{year}{2002}{\natexlab{a}}),
  \bibinfo{note}{{Erratum-ibid~{\bf 71}, 129903(E) (2005)}},
  \eprint{gr-qc/0105098}.

\bibitem[{\citenamefont{Blanchet
  et~al.}(2002{\natexlab{b}})\citenamefont{Blanchet, Faye, Iyer, and
  Joguet}}]{BFIJ02}
\bibinfo{author}{\bibfnamefont{L.}~\bibnamefont{Blanchet}},
  \bibinfo{author}{\bibfnamefont{G.}~\bibnamefont{Faye}},
  \bibinfo{author}{\bibfnamefont{B.~R.} \bibnamefont{Iyer}}, \bibnamefont{and}
  \bibinfo{author}{\bibfnamefont{B.}~\bibnamefont{Joguet}},
  \bibinfo{journal}{Phys. Rev. D} \textbf{\bibinfo{volume}{65}},
  \bibinfo{pages}{061501(R)} (\bibinfo{year}{2002}{\natexlab{b}}),
  \bibinfo{note}{{Erratum-ibid~{\bf 71}, 129902(E) (2005)}},
  \eprint{gr-qc/0105099}.

\bibitem[{\citenamefont{Blanchet
  et~al.}(2004{\natexlab{a}})\citenamefont{Blanchet, Damour,
  Esposito-Far{\`e}se, and Iyer}}]{BDEI04}
\bibinfo{author}{\bibfnamefont{L.}~\bibnamefont{Blanchet}},
  \bibinfo{author}{\bibfnamefont{T.}~\bibnamefont{Damour}},
  \bibinfo{author}{\bibfnamefont{G.}~\bibnamefont{Esposito-Far{\`e}se}},
  \bibnamefont{and} \bibinfo{author}{\bibfnamefont{B.~R.} \bibnamefont{Iyer}},
  \bibinfo{journal}{Phys. Rev. Lett.} \textbf{\bibinfo{volume}{93}},
  \bibinfo{pages}{091101} (\bibinfo{year}{2004}{\natexlab{a}}),
  \eprint{gr-qc/0406012}.

\bibitem[{\citenamefont{Blanchet et~al.}(1996)\citenamefont{Blanchet, Iyer,
  Will, and Wiseman}}]{BIWW96}
\bibinfo{author}{\bibfnamefont{L.}~\bibnamefont{Blanchet}},
  \bibinfo{author}{\bibfnamefont{B.~R.} \bibnamefont{Iyer}},
  \bibinfo{author}{\bibfnamefont{C.~M.} \bibnamefont{Will}}, \bibnamefont{and}
  \bibinfo{author}{\bibfnamefont{A.~G.} \bibnamefont{Wiseman}},
  \bibinfo{journal}{Class. Quantum Grav.} \textbf{\bibinfo{volume}{13}},
  \bibinfo{pages}{575} (\bibinfo{year}{1996}), \eprint{gr-qc/9602024}.

\bibitem[{\citenamefont{Arun et~al.}(2004)\citenamefont{Arun, Blanchet, Iyer,
  and Qusailah}}]{ABIQ04}
\bibinfo{author}{\bibfnamefont{K.~G.} \bibnamefont{Arun}},
  \bibinfo{author}{\bibfnamefont{L.}~\bibnamefont{Blanchet}},
  \bibinfo{author}{\bibfnamefont{B.~R.} \bibnamefont{Iyer}}, \bibnamefont{and}
  \bibinfo{author}{\bibfnamefont{M.~S.~S.} \bibnamefont{Qusailah}},
  \bibinfo{journal}{Class. Quantum Grav.} \textbf{\bibinfo{volume}{21}},
  \bibinfo{pages}{3771} (\bibinfo{year}{2004}), \bibinfo{note}{erratum-ibid.
  {\bf 22}, 3115 (2005)}, \eprint{gr-qc/0404185}.

\bibitem[{\citenamefont{Kidder et~al.}(2007)\citenamefont{Kidder, Blanchet, and
  Iyer}}]{KBI07}
\bibinfo{author}{\bibfnamefont{L.~E.} \bibnamefont{Kidder}},
  \bibinfo{author}{\bibfnamefont{L.}~\bibnamefont{Blanchet}}, \bibnamefont{and}
  \bibinfo{author}{\bibfnamefont{B.~R.} \bibnamefont{Iyer}},
  \bibinfo{journal}{Class. Quant. Grav.} \textbf{\bibinfo{volume}{24}},
  \bibinfo{pages}{5307} (\bibinfo{year}{2007}), \eprint{arXiv:0706.0726}.

\bibitem[{\citenamefont{Kidder}(2008)}]{K07}
\bibinfo{author}{\bibfnamefont{L.~E.} \bibnamefont{Kidder}},
  \bibinfo{journal}{Phys. Rev. D} \textbf{\bibinfo{volume}{D77}},
  \bibinfo{pages}{044016} (\bibinfo{year}{2008}), \eprint{arXiv:0710.0614}.

\bibitem[{\citenamefont{Cutler et~al.}(1993)\citenamefont{Cutler, Apostolatos,
  Bildsten, Finn, Flanagan, Kennefick, Markovic, Ori, Poisson, Sussman
  et~al.}}]{3mn}
\bibinfo{author}{\bibfnamefont{C.}~\bibnamefont{Cutler}},
  \bibinfo{author}{\bibfnamefont{T.}~\bibnamefont{Apostolatos}},
  \bibinfo{author}{\bibfnamefont{L.}~\bibnamefont{Bildsten}},
  \bibinfo{author}{\bibfnamefont{L.}~\bibnamefont{Finn}},
  \bibinfo{author}{\bibfnamefont{E.}~\bibnamefont{Flanagan}},
  \bibinfo{author}{\bibfnamefont{D.}~\bibnamefont{Kennefick}},
  \bibinfo{author}{\bibfnamefont{D.}~\bibnamefont{Markovic}},
  \bibinfo{author}{\bibfnamefont{A.}~\bibnamefont{Ori}},
  \bibinfo{author}{\bibfnamefont{E.}~\bibnamefont{Poisson}},
  \bibinfo{author}{\bibfnamefont{G.}~\bibnamefont{Sussman}},
  \bibnamefont{et~al.}, \bibinfo{journal}{Phys. Rev. Lett.}
  \textbf{\bibinfo{volume}{70}}, \bibinfo{pages}{2984} (\bibinfo{year}{1993}).

\bibitem[{\citenamefont{Cutler and Flanagan}(1994)}]{CF94}
\bibinfo{author}{\bibfnamefont{C.}~\bibnamefont{Cutler}} \bibnamefont{and}
  \bibinfo{author}{\bibfnamefont{E.}~\bibnamefont{Flanagan}},
  \bibinfo{journal}{Phys. Rev. D} \textbf{\bibinfo{volume}{49}},
  \bibinfo{pages}{2658} (\bibinfo{year}{1994}).

\bibitem[{\citenamefont{Sathyaprakash}(1994)}]{SathyaFilter94}
\bibinfo{author}{\bibfnamefont{B.~S.} \bibnamefont{Sathyaprakash}},
  \bibinfo{journal}{Phys. Rev. D} \textbf{\bibinfo{volume}{50}},
  \bibinfo{pages}{R7111} (\bibinfo{year}{1994}).

\bibitem[{\citenamefont{Poisson and Will}(1995)}]{PW95}
\bibinfo{author}{\bibfnamefont{E.}~\bibnamefont{Poisson}} \bibnamefont{and}
  \bibinfo{author}{\bibfnamefont{C.}~\bibnamefont{Will}},
  \bibinfo{journal}{Phys. Rev. D} \textbf{\bibinfo{volume}{52}},
  \bibinfo{pages}{848} (\bibinfo{year}{1995}).

\bibitem[{\citenamefont{Balasubramanian
  et~al.}(1996)\citenamefont{Balasubramanian, Sathyaprakash, and
  Dhurandhar}}]{BalSatDhu96}
\bibinfo{author}{\bibfnamefont{R.}~\bibnamefont{Balasubramanian}},
  \bibinfo{author}{\bibfnamefont{B.~S.} \bibnamefont{Sathyaprakash}},
  \bibnamefont{and} \bibinfo{author}{\bibfnamefont{S.~V.}
  \bibnamefont{Dhurandhar}}, \bibinfo{journal}{Phys.~Rev.~D}
  \textbf{\bibinfo{volume}{53}}, \bibinfo{pages}{3033} (\bibinfo{year}{1996}),
  \bibinfo{note}{erratum-ibid.~D {\bf 54}, 1860 (1996)},
  \eprint{gr-qc/9508011}.

\bibitem[{\citenamefont{Damour et~al.}(1998)\citenamefont{Damour, Iyer, and
  Sathyaprakash}}]{DIS98}
\bibinfo{author}{\bibfnamefont{T.}~\bibnamefont{Damour}},
  \bibinfo{author}{\bibfnamefont{B.~R.} \bibnamefont{Iyer}}, \bibnamefont{and}
  \bibinfo{author}{\bibfnamefont{B.~S.} \bibnamefont{Sathyaprakash}},
  \bibinfo{journal}{Phys. Rev. D} \textbf{\bibinfo{volume}{57}},
  \bibinfo{pages}{885} (\bibinfo{year}{1998}), \eprint{gr-qc/9708034}.

\bibitem[{\citenamefont{Buonanno et~al.}(2003)\citenamefont{Buonanno, Chen, and
  Vallisneri}}]{BCV03a}
\bibinfo{author}{\bibfnamefont{A.}~\bibnamefont{Buonanno}},
  \bibinfo{author}{\bibfnamefont{Y.}~\bibnamefont{Chen}}, \bibnamefont{and}
  \bibinfo{author}{\bibfnamefont{M.}~\bibnamefont{Vallisneri}},
  \bibinfo{journal}{Phys. Rev. D} \textbf{\bibinfo{volume}{67}},
  \bibinfo{pages}{024016} (\bibinfo{year}{2003}), \bibinfo{note}{erratum-ibid.
  {\bf D}~74, 029903(E) (2006)}, \eprint{gr-qc/0205122}.

\bibitem[{\citenamefont{Hellings and Moore}(2003)}]{HM03}
\bibinfo{author}{\bibfnamefont{R.~W.} \bibnamefont{Hellings}} \bibnamefont{and}
  \bibinfo{author}{\bibfnamefont{T.~A.} \bibnamefont{Moore}},
  \bibinfo{journal}{Class.~Quant.~Grav.} \textbf{\bibinfo{volume}{20}},
  \bibinfo{pages}{S181} (\bibinfo{year}{2003}), \eprint{gr-qc/0207102}.

\bibitem[{\citenamefont{Moore and Hellings}(2002)}]{MH02}
\bibinfo{author}{\bibfnamefont{T.~A.} \bibnamefont{Moore}} \bibnamefont{and}
  \bibinfo{author}{\bibfnamefont{R.~W.} \bibnamefont{Hellings}},
  \bibinfo{journal}{Phys. Rev.~D} \textbf{\bibinfo{volume}{65}},
  \bibinfo{pages}{062001} (\bibinfo{year}{2002}), \eprint{gr-qc/9910116}.

\bibitem[{\citenamefont{Sintes and Vecchio}(2000)}]{SinVecc00a}
\bibinfo{author}{\bibfnamefont{A.~M.} \bibnamefont{Sintes}} \bibnamefont{and}
  \bibinfo{author}{\bibfnamefont{A.}~\bibnamefont{Vecchio}}, in
  \emph{\bibinfo{booktitle}{Rencontres de Moriond: Gravitational waves and
  experimental gravity}}, edited by
  \bibinfo{editor}{\bibfnamefont{J.}~\bibnamefont{Dumarchez}}
  (\bibinfo{publisher}{Fronti\`eres, Paris}, \bibinfo{year}{2000}),
  \eprint{gr-qc/0005058}.

\bibitem[{\citenamefont{{Sintes} and {Vecchio}}(2000)}]{SinVecc00b}
\bibinfo{author}{\bibfnamefont{A.~M.} \bibnamefont{{Sintes}}} \bibnamefont{and}
  \bibinfo{author}{\bibfnamefont{A.}~\bibnamefont{{Vecchio}}}, in
  \emph{\bibinfo{booktitle}{Third Amaldi conference on Gravitational Waves}},
  edited by \bibinfo{editor}{\bibfnamefont{S.}~\bibnamefont{{Meshkov}}}
  (\bibinfo{publisher}{American Institute of Physics Conference Series, New
  York}, \bibinfo{year}{2000}), p. \bibinfo{pages}{403},
  \eprint{gr-qc/0005059}.

\bibitem[{\citenamefont{Van Den~Broeck}(2006)}]{Chris06}
\bibinfo{author}{\bibfnamefont{C.}~\bibnamefont{Van Den~Broeck}},
  \bibinfo{journal}{Class.~Quantum Grav.} \textbf{\bibinfo{volume}{23}},
  \bibinfo{pages}{L51} (\bibinfo{year}{2006}), \eprint{gr-qc/0604032}.

\bibitem[{\citenamefont{Van Den~Broeck and Sengupta}(2007)}]{ChrisAnand06}
\bibinfo{author}{\bibfnamefont{C.}~\bibnamefont{Van Den~Broeck}}
  \bibnamefont{and} \bibinfo{author}{\bibfnamefont{A.}~\bibnamefont{Sengupta}},
  \bibinfo{journal}{Class.~Quantum Grav.} \textbf{\bibinfo{volume}{24}},
  \bibinfo{pages}{155} (\bibinfo{year}{2007}), \eprint{gr-qc/0607092}.

\bibitem[{\citenamefont{{Arun} et~al.}(2007)\citenamefont{{Arun}, {Iyer},
  {Sathyaprakash}, and {Sinha}}}]{AISS07}
\bibinfo{author}{\bibfnamefont{K.~G.} \bibnamefont{{Arun}}},
  \bibinfo{author}{\bibfnamefont{B.~R.} \bibnamefont{{Iyer}}},
  \bibinfo{author}{\bibfnamefont{B.~S.} \bibnamefont{{Sathyaprakash}}},
  \bibnamefont{and} \bibinfo{author}{\bibfnamefont{S.}~\bibnamefont{{Sinha}}},
  \bibinfo{journal}{Phys.~Rev.~{\bf D}} \textbf{\bibinfo{volume}{75}},
  \bibinfo{pages}{124002} (\bibinfo{year}{2007}), \eprint{arXiv:0704.1086
  [gr-qc]}.

\bibitem[{\citenamefont{Holz and Hughes}(2005)}]{HolzHugh05}
\bibinfo{author}{\bibfnamefont{D.~E.} \bibnamefont{Holz}} \bibnamefont{and}
  \bibinfo{author}{\bibfnamefont{S.~A.} \bibnamefont{Hughes}},
  \bibinfo{journal}{Astrophys.~J} \textbf{\bibinfo{volume}{629}},
  \bibinfo{pages}{15} (\bibinfo{year}{2005}), \eprint{astro-ph/0504616}.

\bibitem[{\citenamefont{Dalal et~al.}(2006)\citenamefont{Dalal, Holz, Hughes,
  and Jain}}]{DaHHJ06}
\bibinfo{author}{\bibfnamefont{N.}~\bibnamefont{Dalal}},
  \bibinfo{author}{\bibfnamefont{D.~E.} \bibnamefont{Holz}},
  \bibinfo{author}{\bibfnamefont{S.~A.} \bibnamefont{Hughes}},
  \bibnamefont{and} \bibinfo{author}{\bibfnamefont{B.}~\bibnamefont{Jain}},
  \bibinfo{journal}{Phys. Rev. D} \textbf{\bibinfo{volume}{74}},
  \bibinfo{pages}{063006} (\bibinfo{year}{2006}), \eprint{astro-ph/0601275}.

\bibitem[{\citenamefont{Arun et~al.}(2007)\citenamefont{Arun, Iyer,
  Sathyaprakash, Sinha, and Broeck}}]{AISSV}
\bibinfo{author}{\bibfnamefont{K.~G.} \bibnamefont{Arun}},
  \bibinfo{author}{\bibfnamefont{B.~R.} \bibnamefont{Iyer}},
  \bibinfo{author}{\bibfnamefont{B.~S.} \bibnamefont{Sathyaprakash}},
  \bibinfo{author}{\bibfnamefont{S.}~\bibnamefont{Sinha}}, \bibnamefont{and}
  \bibinfo{author}{\bibfnamefont{C.~V.~D.} \bibnamefont{Broeck}},
  \bibinfo{journal}{Phys. Rev.} \textbf{\bibinfo{volume}{D76}},
  \bibinfo{pages}{104016} (\bibinfo{year}{2007}),
  \bibinfo{note}{erratum-ibid.~{\bf 76}, 129903 (2007)},
  \eprint{arXiv:0707.3920 [astro-ph]}.

\bibitem[{\citenamefont{Trias and Sintes}(2008)}]{Trias}
\bibinfo{author}{\bibfnamefont{M.}~\bibnamefont{Trias}} \bibnamefont{and}
  \bibinfo{author}{\bibfnamefont{A.~M.} \bibnamefont{Sintes}},
  \bibinfo{journal}{Phys. Rev. D} \textbf{\bibinfo{volume}{77}},
  \bibinfo{pages}{024030} (\bibinfo{year}{2008}), \eprint{arXiv:0707.4434v1
  [gr-qc]}.

\bibitem[{\citenamefont{Blanchet and Damour}(1986)}]{BD86}
\bibinfo{author}{\bibfnamefont{L.}~\bibnamefont{Blanchet}} \bibnamefont{and}
  \bibinfo{author}{\bibfnamefont{T.}~\bibnamefont{Damour}},
  \bibinfo{journal}{Phil. Trans. Roy. Soc. Lond. A}
  \textbf{\bibinfo{volume}{320}}, \bibinfo{pages}{379} (\bibinfo{year}{1986}).

\bibitem[{\citenamefont{Blanchet}(1987)}]{B87}
\bibinfo{author}{\bibfnamefont{L.}~\bibnamefont{Blanchet}},
  \bibinfo{journal}{Proc. Roy. Soc. Lond. A} \textbf{\bibinfo{volume}{409}},
  \bibinfo{pages}{383} (\bibinfo{year}{1987}).

\bibitem[{\citenamefont{Blanchet and Damour}(1992)}]{BD92}
\bibinfo{author}{\bibfnamefont{L.}~\bibnamefont{Blanchet}} \bibnamefont{and}
  \bibinfo{author}{\bibfnamefont{T.}~\bibnamefont{Damour}},
  \bibinfo{journal}{Phys. Rev. D} \textbf{\bibinfo{volume}{46}},
  \bibinfo{pages}{4304} (\bibinfo{year}{1992}).

\bibitem[{\citenamefont{Blanchet}(1995)}]{B95}
\bibinfo{author}{\bibfnamefont{L.}~\bibnamefont{Blanchet}},
  \bibinfo{journal}{Phys. Rev. D} \textbf{\bibinfo{volume}{51}},
  \bibinfo{pages}{2559} (\bibinfo{year}{1995}), \eprint{gr-qc/9501030}.

\bibitem[{\citenamefont{Blanchet}(1996)}]{B96}
\bibinfo{author}{\bibfnamefont{L.}~\bibnamefont{Blanchet}},
  \bibinfo{journal}{Phys. Rev. D} \textbf{\bibinfo{volume}{54}},
  \bibinfo{pages}{1417} (\bibinfo{year}{1996}),
  \bibinfo{note}{{Erratum-ibid.{\bf 71}, 129904(E) (2005)}},
  \eprint{gr-qc/9603048}.

\bibitem[{\citenamefont{Blanchet}(1998{\natexlab{a}})}]{B98mult}
\bibinfo{author}{\bibfnamefont{L.}~\bibnamefont{Blanchet}},
  \bibinfo{journal}{Class. Quantum Grav.} \textbf{\bibinfo{volume}{15}},
  \bibinfo{pages}{1971} (\bibinfo{year}{1998}{\natexlab{a}}),
  \eprint{gr-qc/9801101}.

\bibitem[{\citenamefont{Will and Wiseman}(1996)}]{WWi96}
\bibinfo{author}{\bibfnamefont{C.}~\bibnamefont{Will}} \bibnamefont{and}
  \bibinfo{author}{\bibfnamefont{A.}~\bibnamefont{Wiseman}},
  \bibinfo{journal}{Phys. Rev. D} \textbf{\bibinfo{volume}{54}},
  \bibinfo{pages}{4813} (\bibinfo{year}{1996}).

\bibitem[{\citenamefont{Pati and Will}(2000)}]{PW00}
\bibinfo{author}{\bibfnamefont{M.~E.} \bibnamefont{Pati}} \bibnamefont{and}
  \bibinfo{author}{\bibfnamefont{C.~M.} \bibnamefont{Will}},
  \bibinfo{journal}{Phys. Rev. D} \textbf{\bibinfo{volume}{62}},
  \bibinfo{pages}{124015} (\bibinfo{year}{2000}), \eprint{gr-qc/0007087}.

\bibitem[{\citenamefont{Pati and Will}(2002)}]{PW02}
\bibinfo{author}{\bibfnamefont{M.~E.} \bibnamefont{Pati}} \bibnamefont{and}
  \bibinfo{author}{\bibfnamefont{C.~M.} \bibnamefont{Will}},
  \bibinfo{journal}{Phys. Rev. D} \textbf{\bibinfo{volume}{65}},
  \bibinfo{pages}{104008} (\bibinfo{year}{2002}), \eprint{gr-qc/0201001}.

\bibitem[{\citenamefont{Thorne}(1980)}]{Th80}
\bibinfo{author}{\bibfnamefont{K.}~\bibnamefont{Thorne}},
  \bibinfo{journal}{Rev. Mod. Phys.} \textbf{\bibinfo{volume}{52}},
  \bibinfo{pages}{299} (\bibinfo{year}{1980}).

\bibitem[{\citenamefont{Blanchet et~al.}(1995)\citenamefont{Blanchet, Damour,
  and Iyer}}]{BDI95}
\bibinfo{author}{\bibfnamefont{L.}~\bibnamefont{Blanchet}},
  \bibinfo{author}{\bibfnamefont{T.}~\bibnamefont{Damour}}, \bibnamefont{and}
  \bibinfo{author}{\bibfnamefont{B.~R.} \bibnamefont{Iyer}},
  \bibinfo{journal}{Phys. Rev. D} \textbf{\bibinfo{volume}{51}},
  \bibinfo{pages}{5360} (\bibinfo{year}{1995}), \eprint{gr-qc/9501029}.

\bibitem[{\citenamefont{Blanchet and Iyer}(2004)}]{BI04mult}
\bibinfo{author}{\bibfnamefont{L.}~\bibnamefont{Blanchet}} \bibnamefont{and}
  \bibinfo{author}{\bibfnamefont{B.~R.} \bibnamefont{Iyer}},
  \bibinfo{journal}{Phys. Rev. D} \textbf{\bibinfo{volume}{71}},
  \bibinfo{pages}{024004} (\bibinfo{year}{2004}), \eprint{gr-qc/0409094}.

\bibitem[{\citenamefont{Blanchet}(1998{\natexlab{b}})}]{B98tail}
\bibinfo{author}{\bibfnamefont{L.}~\bibnamefont{Blanchet}},
  \bibinfo{journal}{Class. Quantum Grav.} \textbf{\bibinfo{volume}{15}},
  \bibinfo{pages}{113} (\bibinfo{year}{1998}{\natexlab{b}}),
  \bibinfo{note}{erratum-ibid~{\bf 22}, 3381 (2005)}, \eprint{gr-qc/9710038}.

\bibitem[{\citenamefont{Blanchet}(1998{\natexlab{c}})}]{B98quad}
\bibinfo{author}{\bibfnamefont{L.}~\bibnamefont{Blanchet}},
  \bibinfo{journal}{Class. Quantum Grav.} \textbf{\bibinfo{volume}{15}},
  \bibinfo{pages}{89} (\bibinfo{year}{1998}{\natexlab{c}}),
  \eprint{gr-qc/9710037}.

\bibitem[{\citenamefont{Christodoulou}(1991)}]{Chr91}
\bibinfo{author}{\bibfnamefont{D.}~\bibnamefont{Christodoulou}},
  \bibinfo{journal}{Phys. Rev. Lett.} \textbf{\bibinfo{volume}{67}},
  \bibinfo{pages}{1486} (\bibinfo{year}{1991}).

\bibitem[{\citenamefont{Wiseman and Will}(1991)}]{WiW91}
\bibinfo{author}{\bibfnamefont{A.}~\bibnamefont{Wiseman}} \bibnamefont{and}
  \bibinfo{author}{\bibfnamefont{C.}~\bibnamefont{Will}},
  \bibinfo{journal}{Phys. Rev. D} \textbf{\bibinfo{volume}{44}},
  \bibinfo{pages}{R2945} (\bibinfo{year}{1991}).

\bibitem[{\citenamefont{Thorne}(1992)}]{Th92}
\bibinfo{author}{\bibfnamefont{K.}~\bibnamefont{Thorne}},
  \bibinfo{journal}{Phys. Rev. D} \textbf{\bibinfo{volume}{45}},
  \bibinfo{pages}{520} (\bibinfo{year}{1992}).

\bibitem[{\citenamefont{Mart\'in-Garc\'ia}(2002)}]{xtensor}
\bibinfo{author}{\bibnamefont{Mart\'in-Garc\'ia}} (\bibinfo{year}{2002}),
  \bibinfo{note}{http://metric.iem.csic.es/Martin-Garcia/xAct/}.

\bibitem[{\citenamefont{Blanchet and Faye}(2000)}]{BF00}
\bibinfo{author}{\bibfnamefont{L.}~\bibnamefont{Blanchet}} \bibnamefont{and}
  \bibinfo{author}{\bibfnamefont{G.}~\bibnamefont{Faye}},
  \bibinfo{journal}{Phys. Lett. A} \textbf{\bibinfo{volume}{271}},
  \bibinfo{pages}{58} (\bibinfo{year}{2000}), \eprint{gr-qc/0004009}.

\bibitem[{\citenamefont{Blanchet and Faye}(2001)}]{BFeom}
\bibinfo{author}{\bibfnamefont{L.}~\bibnamefont{Blanchet}} \bibnamefont{and}
  \bibinfo{author}{\bibfnamefont{G.}~\bibnamefont{Faye}},
  \bibinfo{journal}{Phys. Rev. D} \textbf{\bibinfo{volume}{63}},
  \bibinfo{pages}{062005} (\bibinfo{year}{2001}), \eprint{gr-qc/0007051}.

\bibitem[{\citenamefont{Blanchet
  et~al.}(2004{\natexlab{b}})\citenamefont{Blanchet, Damour, and
  Esposito-Far{\`e}se}}]{BDE04}
\bibinfo{author}{\bibfnamefont{L.}~\bibnamefont{Blanchet}},
  \bibinfo{author}{\bibfnamefont{T.}~\bibnamefont{Damour}}, \bibnamefont{and}
  \bibinfo{author}{\bibfnamefont{G.}~\bibnamefont{Esposito-Far{\`e}se}},
  \bibinfo{journal}{Phys. Rev. D} \textbf{\bibinfo{volume}{69}},
  \bibinfo{pages}{124007} (\bibinfo{year}{2004}{\natexlab{b}}),
  \eprint{gr-qc/0311052}.

\bibitem[{\citenamefont{MacMahon}(1960)}]{MacMahon}
\bibinfo{author}{\bibfnamefont{M.~A.} \bibnamefont{MacMahon}},
  \emph{\bibinfo{title}{Combinatory Analysis}} (\bibinfo{publisher}{Cambridge
  University Press, Cambridge, England}, \bibinfo{year}{1960}), vol.
  \bibinfo{volume}{I, II}.

\bibitem[{\citenamefont{Itoh et~al.}(2001)\citenamefont{Itoh, Futamase, and
  Asada}}]{IFA01}
\bibinfo{author}{\bibfnamefont{Y.}~\bibnamefont{Itoh}},
  \bibinfo{author}{\bibfnamefont{T.}~\bibnamefont{Futamase}}, \bibnamefont{and}
  \bibinfo{author}{\bibfnamefont{H.}~\bibnamefont{Asada}},
  \bibinfo{journal}{Phys. Rev. D} \textbf{\bibinfo{volume}{63}},
  \bibinfo{pages}{064038} (\bibinfo{year}{2001}), \eprint{gr-qc/0101114}.

\bibitem[{\citenamefont{Itoh and Futamase}(2003)}]{itoh1}
\bibinfo{author}{\bibfnamefont{Y.}~\bibnamefont{Itoh}} \bibnamefont{and}
  \bibinfo{author}{\bibfnamefont{T.}~\bibnamefont{Futamase}},
  \bibinfo{journal}{Phys. Rev. D} \textbf{\bibinfo{volume}{68}},
  \bibinfo{pages}{121501(R)} (\bibinfo{year}{2003}), \eprint{gr-qc/0310028}.

\bibitem[{\citenamefont{Itoh}(2004)}]{itoh2}
\bibinfo{author}{\bibfnamefont{Y.}~\bibnamefont{Itoh}}, \bibinfo{journal}{Phys.
  Rev. D} \textbf{\bibinfo{volume}{69}}, \bibinfo{pages}{064018}
  (\bibinfo{year}{2004}), \eprint{gr-qc/0310029}.

\bibitem[{\citenamefont{Jaranowski and Sch{\"a}fer}(1998)}]{JaraS98}
\bibinfo{author}{\bibfnamefont{P.}~\bibnamefont{Jaranowski}} \bibnamefont{and}
  \bibinfo{author}{\bibfnamefont{G.}~\bibnamefont{Sch{\"a}fer}},
  \bibinfo{journal}{Phys. Rev. D} \textbf{\bibinfo{volume}{57}},
  \bibinfo{pages}{7274} (\bibinfo{year}{1998}), \eprint{gr-qc/9712075}.

\bibitem[{\citenamefont{Jaranowski and Sch\"afer}(1999)}]{JS99}
\bibinfo{author}{\bibfnamefont{P.}~\bibnamefont{Jaranowski}} \bibnamefont{and}
  \bibinfo{author}{\bibfnamefont{G.}~\bibnamefont{Sch\"afer}},
  \bibinfo{journal}{Phys. Rev. D} \textbf{\bibinfo{volume}{60}},
  \bibinfo{pages}{124003} (\bibinfo{year}{1999}), \eprint{gr-qc/9906092}.

\bibitem[{\citenamefont{Damour et~al.}(2001)\citenamefont{Damour, Jaranowski,
  and Sch\"afer}}]{DJSdim}
\bibinfo{author}{\bibfnamefont{T.}~\bibnamefont{Damour}},
  \bibinfo{author}{\bibfnamefont{P.}~\bibnamefont{Jaranowski}},
  \bibnamefont{and}
  \bibinfo{author}{\bibfnamefont{G.}~\bibnamefont{Sch\"afer}},
  \bibinfo{journal}{Phys. Lett. B} \textbf{\bibinfo{volume}{513}},
  \bibinfo{pages}{147} (\bibinfo{year}{2001}), \eprint{gr-qc/0105038}.

\bibitem[{\citenamefont{Gradshteyn and Ryzhik}(1980)}]{GZ}
\bibinfo{author}{\bibfnamefont{I.}~\bibnamefont{Gradshteyn}} \bibnamefont{and}
  \bibinfo{author}{\bibfnamefont{I.}~\bibnamefont{Ryzhik}},
  \emph{\bibinfo{title}{Table of Integrals, Series and Products}}
  (\bibinfo{publisher}{Academic Press}, \bibinfo{year}{1980}).

\bibitem[{\citenamefont{Blanchet and Sch{\"a}fer}(1993)}]{BS93}
\bibinfo{author}{\bibfnamefont{L.}~\bibnamefont{Blanchet}} \bibnamefont{and}
  \bibinfo{author}{\bibfnamefont{G.}~\bibnamefont{Sch{\"a}fer}},
  \bibinfo{journal}{Class. Quantum Grav.} \textbf{\bibinfo{volume}{10}},
  \bibinfo{pages}{2699} (\bibinfo{year}{1993}).

\bibitem[{\citenamefont{Sasaki}(1994)}]{Sasa94}
\bibinfo{author}{\bibfnamefont{M.}~\bibnamefont{Sasaki}},
  \bibinfo{journal}{Prog. Theor. Phys.} \textbf{\bibinfo{volume}{92}},
  \bibinfo{pages}{17} (\bibinfo{year}{1994}).

\bibitem[{\citenamefont{Tagoshi and Sasaki}(1994)}]{TSasa94}
\bibinfo{author}{\bibfnamefont{H.}~\bibnamefont{Tagoshi}} \bibnamefont{and}
  \bibinfo{author}{\bibfnamefont{M.}~\bibnamefont{Sasaki}},
  \bibinfo{journal}{Prog. Theor. Phys.} \textbf{\bibinfo{volume}{92}},
  \bibinfo{pages}{745} (\bibinfo{year}{1994}), \eprint{gr-qc/9405062}.

\bibitem[{\citenamefont{Tanaka et~al.}(1996)\citenamefont{Tanaka, Tagoshi, and
  Sasaki}}]{TTS96}
\bibinfo{author}{\bibfnamefont{T.}~\bibnamefont{Tanaka}},
  \bibinfo{author}{\bibfnamefont{H.}~\bibnamefont{Tagoshi}}, \bibnamefont{and}
  \bibinfo{author}{\bibfnamefont{M.}~\bibnamefont{Sasaki}},
  \bibinfo{journal}{Prog. Theor. Phys.} \textbf{\bibinfo{volume}{96}},
  \bibinfo{pages}{1087} (\bibinfo{year}{1996}), \eprint{gr-qc/9701050}.

\end{thebibliography}

\end{document}